\documentclass[twocolumn]{aastex}
\usepackage{style_paper}
\usepackage{soul}
\begin{document}

\title{exoALMA XXI: The Morphology and Dynamics of Vertical Flows}

\newcommand{\InstMPIA}{Max-Planck Institute for Astronomy (MPIA), Königstuhl 17, 69117 Heidelberg, Germany}
\newcommand{\InstMIT}{Department of Earth, Atmospheric, and Planetary Sciences, Massachusetts Institute of Technology, Cambridge, MA 02139, USA}
\newcommand{\InstOCA}{Université Côte d’Azur, Observatoire de la Côte d’Azur, CNRS, Laboratoire Lagrange, France}
\newcommand{\InstMilano}{Dipartimento di Fisica, Università degli Studi di Milano, Via Celoria 16, 20133 Milano, Italy}
\newcommand{\InstNAOJ}{National Astronomical Observatory of Japan, Osawa 2-21-1, Mitaka, Tokyo 181-8588, Japan}
\newcommand{\InstAIJ}{Astronomical Institute, Graduate School of Science, Tohoku University, 6-3 Aoba, Aramaki, Aoba-ku, Sendai, Miyagi 980-8578 Japan}
\newcommand{\InstIPAGGrenoble}{Univ. Grenoble Alpes, CNRS, IPAG, 38000 Grenoble, France}
\newcommand{\InstMonash}{School of Physics and Astronomy, Monash University, Clayton VIC 3800, Australia}
\newcommand{\InstCfA}{Center for Astrophysics | Harvard \& Smithsonian, Cambridge, MA 02138, USA}
\newcommand{\InstFlorida}{Department of Astronomy, University of Florida, Gainesville, FL 32611, USA}
\newcommand{\InstChile}{Departamento de Astronomía, Universidad de Chile, Camino El Observatorio 1515, Las Condes, Santiago, Chile}
\newcommand{\InstStAndrewsPhysics}{School of Physics \& Astronomy, University of St. Andrews, North Haugh, St. Andrews KY16 9SS, UK}
\newcommand{\InstStAndrewsExoplanets}{Centre for Exoplanet Science, University of St. Andrews, North Haugh, St. Andrews, KY16 9SS, UK}
\newcommand{\InstRicePhysics}{Department of Physics and Astronomy, Rice University, Houston, TX 77005, USA}
\newcommand{\InstLANL}{Los Alamos National Laboratory, Los Alamos, NM 87545, USA}
\newcommand{\InstUGAphysics}{Department of Physics and Astronomy, The University of Georgia, Athens, GA 30602, USA}
\newcommand{\InstUGACSP}{Center for Simulational Physics, The University of Georgia, Athens, GA 30602, USA}
\newcommand{\InstUGAIA}{Institute for Artificial Intelligence, The University of Georgia, Athens, GA, 30602, USA}
\newcommand{\InstColumbia}{Department of Astronomy, Columbia University, 538 W. 120th Street, Pupin Hall, New York, NY, USA}
\newcommand{\InstLeeds}{School of Physics and Astronomy, University of Leeds, Leeds, UK, LS2 9JT}
\newcommand{\InstRiceSpace}{Rice Space Institute, Rice University, 6100 Main St, Houston, TX 77005, USA}
\newcommand{\InstLeiden}{Leiden Observatory, Leiden University, P.O. Box 9513, NL-2300 RA Leiden, The Netherlands}
\newcommand{\InstESO}{European Southern Observatory, Karl-Schwarzschild-Str. 2, D-85748 Garching bei München, Germany}
\newcommand{\InstNHFP}{NASA Hubble Fellowship Program Sagan Fellow}
\newcommand{\InstIbaraki}{College of Science, Ibaraki University, 2-1-1 Bunkyo, Mito, Ibaraki 310-8512, Japan}
\newcommand{\InstCambridge}{Institute of Astronomy, University of Cambridge, Madingley Road, CB3 0HA, Cambridge, UK}
\newcommand{\InstNRAO}{National Radio Astronomy Observatory, 520 Edgemont Rd., Charlottesville, VA 22903, USA}
\newcommand{\InstUNAM}{Instituto de Ciencias Físicas, Universidad Nacional Autónoma de México, Av. Universidad s/n, 62210 Cuernavaca, Mor., Mexico}
\newcommand{\InstBologna}{Alma Mater Studiorum Università di Bologna, Dipartimento di Fisica e Astronomia (DIFA), Via Gobetti 93/2, 40129 Bologna, Italy}
\newcommand{\InstArcetri}{INAF-Osservatorio Astrofisico di Arcetri, Largo E. Fermi 5, 50125 Firenze, Italy}
\newcommand{\InstASIAA}{Academia Sinica Institute of Astronomy \& Astrophysics, 11F of Astronomy-Mathematics Building, AS/NTU, No.1, Sec. 4, Roosevelt Rd, Taipei 10617, Taiwan}
\newcommand{\InstWesleyan}{Department of Astronomy, Van Vleck Observatory, Wesleyan University, 96 Foss Hill Drive, Middletown, CT 06459, USA}
\newcommand{\InstPennState}{Department of Astronomy \& Astrophysics, 525 Davey Laboratory, The Pennsylvania State University, University Park, PA 16802, USA}
\newcommand{\InstSOKENDAI}{Department of Astronomical Science, The Graduate University for Advanced Studies, SOKENDAI, 2-21-1 Osawa, Mitaka, Tokyo 181-8588, Japan}
\newcommand{\InstQMUL}{Astronomy Unit, School of Physics and Astronomy, Queen Mary University of London, London E1 4NS, UK}
\newcommand{\InstTokyo}{Department of Astronomy, Graduate School of Science, The University of Tokyo, 7-3-1 Hongo, Bunkyo-ku, Tokyo 113-0033, Japan}

\author[0000-0002-7695-7605]{Myriam Benisty}
\affiliation{\InstMPIA}
\author[0000-0001-8446-3026]{Andrés F. Izquierdo} 
\affiliation{\InstFlorida}
\affiliation{\InstNHFP}
\author[0000-0002-0491-143X]{Jochen Stadler} 
\affiliation{\InstOCA}
\affiliation{\InstESO}
\author[0000-0002-5503-5476]{Maria Galloway-Sprietsma}
\affiliation{\InstFlorida}
\author[0000-0003-4689-2684]{Stefano Facchini}
\affiliation{\InstMilano}
\author[0000-0002-7501-9801]{Andrew J. Winter}
\affiliation{\InstQMUL}
\author[0000-0001-7258-770X]{Jaehan Bae}
\affiliation{\InstFlorida}
\author[0000-0003-1117-9213]{Misato Fukagawa} 
\affiliation{\InstNAOJ}
\affiliation{\InstAIJ}
\author[0000-0003-1534-5186]{Richard Teague}
\affiliation{\InstMIT}
\author[0000-0001-5907-5179]{Christophe Pinte}
\affiliation{\InstIPAGGrenoble}
\author[0009-0000-7872-3493]{Sean M. Andrews}
\affiliation{\InstCfA}
\author[0000-0001-6378-7873]{Marcelo Barraza-Alfaro}
\affiliation{\InstMIT}
\author[0000-0002-2700-9676]{Gianni Cataldi} 
\affiliation{\InstNAOJ}
\author[0000-0003-2045-2154]{Pietro Curone} 
\affiliation{\InstChile}
\author[0000-0002-1483-8811]{Ian Czekala}
\affiliation{\InstStAndrewsPhysics}
\author[0000-0003-4679-4072]{Daniele Fasano} 
\affiliation{\InstOCA}
\author[0000-0002-9298-3029]{Mario Flock} 
\affiliation{\InstMPIA}
\author[0000-0002-5910-4598]{Himanshi Garg}
\affiliation{\InstMonash}
\author[0000-0001-6947-6072]{Jane Huang} 
\affiliation{\InstColumbia}
\author[0000-0003-1008-1142]{John D. Ilee} 
\affiliation{\InstLeeds}
\author[0000-0001-7235-2417]{Kazuhiro Kanagawa} 
\affiliation{\InstIbaraki}
\author[0009-0007-5371-3548]{Jensen Lawrence}
\affiliation{\InstMIT}
\author[0000-0002-8896-9435]{Geoffroy Lesur} 
\affiliation{\InstIPAGGrenoble}
\author[0000-0002-2357-7692]{Giuseppe Lodato} 
\affiliation{\InstMilano}
\author[0000-0003-4663-0318]{Cristiano Longarini} 
\affiliation{\InstCambridge}
\affiliation{\InstMilano}
\author[0000-0002-8932-1219]{Ryan A. Loomis}
\affiliation{\InstNRAO}
\author[0000-0002-1637-7393]{François M\'{e}nard} 
\affiliation{\InstIPAGGrenoble}
\author[0000-0003-4039-8933]{Ryuta Orihara} 
\affiliation{\InstTokyo}
\author[0000-0002-4716-4235]{Daniel J. Price} 
\affiliation{\InstMonash}

\author[0000-0003-4853-5736]{Giovanni Rosotti} 
\affiliation{\InstMilano}

\author[0000-0002-3468-9577]{Gaylor Wafflard-Fernandez} 
\affiliation{\InstIPAGGrenoble}

\author[0000-0003-1526-7587]{David J. Wilner} 
\affiliation{\InstCfA}
\author[0000-0002-7212-2416]{Lisa Wölfer} 
\affiliation{\InstMIT}

\author[0000-0003-1412-893X]{Hsi-Wei Yen} 
\affiliation{\InstASIAA}

\author[0000-0001-8002-8473]{Tomohiro C. Yoshida} 
\affiliation{\InstNAOJ}
\affiliation{\InstSOKENDAI}


\author[0000-0001-9319-1296]{Brianna Zawadzki} 
\affiliation{\InstWesleyan}


\begin{abstract}
Vertical gas flows, such as winds and meridional circulations, are natural outcomes of protoplanetary disk processes and play a critical role in the earliest stages of planet formation. We analyze vertical gas motions in 14 disks as part of the exoALMA Large Program, focusing on the \twCOfull{} and \thCOfull{} emission lines. Using \discminer{} to model the Keplerian velocity field, we extract line-of-sight velocity residuals and measure the radial and vertical components of the gas motion. Vertical motions are detected in most disks. Two types of patterns emerge: (1) oscillatory up/down flows, likely linked to instabilities, and (2) transitions from downward to upward motions that we interpret as the base of a disk wind. In most cases, the velocity amplitudes are of a few tens of m/s. Two disks, however, MWC758 and CQ Tau, show two spiral velocity features in their residual maps, red- and blue-shifted, which we interpret as vertical velocities reaching up to $\sim$350 m/s ($\sim$0.7$c_s$), consistent with gas motion in eccentric disks. Fast upward motions (up to 500 m/s; $\sim$1.8$c_s$) is also detected in the outer disk of MWC758. Synthetic observations from (magneto)hydrodynamic simulations validate the reliability of our method. Although strong molecular winds appear to be relatively rare in \twCO{} and \thCO{}, our study shows that, when traced by deep high spectral resolution line data, protoplanetary disks exhibit ubiquitous vertical flows.  However, their overall velocity structure is highly complex, preventing to identify 
a coherent, dominant physical mechanism driving the vertical motions across all disks, thus requiring further theoretical investigation. 
\end{abstract}



\section{Introduction} \label{sec:intro}
The gas dynamics in a protoplanetary disk is dominated by  Keplerian rotation around the central young star, and such a velocity structure has been used to identify disks themselves \citep{Koerner1993,Guilloteau1994}. Keplerian rotation has also routinely been used to obtain dynamical mass estimates for the central star and determine global disk properties \citep{Dutrey1998,Isella2007,Andrews2024}. In recent years, deviations from Keplerian rotation were proposed as a method to identify companions or planets still embedded in their host disks \citep{Teague_ea_2018b, Pinte_ea_2018b} as a range of non-Keplerian motions, including vertical and radial components, can naturally emerge from planet–disk interactions \citep[e.g.,][]{Bae2025}. Once the Keplerian contribution is removed from the disk velocity field, the velocity residuals can be decomposed in three azimuthally symmetric orthogonal velocity components (azimuthal, vertical and radial velocities).  Azimuthal velocity residuals can for example  result from  pressure variations, locally (e.g., gaps) or globally (pressure drop-off in outer disk, self-gravity) \citep[e.g.,][]{Stadler2025, Longarini2025}. 

Vertical motions in protoplanetary disks can be induced by various processes that result from the local combined effects of gravity, pressure, turbulence and magnetic fields. For example, while the vertical shear instability will result
in clear upward and downward motions, the magneto-rotational
instability will induce uncorrelated motions \citep{Flock2015,Barraza2024,Barraza2025}. Winds, of magneto-hydrodynamical (MHD) or photo-evaporative (PE) nature, naturally arise depending on the strength of the disk magnetic field and the level of stellar activity. MHD winds play a fundamental role in the evolution of protoplanetary disks as they regulate angular momentum and allow material to accrete onto the star \citep{Lesur2023,Hu2024}. PE winds in turn play an important role in the late-stage dispersal of disk material \citep{Alexander2014}. Both have a dramatic impact on the timescale and local conditions for planet formation and migration \citep{Wafflard2025,Hu2025}. Upward and downward flows such as meridional circulation flows are expected to occur in planetary gaps \citep{Morbidelli2014,Lega2024} and to deliver the material that will constitute the planetary atmospheres and define its chemical complexity \citep{Cridland2025}. 
All of these processes overall directly affect the disk thermal and chemical structure, dust mixing, grain growth and settling, and can lead to substructure formation \citep[e.g.,][]{Riols2020}. 

Observational evidence for winds launched in regions closed to the star has accumulated for decades at optical and infrared wavelengths \citep{Pascucci2023,Pascucci2025}. This is particularly clear in the forbidden lines that show a low-velocity blue-shifted emission \citep[e.g.,][]{Pascucci2020}. Additional evidence for winds was found in the CO fundamental M-band emission line that shows broadened line wings and low velocity component \citep{Bast2011,Banzatti2022}. At millimeter wavelengths, large scale outflows are also routinely found at the Class I/II stage \citep{tabone2017,Fernandez2020,Bacciotti2025}, sometimes with clear evidence of co-rotation with the disk  \citep{Louvet2018,devalon2020,tabone2020}. 

Evidence for disk winds in Class II objects are, on the other hand, rather sparse at these wavelengths, even though models incorporating MHD wind-driven accretion better match the observed disk properties than turbulent models \citep{Tabone2025}. Recent  high angular and spectral resolution observations of disks with substructures allowed for vertical motions to be mapped. In HD100546, \citet{Casassus_ea_2022} found a so-called 'Doppler flip' (local sign change in velocity residuals) located in the dust continuum emission  \citep[see also, ][]{Norfolk2022}. Using multiple CO isotopologues, \citet{Teague_ea_2019b}, \citet{Yu2021}, \citet{Izquierdo_ea_2023} and \citet{Galloway2023} found strong upward flows (so-called 'meridional fountain') in gas gaps located at distances larger than 100 au in two Herbig disks, HD163296 and HD169142, and the T Tauri disk, AS209. Similarly, \citet{Yu2021} identified a vertical upward flow at 125\,au in the outer disk of the Herbig star HD169142, beyond its main dust structures.  

Interestingly, \cite{Teague2021} and \citet{Izquierdo_ea_2023} found both upward and downward vertical motions of a few tens of m/s in MWC480 in the \twCOmaps{} line while the other isotopologues tracing lower emitting heights did not show similar behaviors. 

In this paper, we expand such an analysis by considering the \twCOfull{} and \thCOfull{} emission line observations for 14 out of the 15 disks in the exoALMA program (AA Tau being excluded from the analysis, see below). exoALMA constitutes an unprecedented dataset specifically designed for detailed study of kinematics \citep{Teague2025}. The paper is structured as follows: we first present the data that we use (Sect.\,\ref{sec:data}) and the method employed to extract vertical motions (Sect.\,\ref{sec:extraction}) that we benchmark in Section \ref{sec:benchmark}. In  Sect.\,
\ref{sec:results}, we report the vertical motions measured in an axisymmetric analysis for 11 targets, and those extracted along spiral spines for 3 targets. We discuss our findings in the Discussion section (Sect.\,\ref{sec:discussion}), and conclude.

\begin{figure*}[t]
    \centering
    \includegraphics[width=1.0\linewidth]{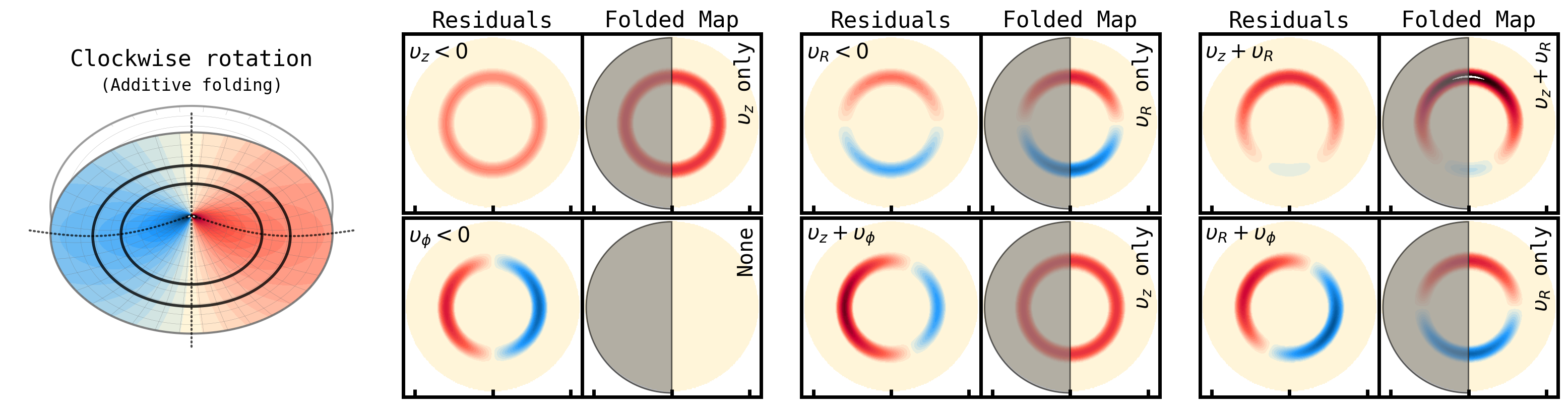}  
    \caption{Sketch explaining the types of velocity residuals expected after subtraction of a Keplerian disk model, in 2D residual maps (labeled 'residuals') and in folded residual maps (labeled 'folded'). A more comprehensive sketch encompassing more cases can be found in exoALMA XX/Izquierdo et al. subm.}
    \label{fig:sketch}
\end{figure*}

\section{Data} \label{sec:data}
We refer to \cite{Loomis2025} for the details of the calibration and imaging of the exoALMA observations. In this paper, we focus on the \twCOfull{} and \thCOfull{} continuum-subtracted  cubes. Similar to \citet{Stadler2025}, we consider cubes with different angular resolutions depending on the signal-to-noise ratio. We focus mainly  on the so-called `Fiducial Images', with beam size 0.15\arcsec{} and channel spacing of 100 m/s, and on the `High Surface Brightness Sensitivity Images' (beam size 0.30\arcsec{}, channel spacing 100 m/s; see \cite{Teague2025} for details). As we are mostly interested in following the large scale features in this paper, we favor high signal to noise ratio at the cost of angular resolution when selecting the imaging setup. We report the cube angular resolution for all targets in the Appendix (Table~\ref{tab:asymmfactor}). As shown in \citet{Stadler2025}, using cubes at different angular resolution from the exoALMA images does not  significantly affect the overall velocity extraction. 
We exclude data within the central beam from the analysis. 

We also consider three \twCOmaps{} data cubes for HD163296, MWC480 and AS209 from the MAPS Large Program \citep{oberg2021}, with 0.15\arcsec{} angular resolution and 200 m/s channel spacing, for which kinematical analysis was previously published in \citet{Izquierdo_ea_2023}. In addition, and only for the purpose of discussing the targets' systemic velocities, we include the \CSfull{} cubes from exoALMA, with 0.3\arcsec{} and 200 m/s spatial and spectral resolutions, as well as the \eiCO{} cubes from MAPS with similar properties as the \twCO{} cubes.  The full analysis of the \CSfull{} cubes will appear in a forthcoming publication from the collaboration (Cataldi et al. in prep).

\section{Method} \label{sec:extraction}
\subsection{Velocity extraction}
The line-of-sight velocity can be expressed in terms of its three orthogonal velocity components, \vphi{}, \vrad{} and \vz{}, defined in the disk frame of reference. 
\begin{equation}
    \upsilon_{\rm los}=\upsilon_{\phi, \text { proj }} + \upsilon_{r, \text { proj }} + \upsilon_{z, \text { proj }} + \upsilon_{\rm sys} \label{eq:v_los}
\end{equation}
with: 
\begin{equation} 
     \upsilon_{\phi, \text { proj }}=\upsilon_\phi \cos (\phi) \sin (i) \cdot sgn_\mathrm{rot} \label{eq:vphi_los}
\end{equation} 
\begin{equation}
    \upsilon_{r, \text { proj }}=-\upsilon_r \sin (\phi) \sin (i) \label{eq:vrad_los}
\end{equation}
\begin{equation}
    \upsilon_{z, \text { proj }}=-\upsilon_z \cos (i) \label{eq:vz_los}
\end{equation}

and $\upsilon_{\rm sys}$ the systemic velocity, $\phi$ the azimuthal angle of the disk measured anticlockwise from the red-shifted major axis, ${i\in[-90\degr, 90\degr]}$ the disk inclination, and $sgn_\mathrm{rot}$ denotes the sign of the disk rotation direction, with clockwise rotation defined as positive  \citep[see ][]{Izquierdo2025}. Following these equations, an axisymmetric azimuthal (radial) velocity perturbation will appear in a 2D $\upsilon_{\rm los}$ map as two symmetric arcs with opposite velocity signs relative to the projected minor (major) axis of the disk, while an axisymmetric vertical perturbation will manifest as a full annulus across all disk axes without changing signs. Figure\,\ref{fig:sketch}, left panels, illustrate these three examples (\vz{} only, \vphi{} only, \vrad{} only).

We follow the procedure outlined in \citet{Izquierdo2025} and use the \discminer code that fits a  Keplerian disk model to the channel maps of a line intensity cube, assuming radial parametric forms for the peak intensity, line width, and emission surface height of the disk \citep{Izquierdo_ea_2021}. In the analysis, we consider only smooth disks with no warped regions. As in all exoALMA papers, the geometrical parameters and the systemic velocity used in the \thCO{} models are fixed to those derived from the \twCO{} analysis. The best fit models for all targets are presented in \citet{Izquierdo2025}. At each location in the disk, the line-of-sight velocity is estimated by fitting the line profile with either a Gaussian, a Bell, or a double-Bell function, depending on whether there is significant contribution from the back side of the disk to the observed emission. A Keplerian model is then subtracted from the data to produce a 2D residual $\upsilon_{\rm los}$ map, encoding velocity fluctuations in all three dimensions: ${\delta\upsilon_{\phi}}$, \vrad{} and \vz{}. An analysis of the radial ${\delta\upsilon_{\phi}}$ profiles and their relation to gas pressure substructures can be found in \citet{Stadler2025}. 

The analysis of the velocity components have been commonly performed following a 1D axisymmetric extraction \citep[e.g.,][]{Teague_ea_2019b, Yu2021, Galloway2023}. Radial profiles of the azimuthal and vertical components are obtained by assuming that they are axisymmetric and orthogonal \citep[for more details, see][]{Izquierdo_ea_2023}. Assuming azimuthal symmetry of the azimuthal and radial velocity components, the vertical velocity component is simply the azimuthal average of the residual $\upsilon_{\rm los}$ map divided by $-sec(i)$ \citep{Izquierdo_ea_2022}. 
The radial velocities are obtained through an azimuthal average of all remaining velocity residuals that include \vrad{} as well as the remaining non-axisymmetric contributions from ${\delta\upsilon_{\phi}}$ and \vz{}. 
However, as noted in \citet{Izquierdo_ea_2023}, when velocity perturbations strongly deviate from axisymmetry, such decomposition methods fail at properly retrieving the right amplitude for each velocity component. This motivates two different approaches for the axisymmetric and non-axisymmetric residuals that we will explore in the following sections. 


\begin{figure*}[t]
    \centering
    \begin{tabular}{cccc}
        \includegraphics[width=0.25\textwidth]{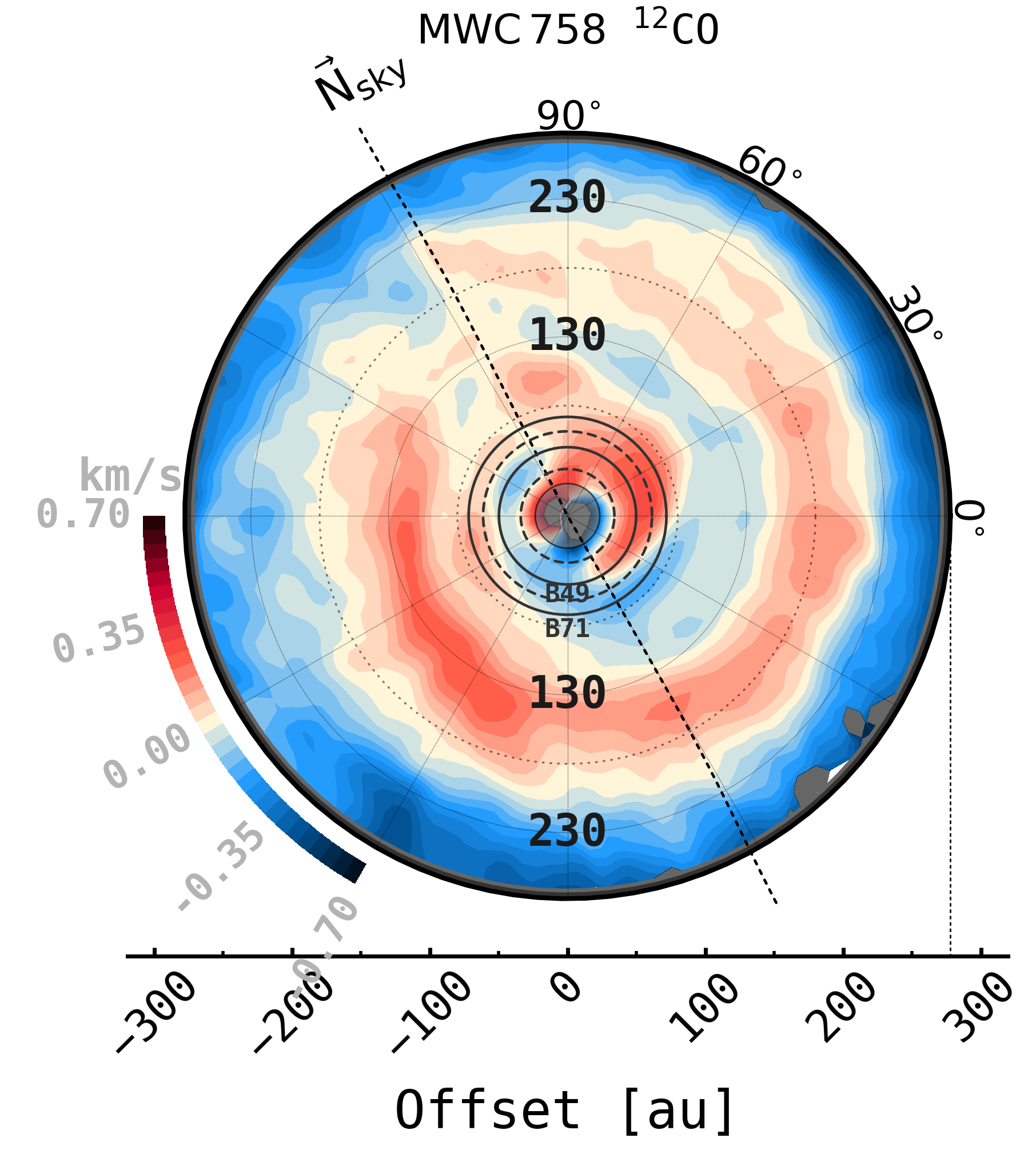} & 
        \includegraphics[width=0.25\textwidth]{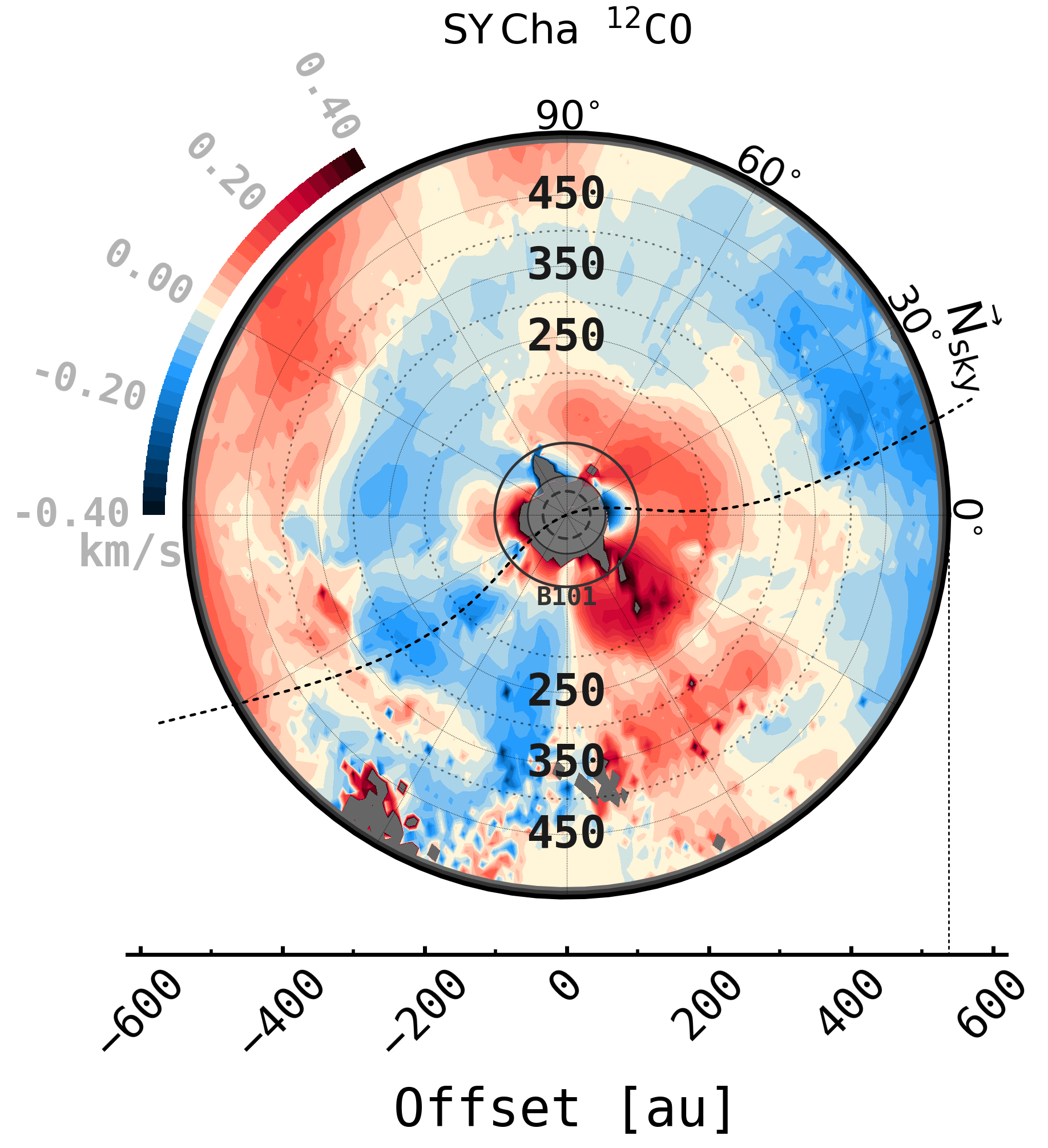} & 
        \includegraphics[width=0.25\textwidth]{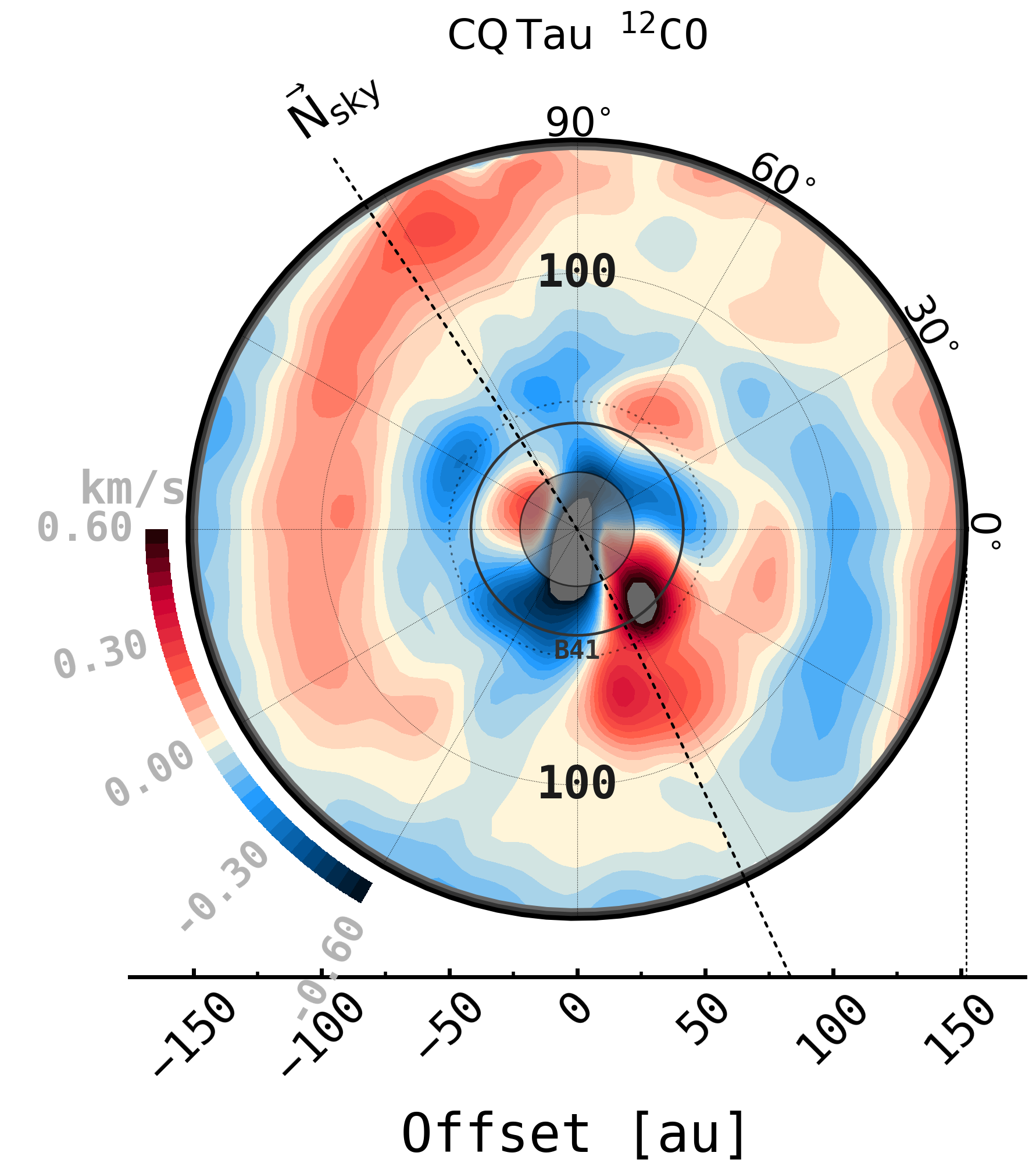} &
        \includegraphics[width=0.25\textwidth]{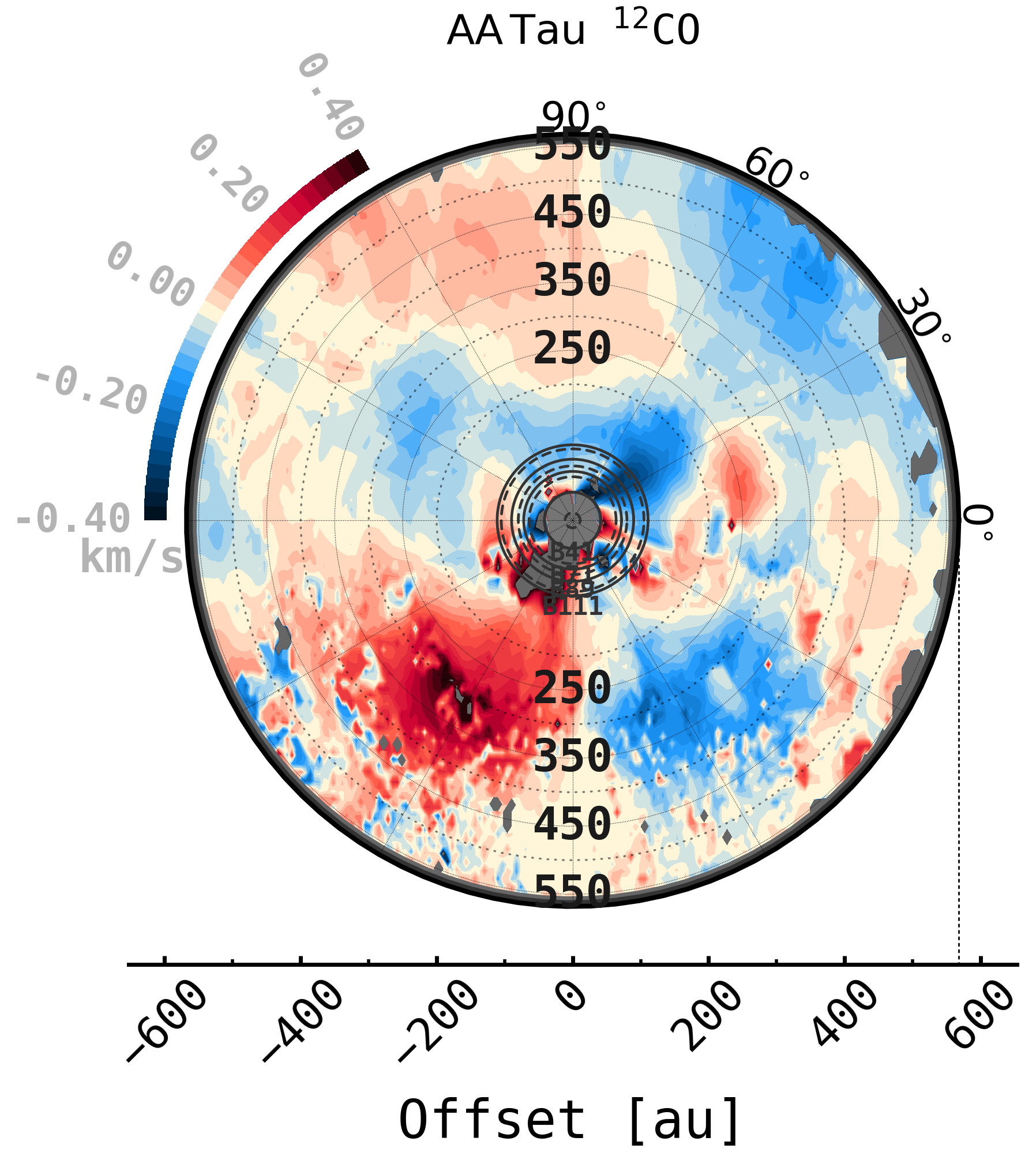} \\
        \includegraphics[width=0.25\textwidth]{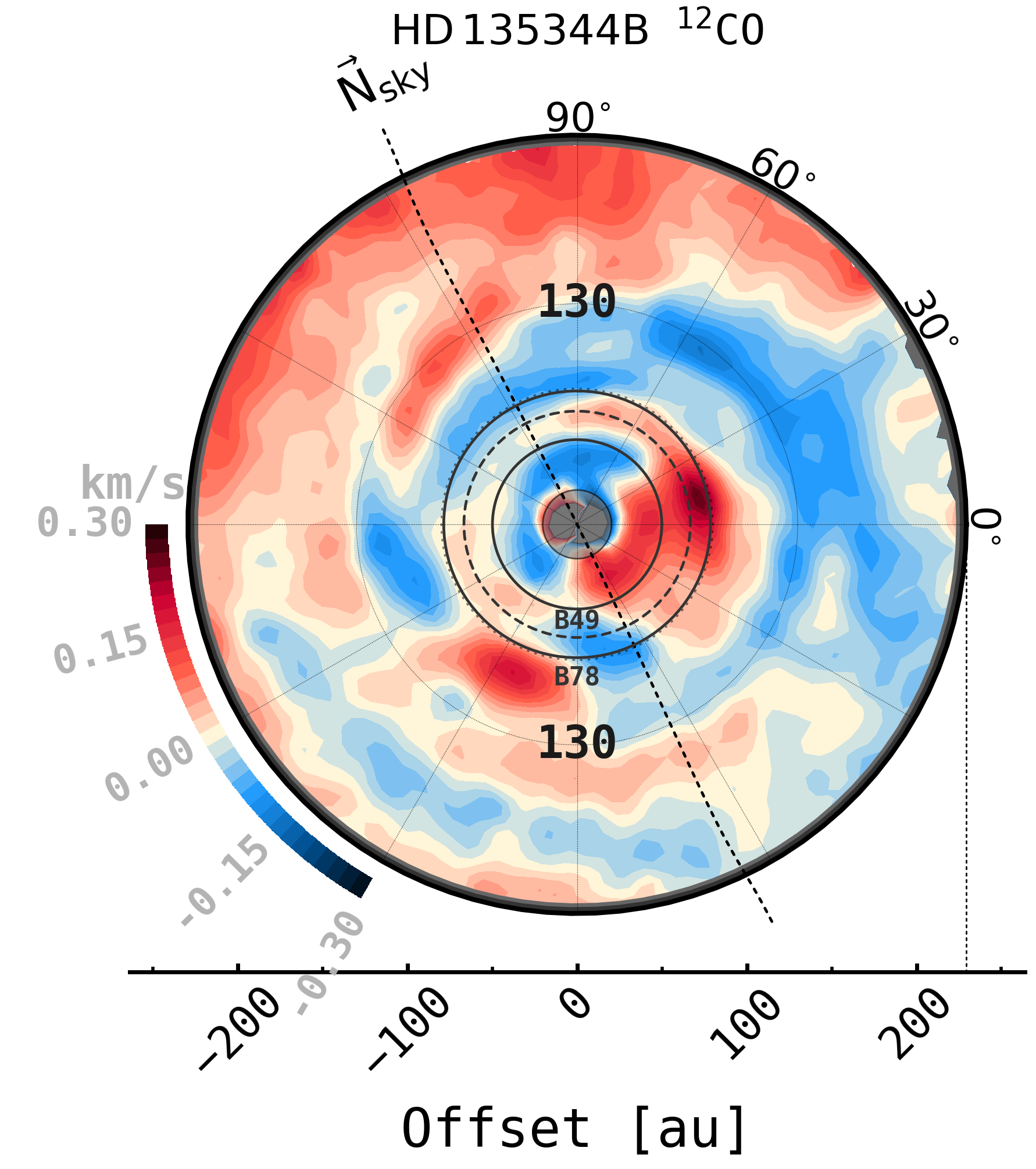} & 
        \includegraphics[width=0.25\textwidth]{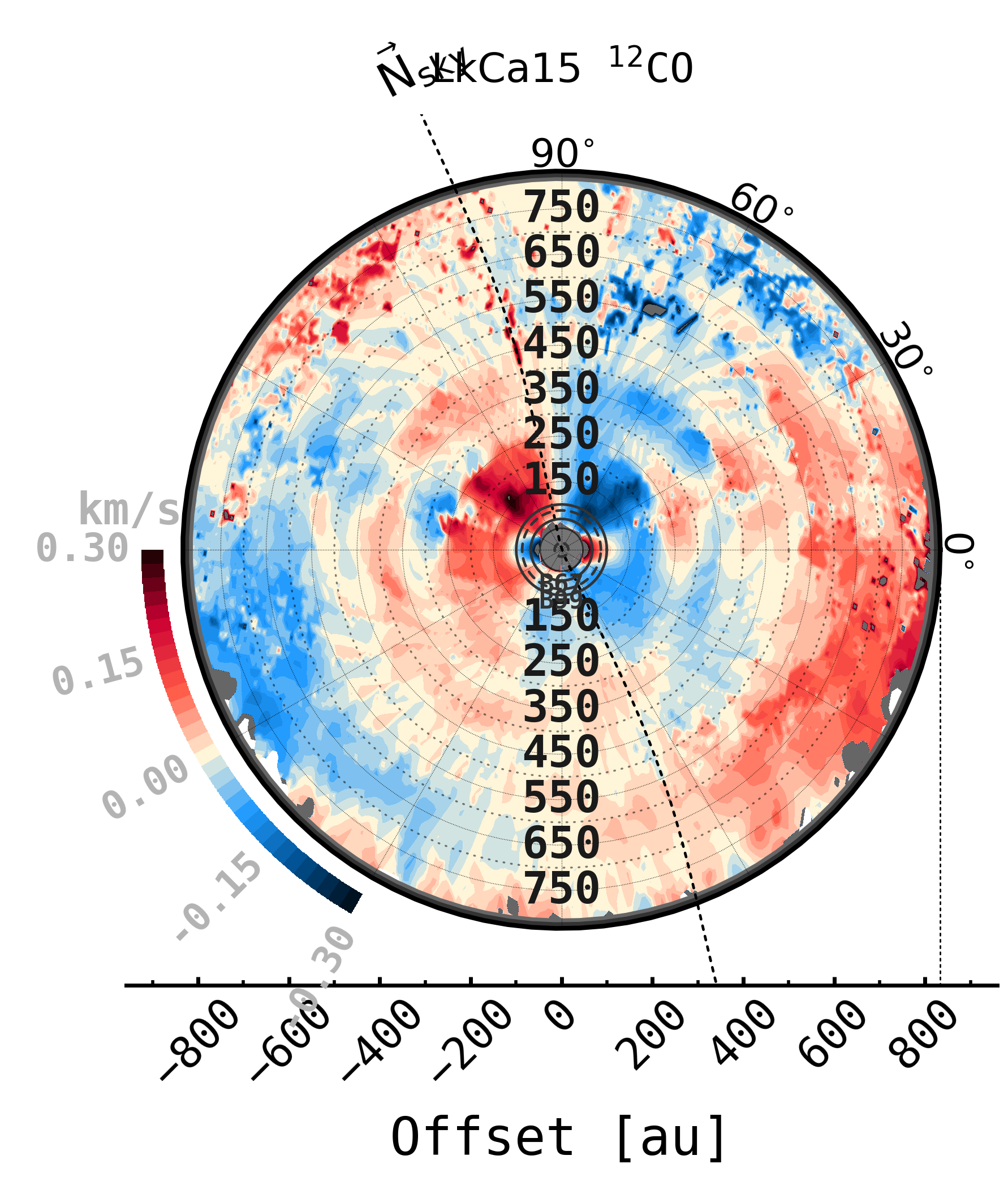} &  
        \includegraphics[width=0.25\textwidth]{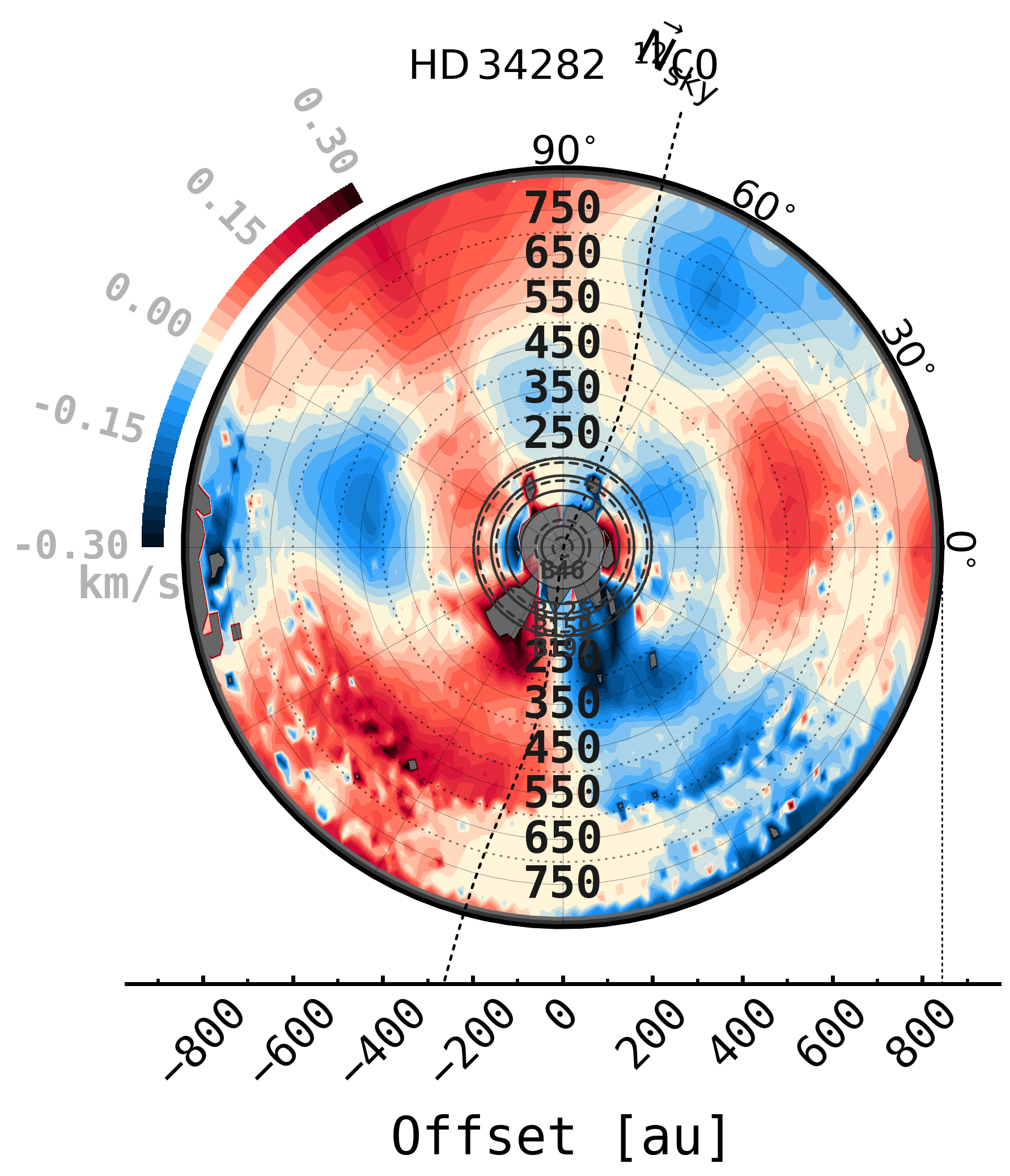} & 
        \includegraphics[width=0.25\textwidth]{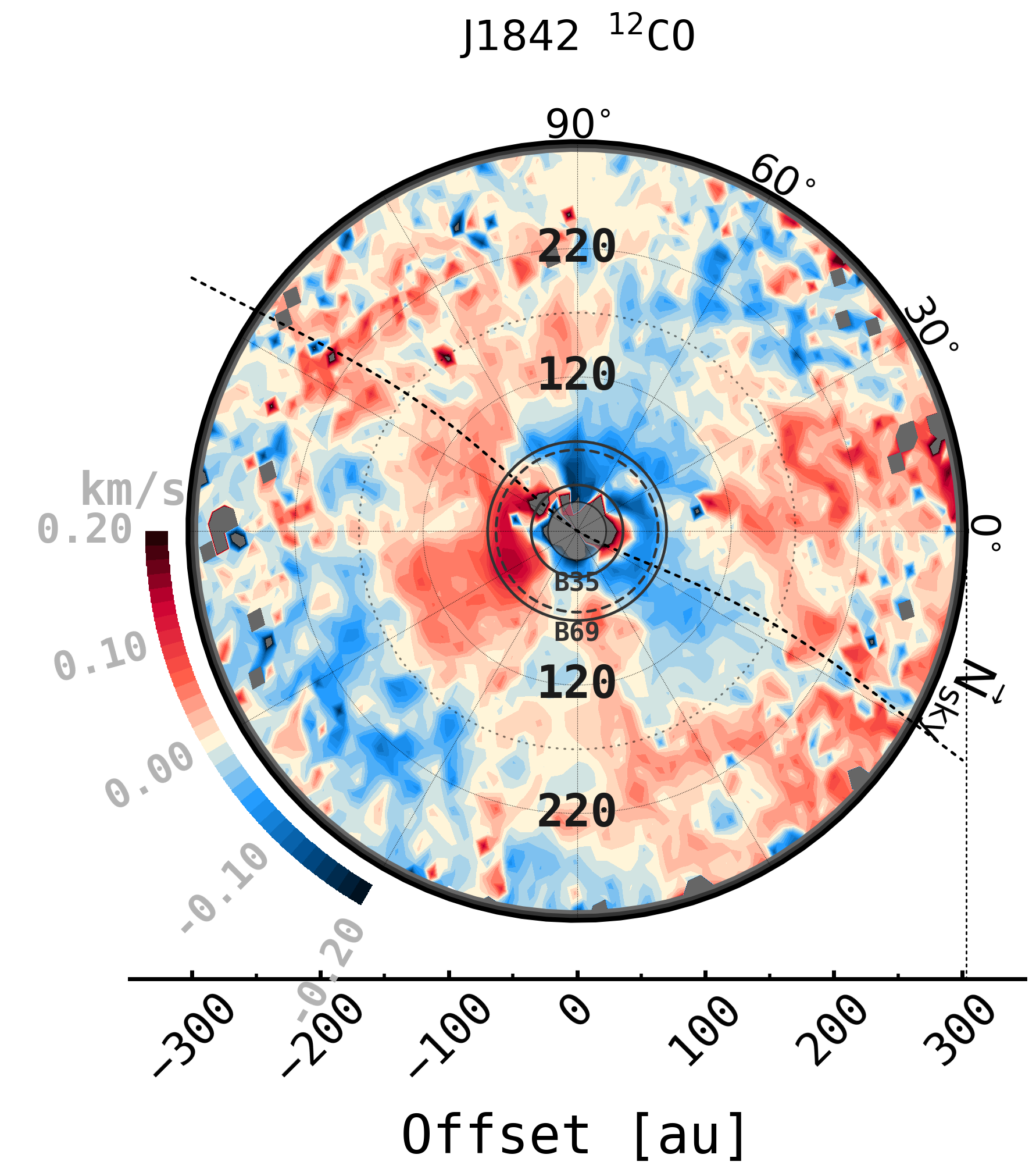} \\
        \includegraphics[width=0.25\textwidth]{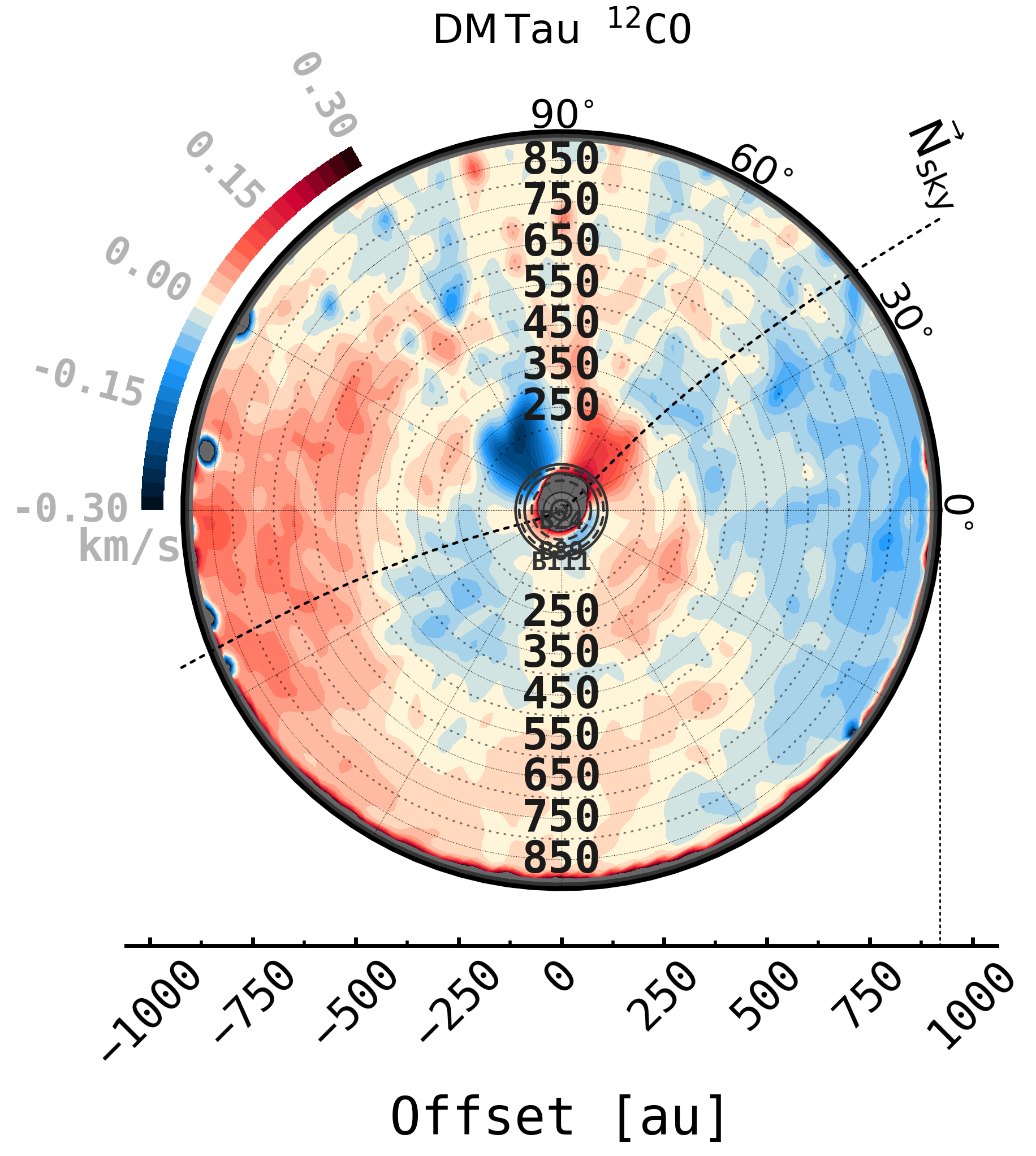} &
        \includegraphics[width=0.25\textwidth]{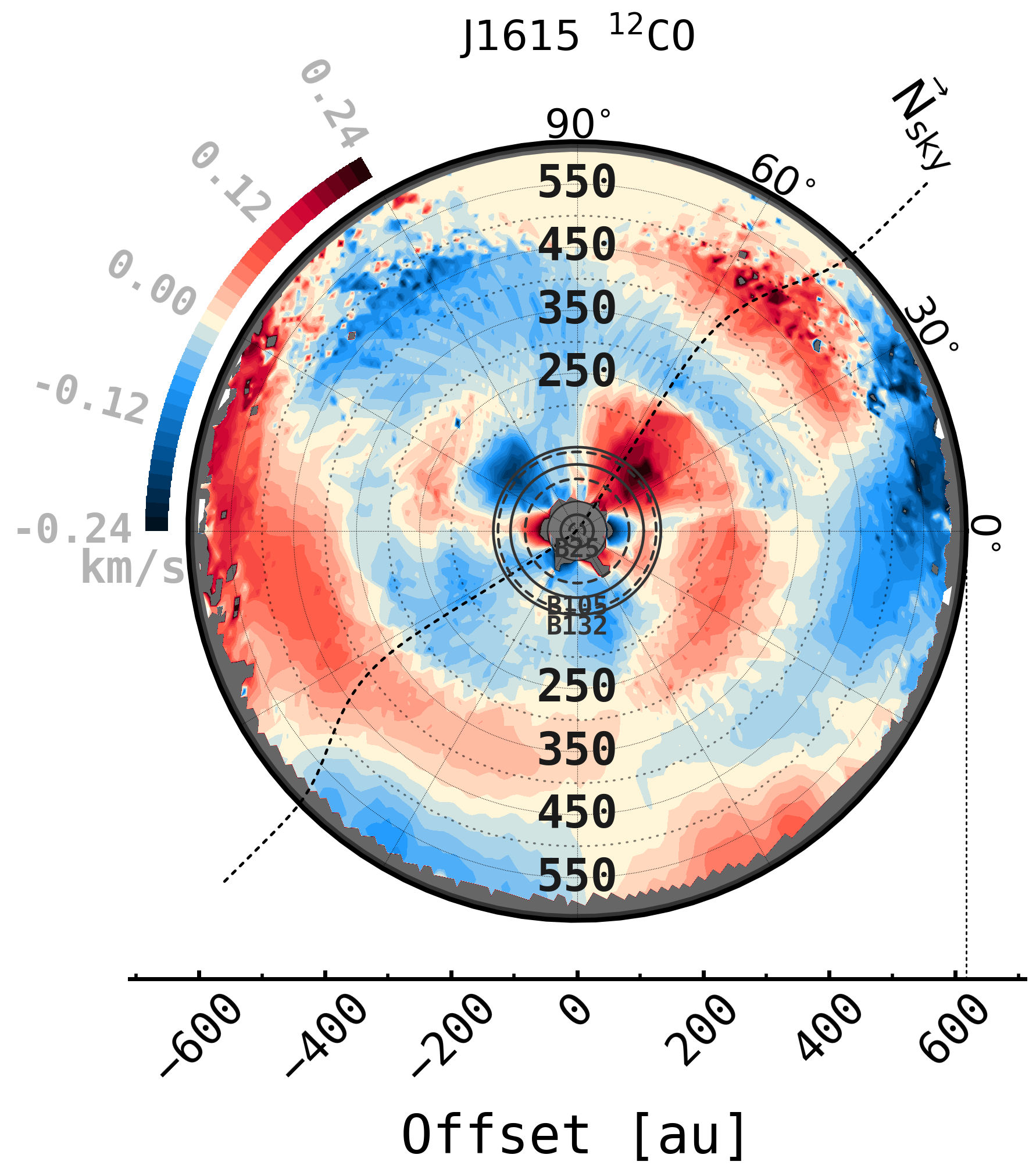} & 
        \includegraphics[width=0.25\textwidth]{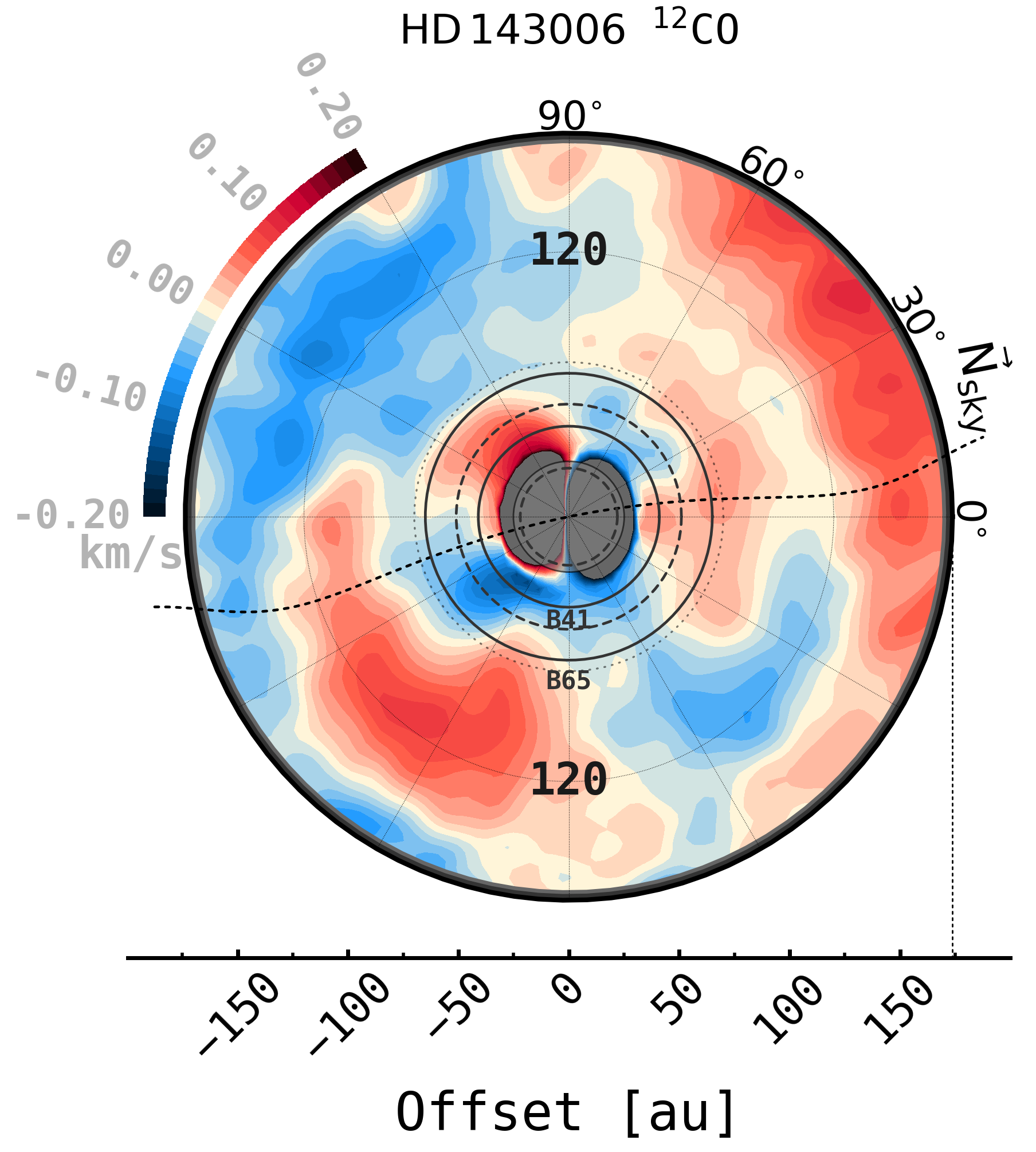} & 
        \includegraphics[width=0.25\textwidth]{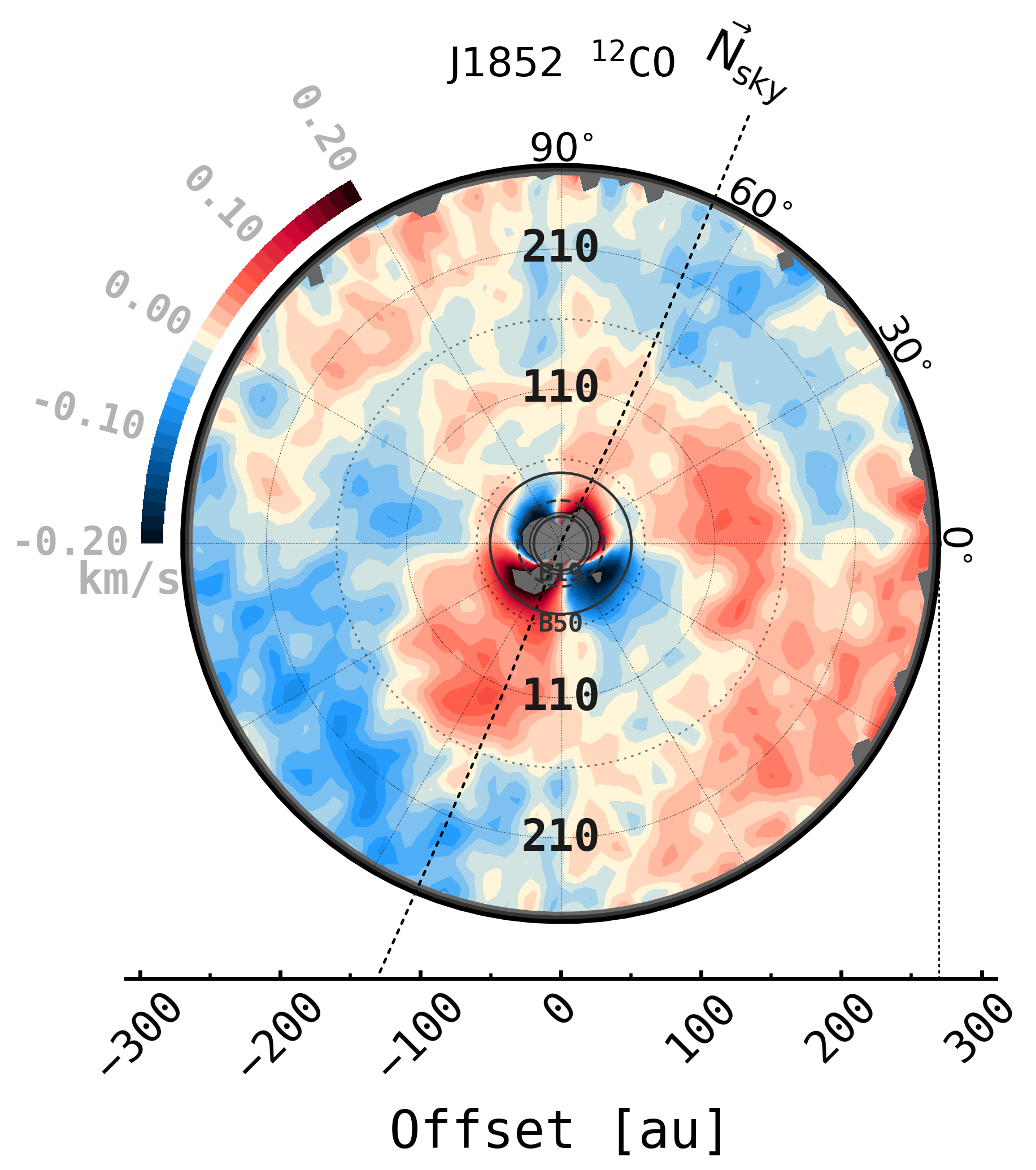} \\
        \includegraphics[width=0.25\textwidth]{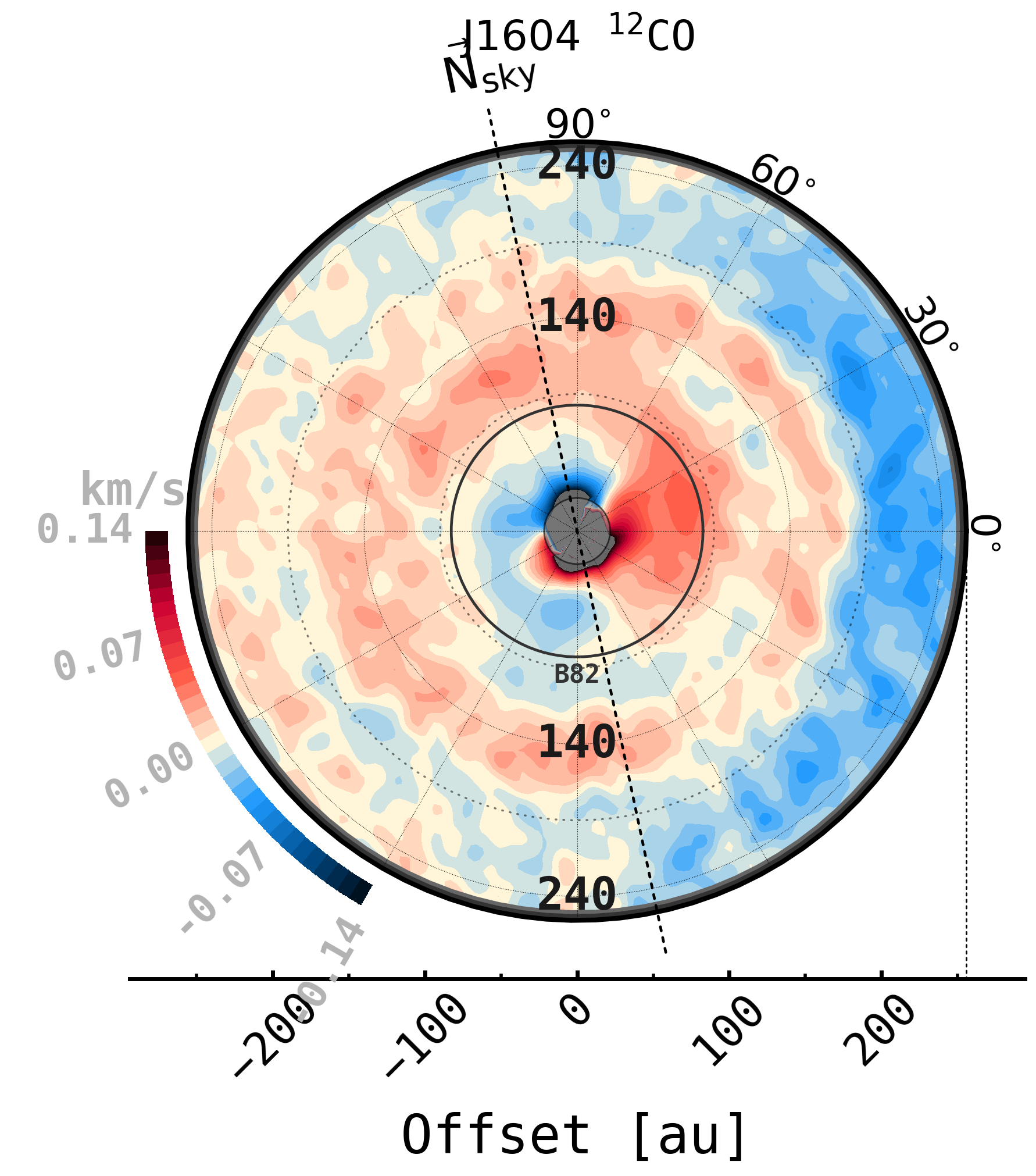} & 
        \includegraphics[width=0.25\textwidth]{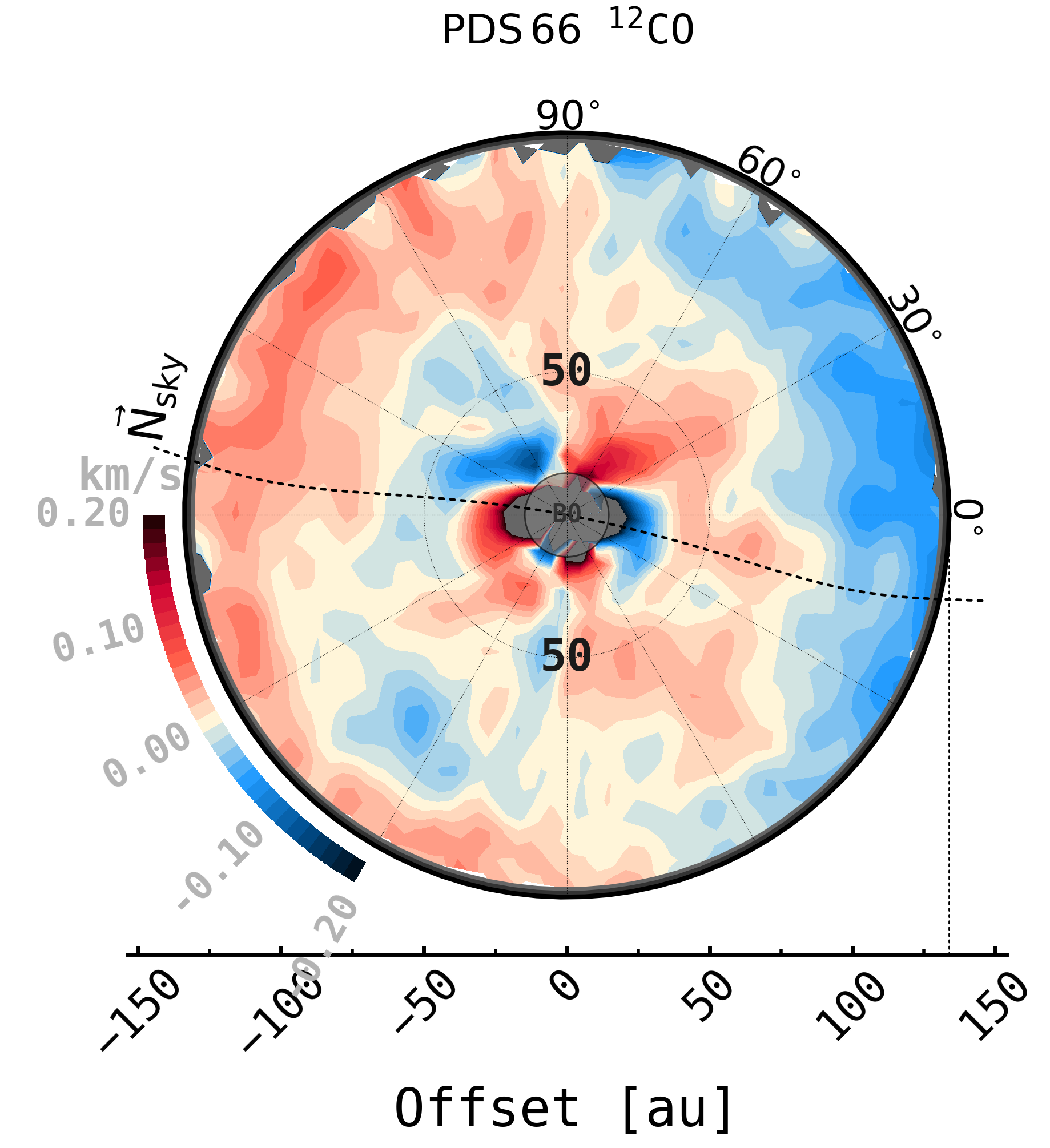} & 
        \includegraphics[width=0.25\textwidth]{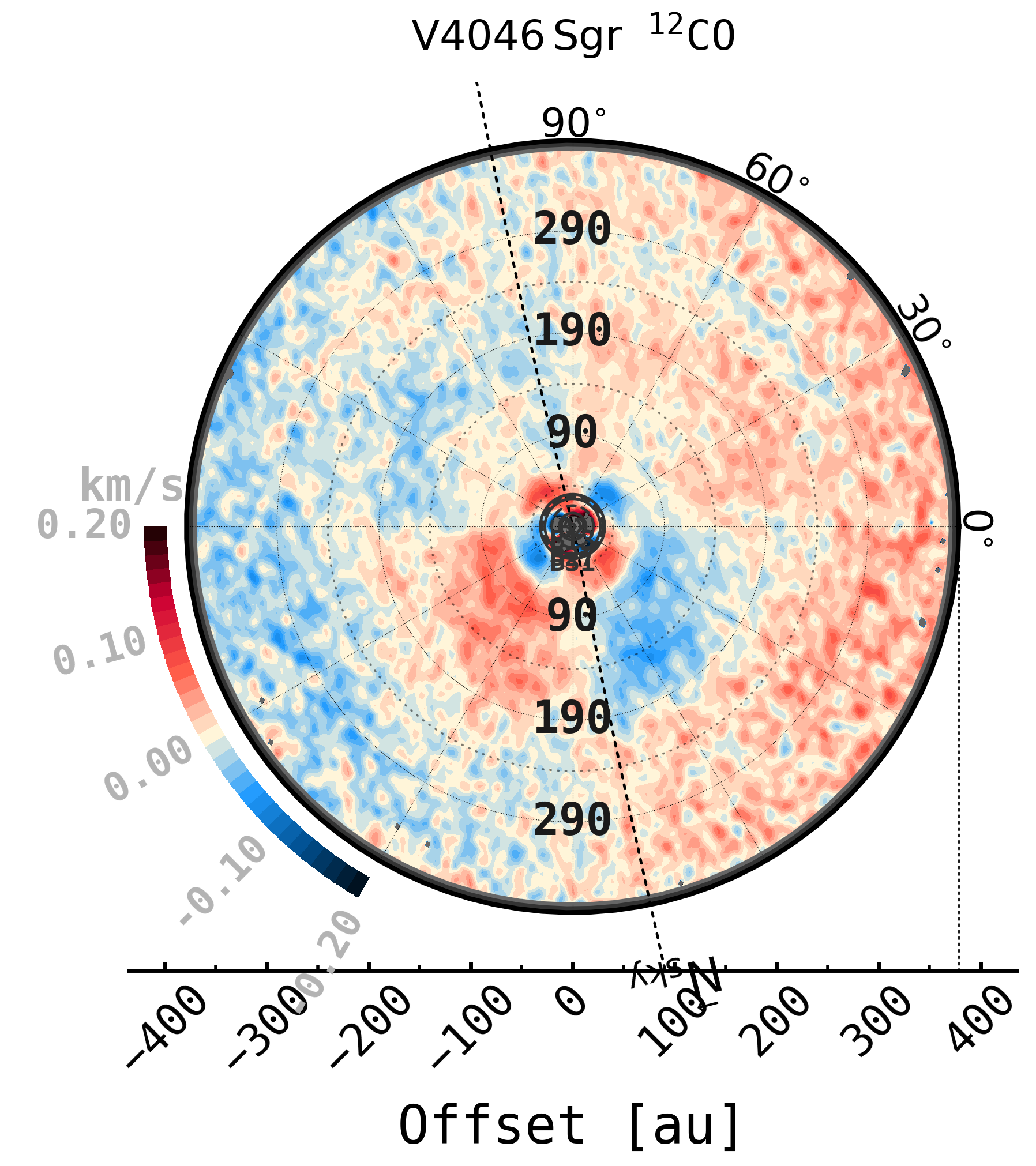} & \\
    \end{tabular}
    \caption{\twCO{} line of sight residual maps for exoALMA targets. Note the different range of velocity amplitudes.  }
    \label{fig:2dmaps}
\end{figure*}

\subsection{Residuals maps}
\label{sec:method}
\paragraph{2D residual maps.} Among the 15 exoALMA targets, a large variety of line of sight velocity residuals is observed, as indicated in Figure\,\ref{fig:2dmaps}, some showing significant non-axisymmetric features (see also exoALMA XXII/Fukagawa et al.  subm). Roughly speaking, there are three types  of residual features: large spirals (as in MWC758, top left in Figure\,\ref{fig:2dmaps}), ring-like features (as in J1604, bottom left) and patches/arcs (as in HD135344B, second row, left). 

\paragraph{Folded residual maps.} To quantitatively assess the presence and prominence of asymmetries and vertical velocity flows in our sources, we use folded residual maps, in which the line centroid velocity residuals from one half of the disk are added to those of the opposite half relative to the disk minor axis (see exoALMA XX/Izquierdo et al., for details).  This procedure removes the typically dominant symmetric azimuthal velocity perturbations, such as those caused by pressure drop-offs in the outer disk or pressure perturbations due to gaps \citep{Stadler2025}, leaving only non-axisymmetric motions and the axisymmetric component of \vz{} in the resulting map. Figure \ref{fig:sketch} illustrates the resulting folded maps for cases with only vertical, only azimuthal, and only radial velocity perturbations, and for combinations of those. In folded residual maps, any coherent vertical flow will appear as a single-colored semi-circle.  

We compute the folded residual maps for all of our targets and order them by the mean and standard deviation of their folded residual maps as a very simple metrics to estimate their dynamical state. We report the values for the \twCO{} and \thCO{} maps in Table \ref{tab:asymmfactor} of the Appendix. We note that the targets that exhibit clear spiral features in their velocity residual maps (MWC758, CQ Tau, SY\,Cha) show both the highest mean and standard deviation of residual velocities. In contrast, J1604, PDS66, and V4046\,Sgr yield the smallest mean and standard deviation values, indicative of uniform and low-amplitude flows, suggesting that these sources are comparatively less dynamically active.

\paragraph{Samples for 2D and 1D analysis.} In Sect.\,\ref{sec:results}, we will present the velocity extraction for our sample. For a sub-set of 3 targets, for which clear velocity spirals are observed in the 2D residual maps, we  trace coherent sub- and super-Keplerian substructures (filaments) using \filfinder \citep{Koch2015}. We  then extract the peak velocity residual along the spine of the filaments as well as the distribution of velocities across the spiral width. We consider a filament to trace vertical motions when the velocity does not change sign across neither the minor nor the major axis, as azimuthal and radial velocity components would \cite[see Eq.\,\ref{eq:vphi_los} and \ref{eq:vrad_los} and][]{Teague_ea_2022b}. This applies to MWC758, CQ~Tau, and SY~Cha, though only marginally in the latter. As explained previously, in the 2D and folded residual maps, such a filament would cover a large azimuthal extent without changing velocity sign. 

For all the remaining targets, except AA Tau (see below), we use the 1D axisymmetric extraction method as in previous studies. We note that HD135344B shows a variety of arcs similar to short spirals, but since the velocity spirals change sign when crossing major and minor axis, we do not consider those as coherent vertical motions. This target was explored in \citet{Wolfer2025} and a follow-up study will explore these motions in more details in the context of planet-disk interactions (exoALMA XX/Izquierdo et al. subm). 

Surface height retrievals from the channel maps performed by \citet{Galloway2025} show that the emission in the outer regions of SY\,Cha and AA\,Tau is diffuse and may originate from a distinct physical component or result from a pronounced pressure drop. We therefore exclude these outer regions (specifically, radii larger than 360 and 269 au, respectively). In the case of AA Tau, the remaining velocity residual map after removing the diffuse emission (see Figures \ref{fig:2dmaps} and \ref{fig:folded_gallery}) displays a characteristic pattern consistent with disk back-side contamination \citep{Izquierdo2025}. Given the combination of compact line emission and very high inclination, we exclude AA Tau from further analysis.

\section{Benchmarking against models}
\label{sec:benchmark}
\begin{figure*}[h]
    \centering
    \includegraphics[width=1.0\linewidth]{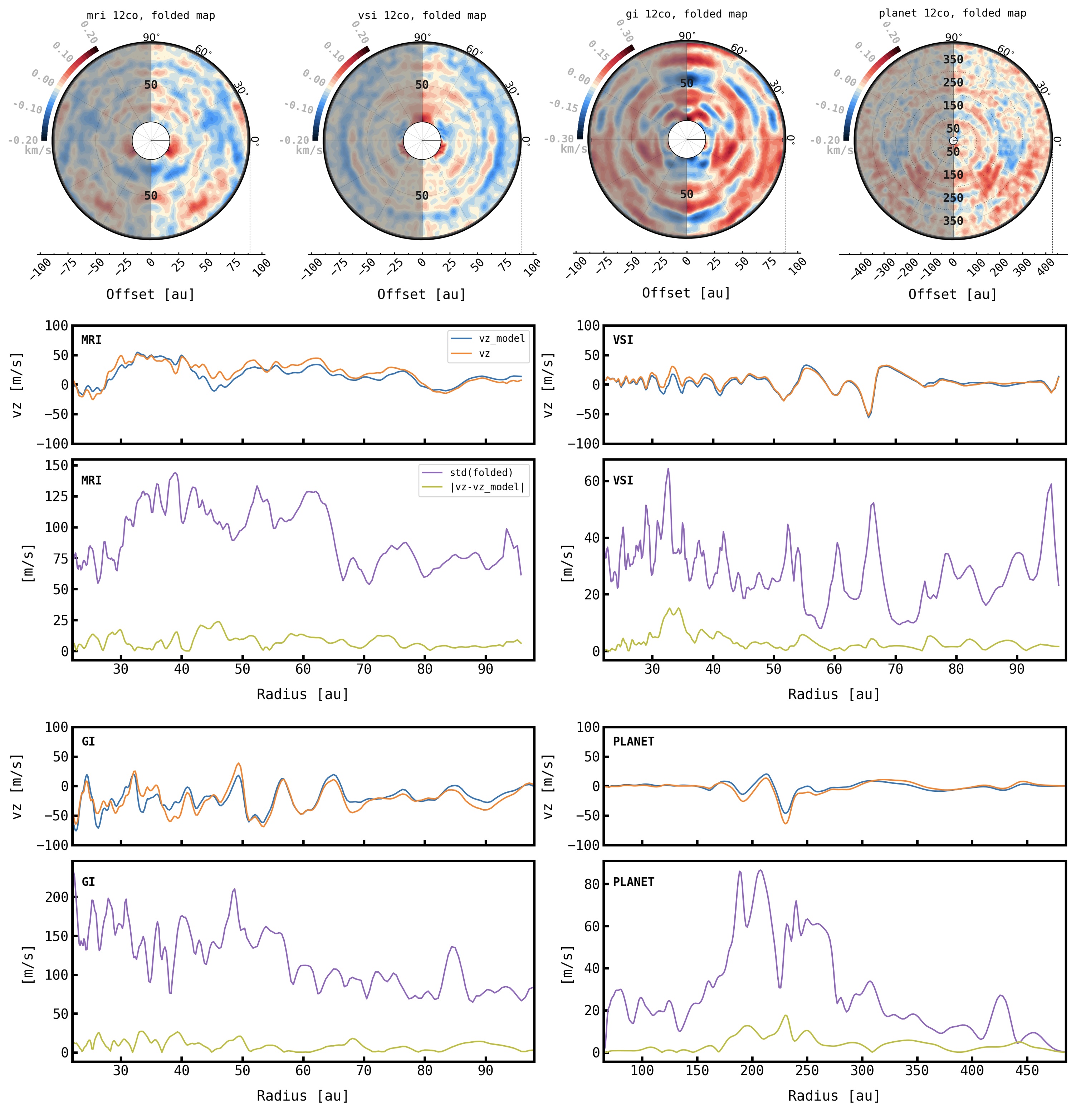}
    \caption{Benchmark of our \vz{} retrieval method with hydrodynamical models. In the top, we show folded residual maps from MRI, VSI, GI and planet-disk interaction models. For each subset of two panels, we present \vz{} from the model and the ones retrieved with \discminer (top), as well as the standard deviation of the folded map and the difference between the model and measured \vz{} (bottom), all against  radius.}
    \label{fig:hydro}
\end{figure*}

To understand our ability to recover vertical motions in dynamically perturbed disks as those in our sample, we first test our retrieval methodology on synthetic cubes generated from hydrodynamical simulations for which we know the exact velocities. We consider four models. First, a planet-disk interaction 3D hydrodynamical model, using the same surface density and temperature profile as in \citet{Teague_ea_2019b}, but considering a 4 M$_{\rm{Jup}}$ planet at 240 au. Then, the 3D (magneto)hydrodynamical simulations for (1) the vertical shear instability (VSI), (2) the magnetorotational instability (MRI), and (3) the gravitational instability (GI) computed with the \texttt{PLUTO} code \citep{Mignone2012} and presented in \citet{Barraza2025}. We first produce synthetic observations after radiative transfer with \texttt{RADMC3D} \citep{Dullemond_2012}, with a 100 m/s channel spacing, and then fit them with \discminer, using the same procedure as for our observations. We then fold the line-of-sight velocity residual maps, that are presented in Figure~\ref{fig:hydro}, top panels. Just like in our observations, the folded maps include any non-axisymmetric ${\delta\upsilon_{\phi}}$, \vrad{}, as well as \vz{}, for which any coherent vertical flow will appear as a single-colored semi-circle.  

In the panels below the folded maps in Figure~\ref{fig:hydro}, we show four groups of 2 panels for each simulation. In the top panel, the vertical motions retrieved by our method (orange lines, labeled 'vz') and the ones that are encoded in the hydrodynamical simulation at a given $Z/R$ that correspond to $\sim$2.5 pressure scale heights (blue lines, labeled 'vz model'). In the bottom panel, we present a radial profile of the standard deviation of the folded map, used as a measure of the non-axisymmetric motions in our sources. 
We find that even in cases where the standard deviation of the folded map is high (e.g., in MRI and GI simulations), indicating a significant degree of asymmetry, the true model vertical velocities are consistently recovered within a few tens m/s. This indicates that in the context of these four simulations, the contributions from residual ${\delta\upsilon_{\phi}}$ and \vrad{} do not add coherently to mimic the residual pattern characteristic of a vertical flow. 

While these simulations might not be fully representative of the dynamical states of the exoALMA sample, we find that there is likely a floor value of a about 10 m/s below which it is difficult to retrieve accurate velocities. 
Appendix \ref{sec:beamconvolution} shows a comparison between the vertical velocities extracted from the VSI simulation at a given height $Z/R=0.24$ with those extracted after post-processing and beam convolution, for infinite and exoALMA-like angular resolution. We find that beam convolution  naturally limits the amplitude of the velocities measured in our observations \citep[see also,][]{Hilder2025}, implying that the retrieved values in the analysis presented in this paper likely underestimate the true vertical motions occurring in the disks of our sample, and that multiple layers around the $\tau$=1 surface contribute to the measured velocity. Having benchmarked and validated our methodology on models, we apply the same approach to the exoALMA observations in the following sections.

\section{Results}\label{sec:results}

\begin{figure}[t]
    \centering
    \includegraphics[width=1.0\linewidth]{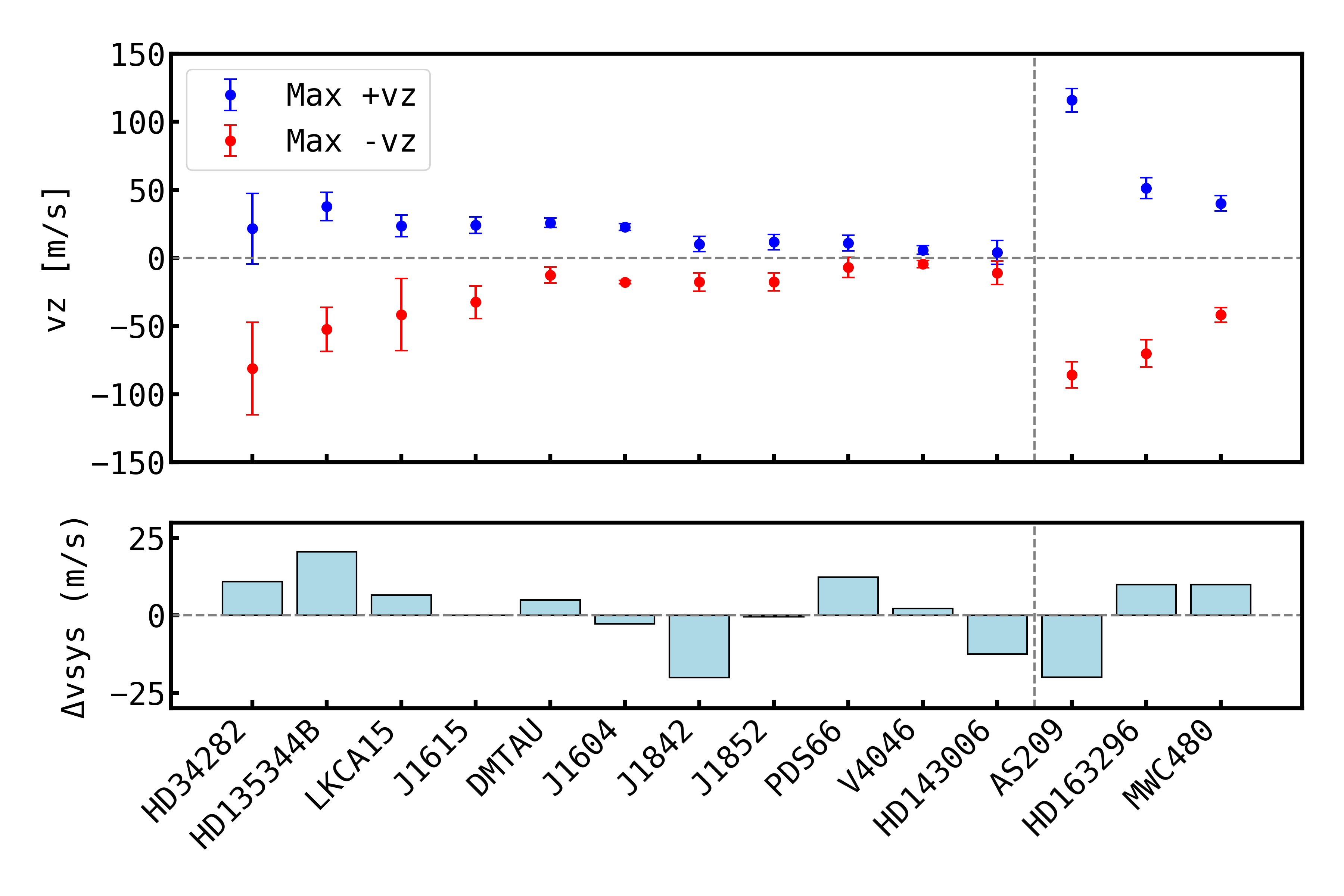}
    \caption{Top: maximum upward/downward \twCO{} vertical velocities for the exoALMA targets for which we apply a 1D analysis. On the right, the 3 MAPS targets. Bottom: differences in \vsys computed from the \twCO{} line and CS (for exoALMA), and \twCO{} and \eiCO{} (for MAPS). }
    \label{fig:maxvz}
\end{figure}

\subsection{Systemic velocities}
\label{subsec:vsys}

To get insights for the amplitude of the \vz{} motions involved, we first compute the maximum vertical motion in upward / downward directions for all targets considered in the 1D analysis. Since we exclude AA Tau, and the spiral targets (MWC758, CQ Tau and SY Cha), we are left with 11 targets for which we will analyze the azimuthally averaged radial profiles: HD34282, HD135344B, LkCa15, J1615, DM Tau, J1604, J1842, HD143006, PDS66, V4046 Sgr. For comparison, we also include HD163296, AS209, and MWC480 from the MAPS Large Program within the same analysis pipeline. 

Figure \ref{fig:maxvz} presents the maximum upward and downward vertical velocities retrieved in our analysis (see also Table~\ref{tab:vsys}). Within the exoALMA sample, peak upward velocities range from $4.0 \pm 8.8$ m/s (HD143006) to $37.8 \pm 10.4$ m/s (HD135344B), while peak downward velocities span from $-4.5 \pm 2.6$ m/s (V4046 Sgr) to $-81.3 \pm 34.0$ m/s (HD34282). Here we consider the errors as the standard deviation in \vz{} along the azimuth at the radial location of the maximum \vz{} to provide an idea of velocity dispersion along the azimuth. In comparison, the MAPS sources exhibit significantly stronger vertical motions, with AS 209 displaying the highest upward velocity of $115.9 \pm 8.7$\,m/s and downward velocity of $-86.0 \pm 9.6$\,m/s \citep[see also][]{Galloway2023, Izquierdo_ea_2023}, which can not be attributed to the different spectral resolution (200 and 100 m/s for MAPS/exoALMA, respectively).  

Interestingly, as immediately clear from Figure\,\ref{fig:maxvz}, we find that the amplitudes of maximum upward and downward motions are generally comparable. The systemic velocities derived from our model fitting thefore represent a midpoint between positive and negative extremes, and this implies a degeneracy in the Keplerian fitting process whereby actual vertical motions in the disk may be partially interpreted as a bulk velocity of the source relative to the observer, hence introducing a systematic offset in the measured \vz{}.

To assess such a possible offset in the vertical velocity measurements, we examine the systemic velocities derived from another molecular line that traces a different vertical layer of the disk, providing an independent measurement. Specifically, we model the \CSfull{} emission line, which, in most systems, predominantly probes regions closer to the midplane \citep[see, ][]{Galloway2025}. 
We proceed in a similar way than for the \twCO{} cubes, and fit a Keplerian model to the channel maps using \discminer{} with the systemic velocity as a free parameter. If the systemic velocities retrieved from both lines agree, the inferred vertical velocities likely reflect the true absolute gas motions. However, any significant discrepancy between them would likely point toward an absolute offset affecting the \vz{} retrieval. 

Table \ref{tab:vsys} summarizes the systemic velocities derived with \discminer{} for \twCO{} and CS, also shown in Figure\,\ref{fig:maxvz}. Among the targets exhibiting clear vertical motions in our retrieval, we find systemic velocity differences between the two lines ranging from 0.5 m/s (J1852) to 20.5 m/s (HD135344B). For the MAPS sources, these differences, computed from the \twCO{} and \eiCO{} lines, are 10\,m/s (HD163296 and MWC480) and 20\,m/s (AS209) \citep{Izquierdo_ea_2023}. These values provide some estimate of the potential systematic offset affecting each \vz{} measurement discussed in this work. 
Applying such offsets in \vz{}, targets such as HD34282 and HD135344B exhibit even stronger downward motions up to 92$\pm$34 and 72$\pm$16 m/s respectively, while J1842 and AS209 show enhanced upward motions up to 30 $\pm$ 6 and 136 $\pm$ 8~m/s, respectively. Targets as PDS66 and HD143006, would have a constant downward and upward flow, of about 10 m/s, respectively. However, for most targets, the offset in \vsys is rather marginal and does not affect our interpretation of the direction of vertical motion. We acknowledge the limits of this simple analysis, as the measured offset traces the intrinsic difference in vertical motions between lines tracing different heights, which are very much dependent on the ongoing physical processes and the vertical thermal structure of the disks. However, we use the best currently available observational datasets to estimate the reliability of our derived vertical motions. Independently from the offset, the peak-to-trough values for each target represent the maximum amplitude of vertical motion that can be measured  in those targets, which in \twCO{} ranges from $\sim$30 m/s (J1852) to $\sim$100 m/s (HD34282) in the exoALMA sample and from 80 m/s (MWC480) to 200 m/s (AS209) in the MAPS targets.

\begin{table}[ht]
\centering
\caption{Maximum positive and negative \twCO{} \vz{} values, and \vsys estimated from \twCO{} and CS for exoALMA targets (this work), and from \twCO{} and \eiCO for MAPS \citep{Izquierdo_ea_2023}. Values are in m/s. 
\label{tab:vsys}}
\begin{tabular}{lcc|cc}
\hline
\multicolumn{5}{c}{\textbf{1D analysis} - exoALMA } \\
\hline
\textbf{Target} &  $\uparrow$ \vz{} &  $\downarrow$ \vz{}  & \vsys \twCO{} &  CS  \\
\hline
HD34282    & $21.5 \pm 26.0$   & $-81.3 \pm 34.0$  & -2326.3 & -2337.3 \\
HD135344B  & $37.8 \pm 10.4$   & $-52.5 \pm 16.2$  & 7086.5 & 7066.0 \\
LkCa15     & $23.6 \pm 7.9$    & $-41.6 \pm 26.5$  & 6288.0 & 6281.5 \\
J1615      & $24.0 \pm 6.0$    & $-32.5 \pm 11.9$  & 4751.0 & 4751.0 \\
DM Tau     & $25.8 \pm 3.5$    & $-12.5 \pm 6.0$   & 6033.6 & 6028.6 \\
J1604      & $22.7 \pm 2.6$    & $-17.7 \pm 1.3$   & 4618.4 & 4621.0 \\
J1842      & $10.2 \pm 5.7$    & $-17.7 \pm 6.8$   & 5940.6 & 5960.7 \\
J1852      & $11.6 \pm 5.5$    & $-17.5 \pm 6.6$   & 5472.6 & 5473.1 \\
HD143006   & $4.0 \pm 8.8$     & $-10.9 \pm 8.7$   & 7715.9 & 7728.4 \\
PDS66      & $10.9 \pm 5.8$    & $-6.9 \pm 7.5$    & 3964.2 & 3951.7 \\
V4046 Sgr  & $5.8 \pm 3.2$     & $-4.5 \pm 2.6$    & 2929.8 & 2927.6 \\
\hline
\multicolumn{5}{c}{\textbf{1D analysis} -  MAPS} \\
\hline
\textbf{Target} &  $\uparrow$ \vz{} &  $\downarrow$ \vz{}  & \vsys \twCO{} &  \eiCO{}  \\
\hline
AS209      & $115.9 \pm 8.7$   & $-86.0 \pm 9.6$   & 4640.0 & 4660.0 \\
HD163296   & $51.2 \pm 7.7$    & $-70.2 \pm 10.1$  & 5770.0 & 5760.0 \\
MWC480     & $40.1 \pm 5.7$    & $-41.8 \pm 5.4$   & 5100.0 & 5090.0 \\
\hline
\multicolumn{5}{c}{\textbf{Spirals} - exoALMA} \\
\hline
\textbf{Target} &  $\uparrow$ \vz{} &  $\downarrow$ \vz{}  & \vsys \twCO{} &  CS  \\
\hline
MWC758     & $\sim$500          &   $\sim-265$      & 5894.4 & 5899.8 \\
CQ Tau     & $\sim$352          &   $\sim-290$      & 6186.9 & 6180.2 \\
SY Cha     & $\sim$259          &                   & 4103.6 & 4113.5 \\
\hline
\end{tabular}
\end{table}

\begin{figure*}[t]
    \centering
    \begin{tabular}{c}
        \includegraphics[width=0.8\linewidth]{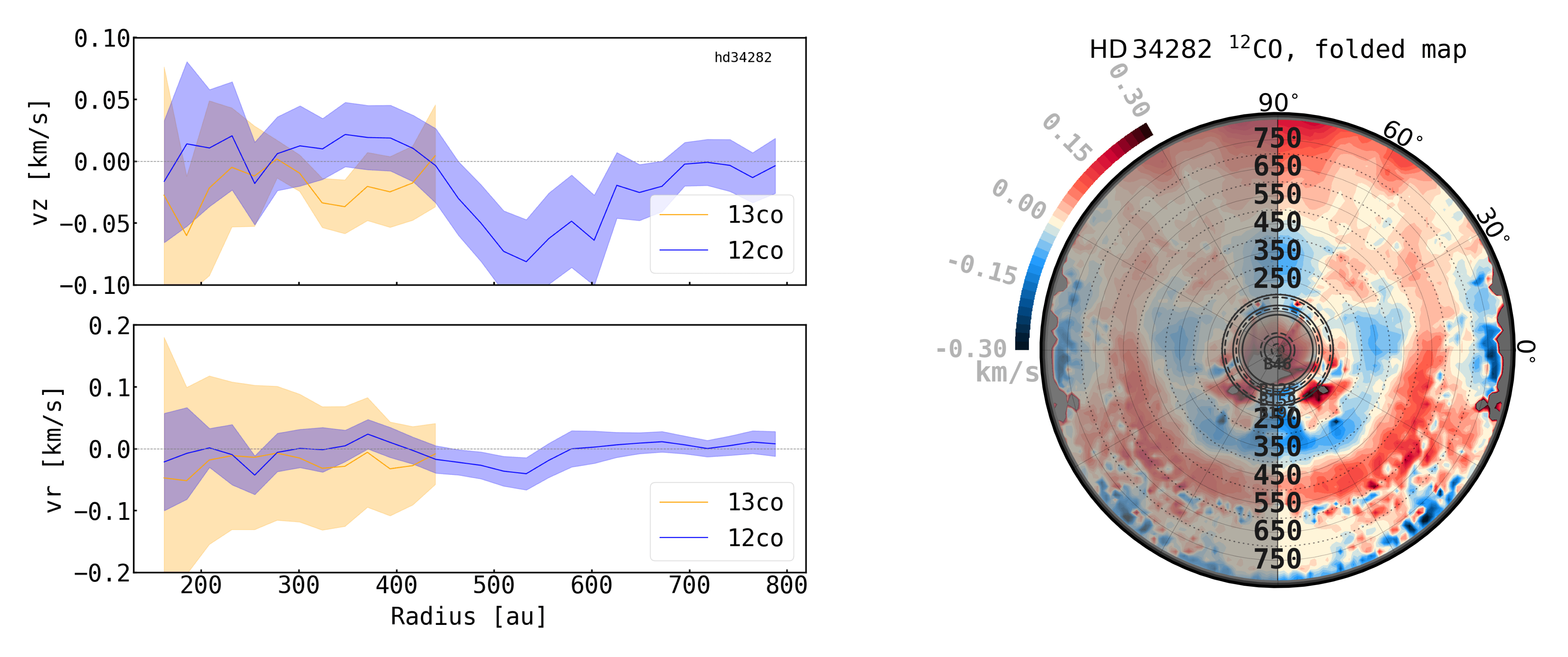} \\[-3ex]
        \includegraphics[width=0.8\linewidth]{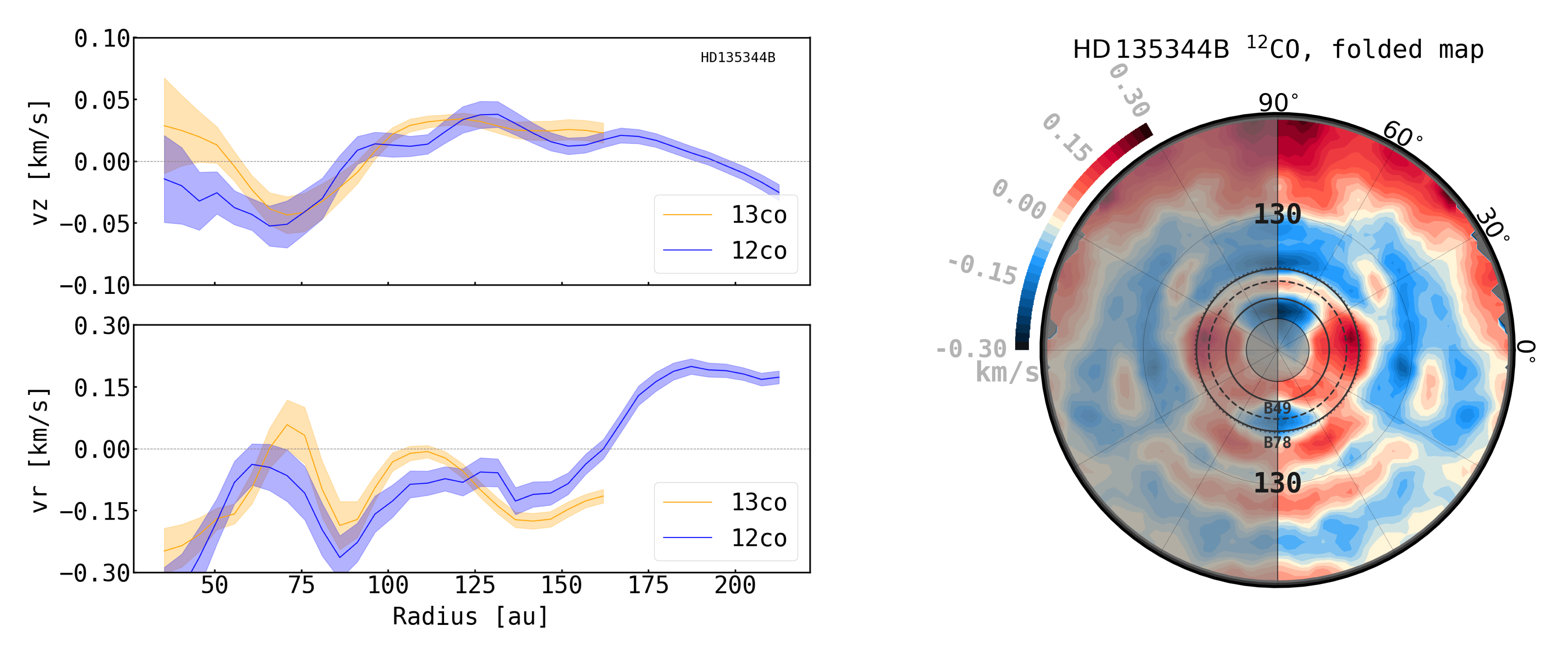} \\[-3ex]
        \includegraphics[width=0.8\linewidth]{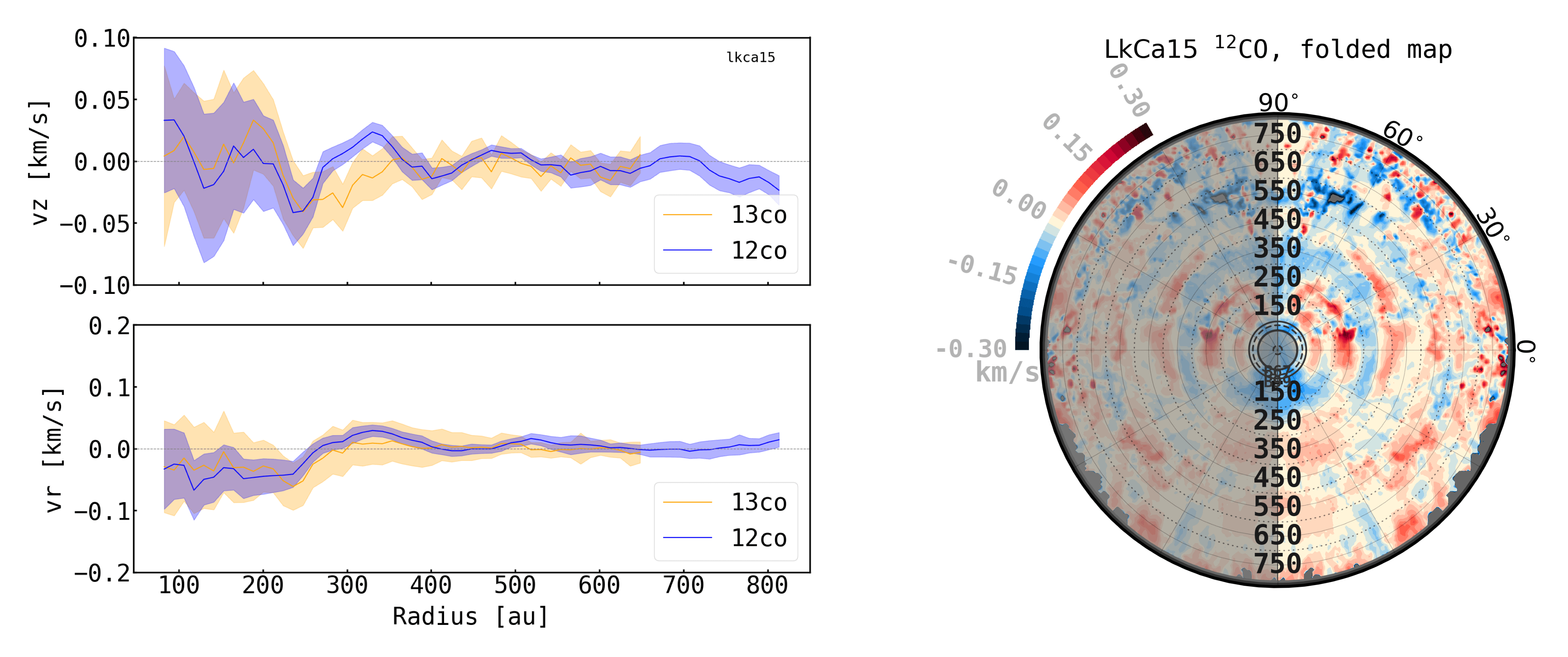} \\[-3ex]
        \includegraphics[width=0.8\linewidth]{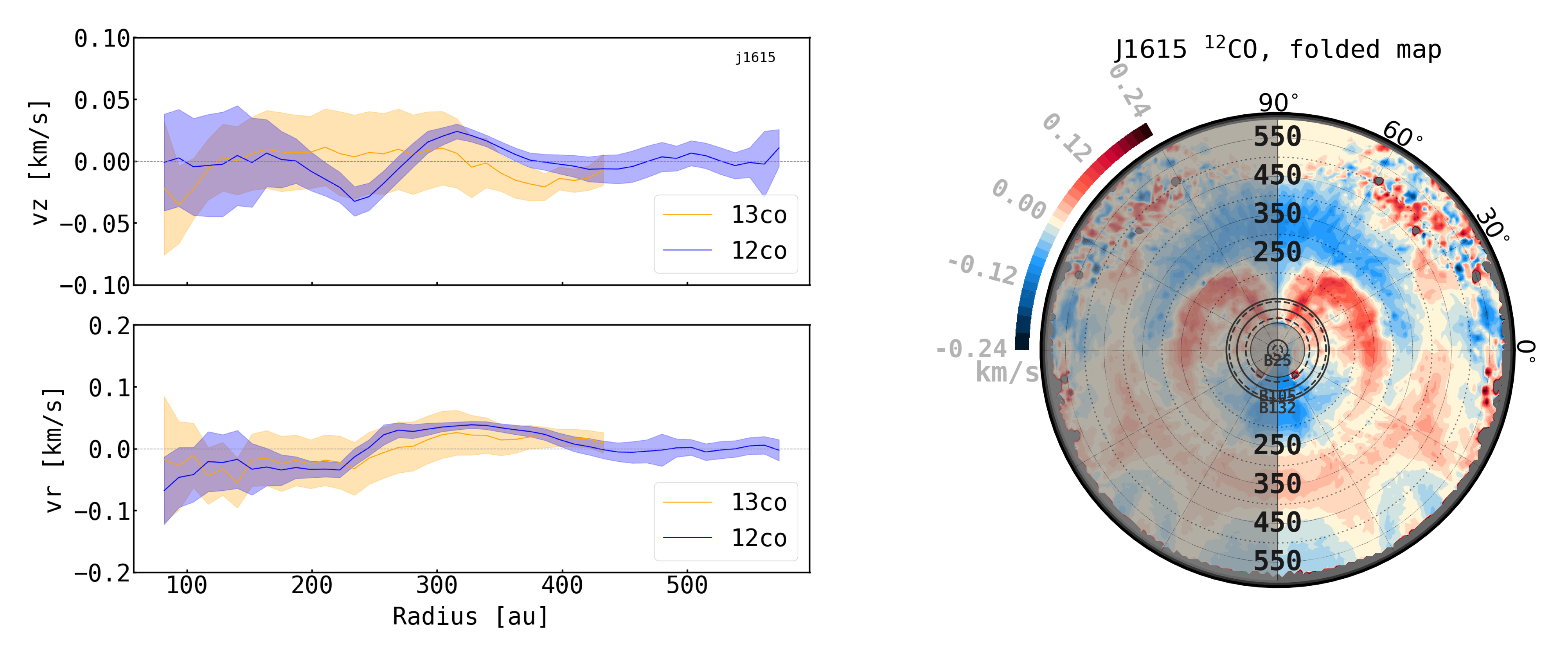} 
\end{tabular}
    \caption{Gallery of azimuthally averaged vertical (top panels) and radial velocity (bottom panels) profiles. Blue lines are \twCO{} and yellow lines are \thCO{}. The errors are the standard deviation along each annulus. On the right, the folded residual maps for \twCO{} are presented, in which any significant coherent vertical flow appears as an arc of uniform color (blue or red). 
    }
    \label{fig:vz1D}
\end{figure*}

\begin{figure*}[t]
    \centering
    \begin{tabular}{c}
        \includegraphics[width=0.8\linewidth]{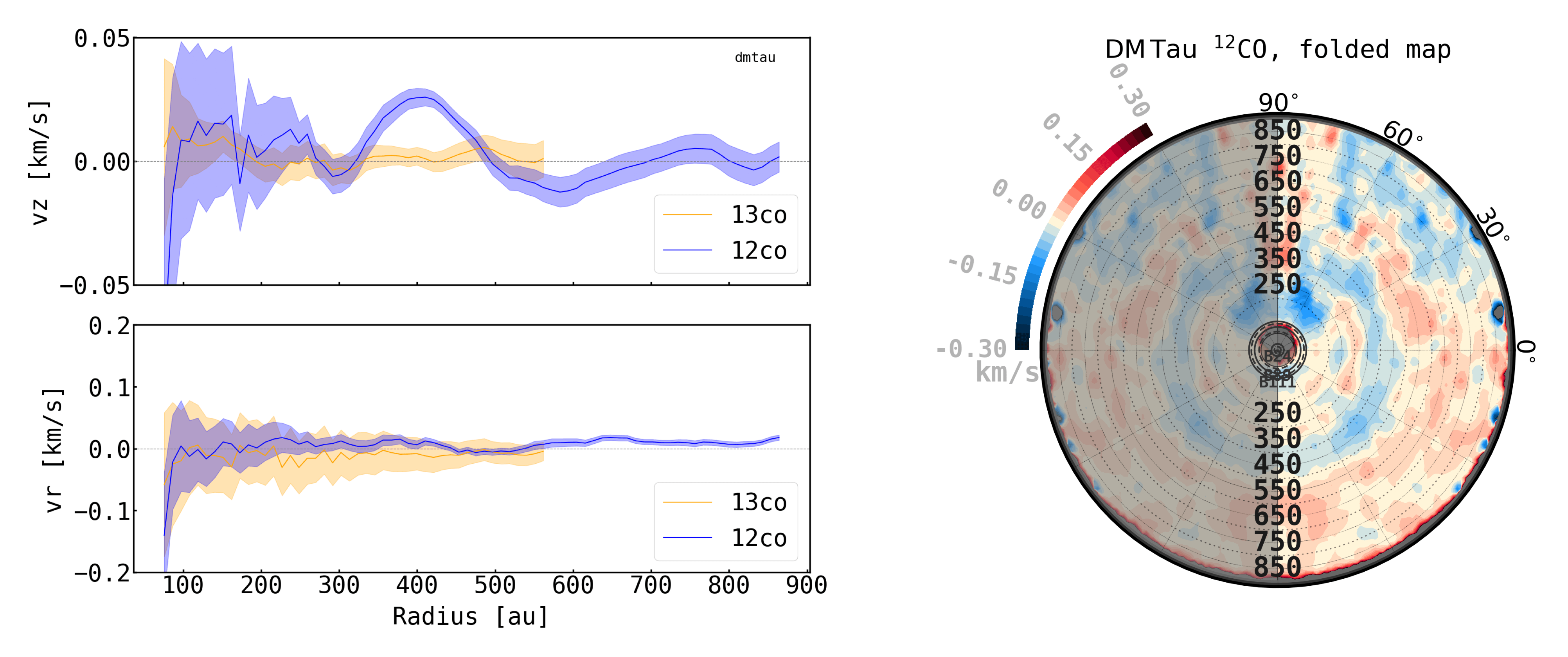} \\[-3ex]
        \includegraphics[width=0.8\linewidth]{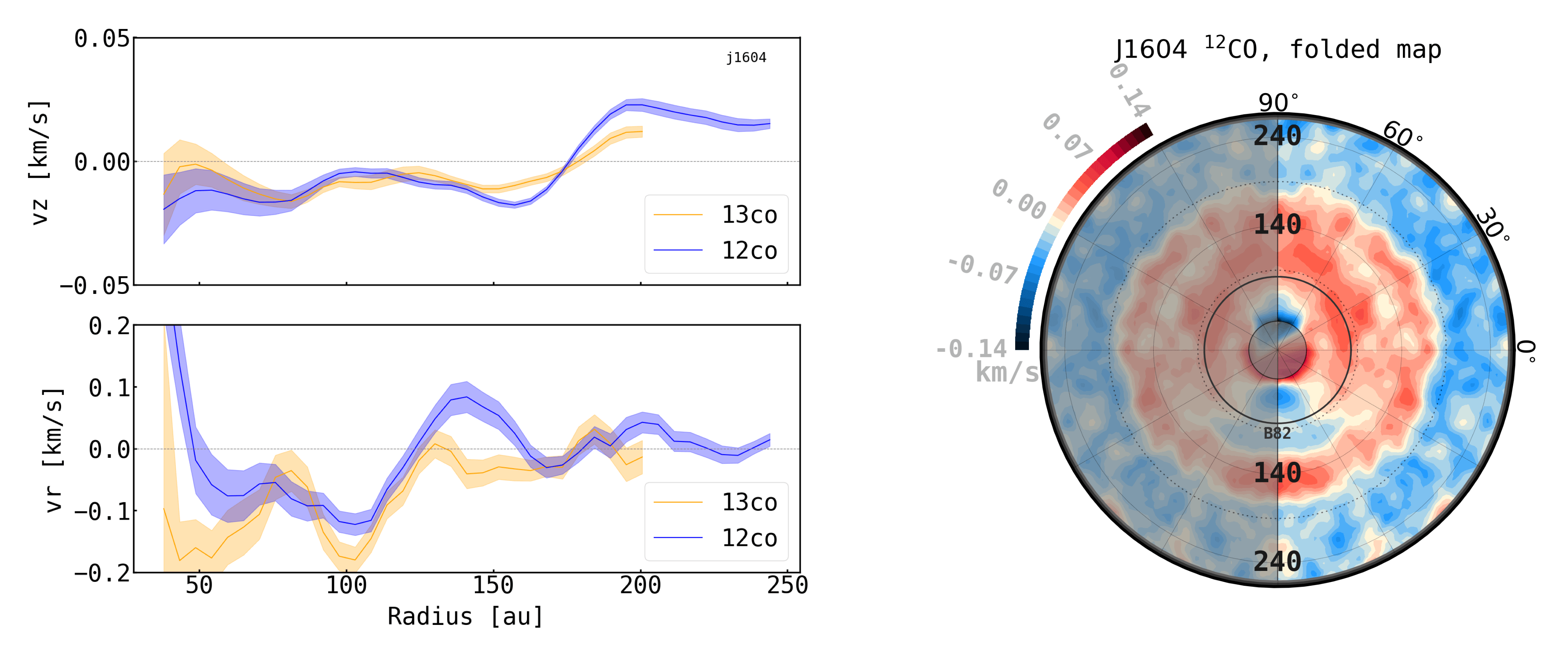} \\[-3ex]
        \includegraphics[width=0.8\linewidth]{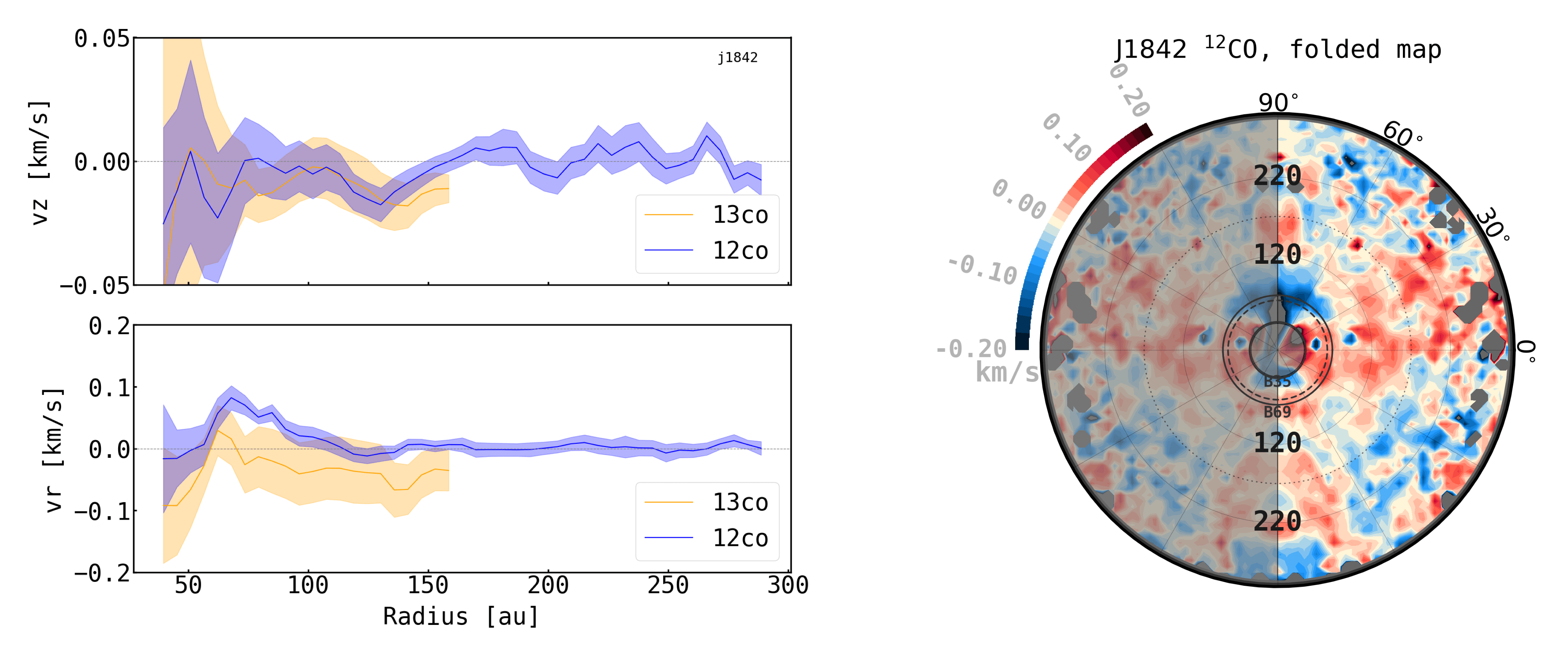} \\[-3ex]
        \includegraphics[width=0.8\linewidth]{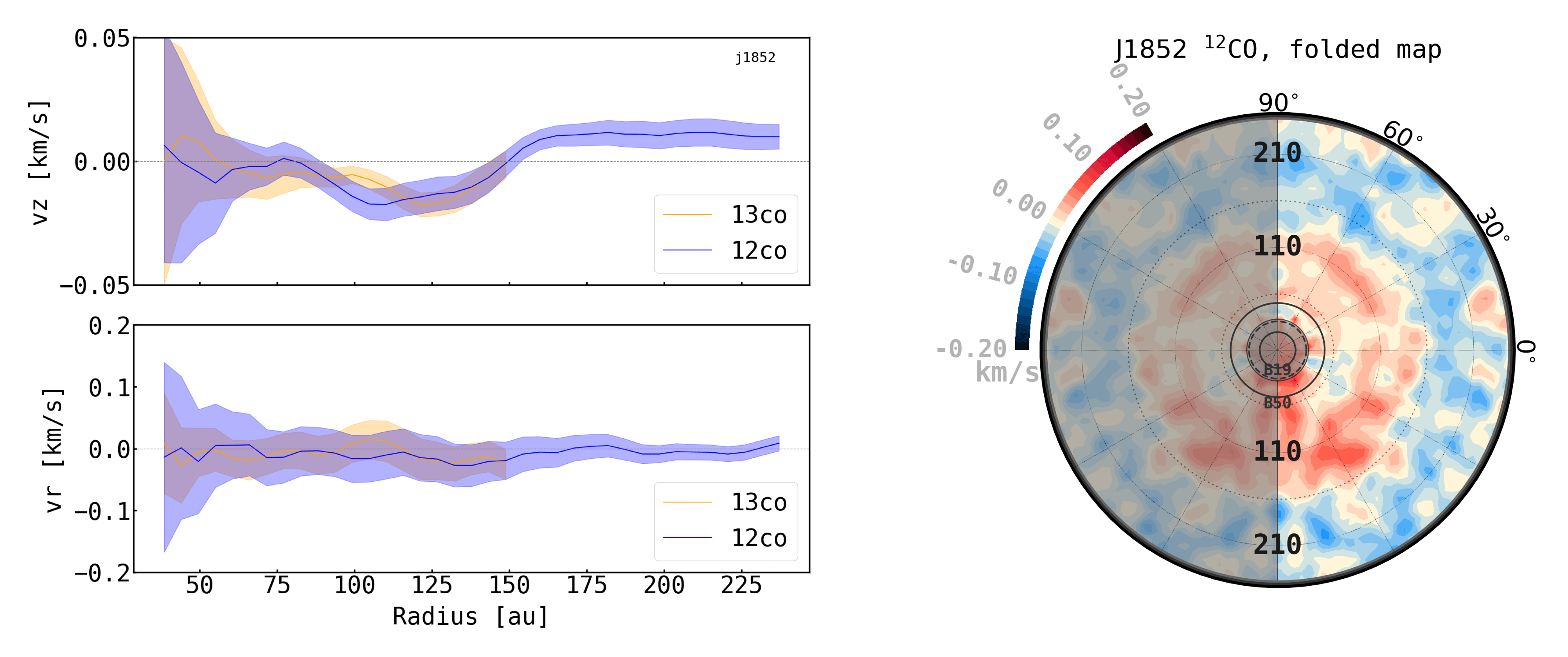} \\
    \end{tabular}
    \caption{Same as in Figure~\ref{fig:vz1D}. The folded residual maps for \twCO{} (in the right) show any significant coherent vertical flow appears as an arc of uniform color (blue or red). DM Tau, J1604 and J1852 provide the clearest examples.  
    }
    \label{fig:vz1D_2}
\end{figure*}

\subsection{Radial profiles}
As explained in Sec. \ref{sec:method}, we compute the radial profiles of the azimuthally averaged vertical velocities for 11 targets. Figure \ref{fig:vz1D} and \ref{fig:vz1D_2} show these profiles for the 8 targets exhibiting significant upward or downward motion in the \twCO{} line, defined as cases where the maximum velocity exceeds the standard deviation along the azimuth. The corresponding \thCO{} profiles are also shown. For completeness, as the ${\delta\upsilon_{\phi}}$ profiles are reported in \citet{Stadler2025}, we provide the \vrad{} component for both lines in the lower panels of the same figures. In the Appendix, we present the \vz{} and \vrad{} profiles for PDS 66, V4046 Sgr, and HD 143006, targets showing little to no vertical motion (Figure~\ref{fig:vz1Dno}). As noted earlier, correcting for a potential offset due to an inaccurate \vsys{} would yield consistently low-velocity flows. The same profiles are provided for the MAPS targets (Figure~\ref{fig:vz1DMAPS}). In all figures, the plotted error bars represent the azimuthal standard deviation at each radius, computed from the line-of-sight velocity maps  and from the 2D \vrad{} maps  (see Figure \ref{fig:vr_leftover_gallery} for the latter, whose complex structure interpretation is beyond the scope of this paper). In Figures \ref{fig:vz1D}, \ref{fig:vz1D_2}, \ref{fig:vz1Dno} and \ref{fig:vz1DMAPS}, we also show the \twCO{} folded residual maps, to the right of the radial profiles. They provide a useful visual diagnostic as coherent vertical motions appear as azimuthally extended arcs of a single color, as illustrated in Figure \ref{fig:sketch}. The clearest examples of such coherent flows in the folded residual maps are J1604, J1852 and DM Tau. 

For the following discussion, we consider the radial profiles as retrieved from our analysis, since the offset in \vsys does not affect our conclusions. Looking at the \twCO{} radial profiles, we find that HD135344B, J1615, J1604 and DM Tau, as well as the three MAPS targets, show vertical motions with values above three times the azimuthal standard deviation. HD34282, LkCa15, J1842, and J1852 also show vertical motions, although with values above twice the azimuthal standard deviation. We note that V4046 Sgr shows a small upward flow at 160 au (5.8\,m/s). 

\begin{figure*}[t]
    \centering
    \includegraphics[width=0.8\linewidth]{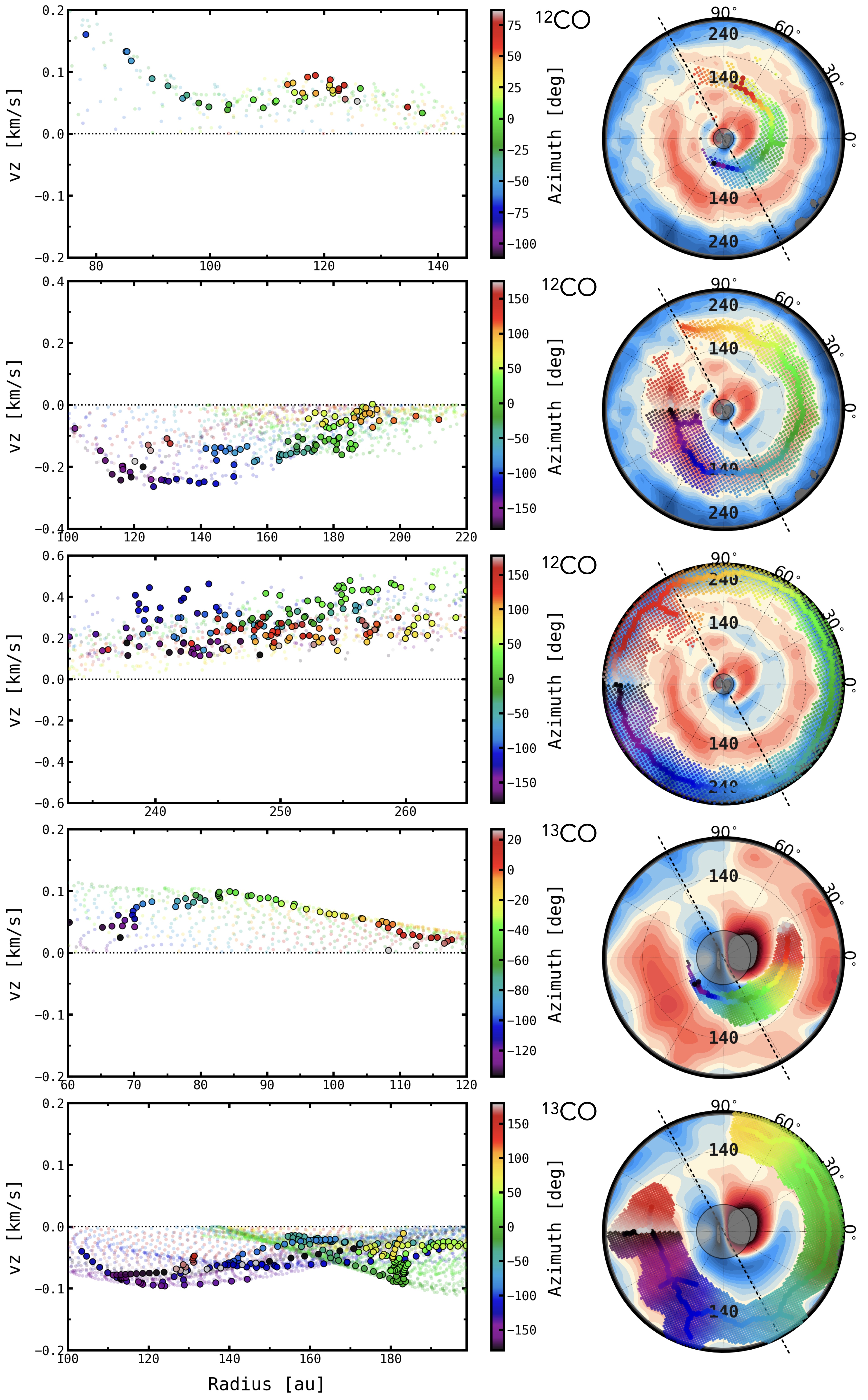}  
    \caption{Line of sight velocities, that we interpret as vertical motion \vz{}, measured along filaments for MWC758 in \twCO{} (top 3 panels) and \thCO{} (bottom 2 panels). Dark colored points are velocities extracted along the spine of the filaments depicted in the residual maps presented on the right of each panel, while pale dots trace velocities measured along the width of the filaments. Note the different y-axis scales in the panels, and the smaller radial extent of the residual maps in \thCO{}. 
    }
    \label{fig:filmwc}
\end{figure*}

The vertical velocity profiles exhibit two types of behavior. The first type of \vz{} modulation is characterized by a single transition from downward to upward motion, with a broad downward plateau that reverses into an upward plateau at larger radii. We note that there is no case where the transition occurs from upward to downward. This behavior is observed in J1604 and J1852, with the transition occurring at approximately 179 and 150, respectively. AS 209 also exhibits this pattern, with the transition point located at $\sim$157 au.

The second, observed in the majority of the disks such as LkCa15, J1615, and DM Tau, is alternating upward and downward motions. To characterize this behavior, we identify peaks and valleys in the profiles using \verb|scipy.signal.find_peaks| function. We discard any peaks and valleys whose error bars make the feature consistent with zero. The radial locations of the peaks and valleys for the $v_z$ profiles are shown in Table~\ref{tab:vz_structure} and will be discussed further in Sec. \ref{sec:correlation}. For instance, LkCa15 displays two upward and two downward velocity 'bumps', with zero-crossing points at approximately 280, 368, 445, 522 au, that is, spaced nearly every $\sim$80 au (i.e., about two beams). DM Tau exhibits upward motion from 322 to 494 au, with the peak at 410 au, followed by downward motion from 494 to 697 au, with the valley at 583 au. Similarly, J1615 shows downward motion from 187 to 275 au, followed by upward motion extending from 275 to 380 au.  Among the MAPS sources, HD163296 and MWC480 display comparable `oscillatory' \vz{} behavior as reported previously  \citep{Teague2021,Izquierdo_ea_2023}. In particular, MWC480 reveals multiple alternating upward and downward motions spanning the full radial extent of the disk, spaced by approximately 30, 40, 50, and 70 au. In contrast, HD34282 and J1842 only show a single broad downward velocity bump radially extending over 265 au and 48 au, respectively. Regardless of the specific radial profile behavior, the amplitude of the detected vertical motions in these cases remains on the order of a few tens of m/s. All together, our results indicate that upward motions, possibly tracing winds, that span the entire radial range are not commonly observed in the \twCO{} emission, but instead tend to dominate in the outer regions of the disk.

The \thCO{} radial profiles are generally noisier and extend over a smaller radial range, reflecting the lower signal to noise ratio of the data cubes. Nevertheless, in certain targets, including J1604, HD135344B, J1852, and LkCa15, the \thCO{} profiles exhibit comparable amplitudes and similar radial structures to those seen in \twCO{}, suggesting that the vertical layers traced by the two lines undergo similar vertical motions. For HD135344B, the vertical velocity crosses zero at approximately the same location for both \twCO{} and \thCO{}.

We also note that in some targets, such as HD34282, J1615, LkCa15 (Figure\,\ref{fig:vz1D}), and most strikingly MWC480 (Figure\,\ref{fig:vz1DMAPS}), there is a positive correlation between \vz{} and \vrad{} motions, indicating the presence of upward+outward or downward+inward flows. Interestingly, HD163296 shows an anti-correlation between \vz{} and \vrad{}, suggesting instead a combination of upward+inward and downward+outward motions. These patterns may point to different circulation or transport mechanisms at play in these disks. 

\begin{figure*}[t]
    \centering
\includegraphics[width=0.8\linewidth]{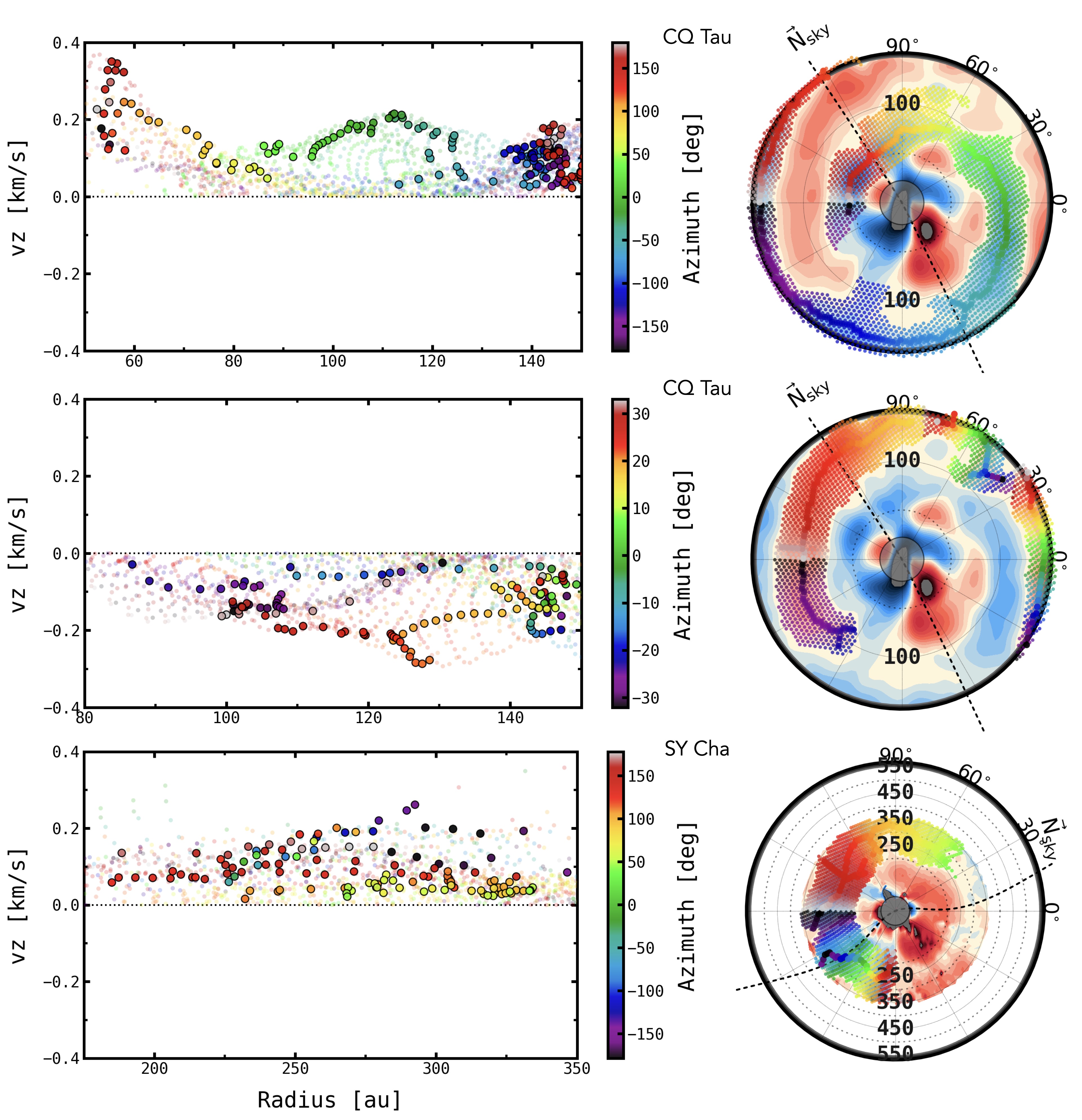}
    \caption{Same as Figure\,\ref{fig:filmwc} for CQ Tau and SY Cha in \twCO{}. These targets do not show clear spiral counterparts in \thCO{}.}
    \label{fig:filcqsy}
\end{figure*}

\subsection{Spiral targets}
\label{sec:spirals}

As discussed in Section~\ref{sec:method}, to quantify vertical motions in targets exhibiting prominent spiral structures in the line-of-sight velocity residuals, we do not perform any azimuthal averaging of \vz{} but instead use \filfinder{} to trace the velocities along the spiral spines. This approach ensures a more accurate characterization of the vertical velocities at a given radial separation and prevents the underestimation typically caused by azimuthal averaging.

In MWC758, we identify two prominent spiral structures in the velocity residuals as illustrated in Figure~\ref{fig:filmwc}. Moving outward from the inner disk, the first spiral (top panel) traces blue-shifted residuals  corresponding to upward motions, extending from the inner regions out to approximately 165\,au, with peak velocities reaching 135\,m/s. The second spiral (second panel) traces red-shifted residuals indicative of downward motions, spanning from roughly 60\,au to 230\,au, with velocities up to $-265$ m/s. The red-shifted spiral shows clear azimuthal dependence with higher velocities along the eastern side of the disk. To highlight the necessity of this approach over simple azimuthal averaging, Figure~\ref{fig:comp1d} compares the two methods for MWC 758.

In addition to these spirals, we detect an outer annulus (third panel) characterized by strong blue-shifted velocities, which we interpret as a signature of a disk wind. This component extends to the outermost detectable radius in our data and exhibits velocities up to 500 m/s, significantly higher than those observed in any other exoALMA or MAPS target. This filament seems to extend slightly further in a spiral-like shape, as seen in lower resolution \twCO{} cubes (see Figure 1 of \citealt{Galloway2025}). 
We find that the velocity residual spirals exhibit significant width (as in, transverse to the filament). Both the blue- and red-shifted spirals are also detected in the \thCO{} velocity residuals, albeit with lower amplitudes reaching approximately $\pm$100 m/s (Figure \ref{fig:filmwc}, fourth and fifth panels).

Figure\,\ref{fig:filcqsy} shows the filaments extracted for CQ~Tau and SY~Cha. CQ~Tau also shows two spirals, blue and red-shifted as in MWC~758. From the inside out, the first spiral (top left) characterised by blue-shifted residuals, corresponds to upward motions extending from the inner disk to the outermost detectable radii, with peak velocities reaching 352\,m/s in the innermost regions. The second spiral, traced along red-shifted residuals, extends from 100\,au to the outer disk, with peak downward velocities of $-290$\,m/s. In addition to the spirals, we observe a Doppler flip in the southeast region. 
The velocity extraction for SY~Cha reveals a single blue-shifted filament, extending out to the radius where the diffuse emission begins \citep{Galloway2025}, with peak velocities of 259 m/s. The velocity extraction for SY\,Cha should be interpreted with caution due to the lack of a clearly identifiable red-shifted spiral counterpart and the complex structure of the velocity residuals.

For these three targets with spiral features in their residual maps, we find differences in the systemic velocity between \twCO{} and CS ranging from $-9.9$\,m/s (SY~Cha) to 6.7\,m/s (CQ~Tau), indicating only a minor offset relative to the amplitudes of the blue- and red-shifted velocities measured.

\section{Discussion}\label{sec:discussion}

\subsection{Correlation with surface and intensity profiles}
\label{sec:correlation}
\begin{figure*}
    \centering
\includegraphics[width=1\linewidth]{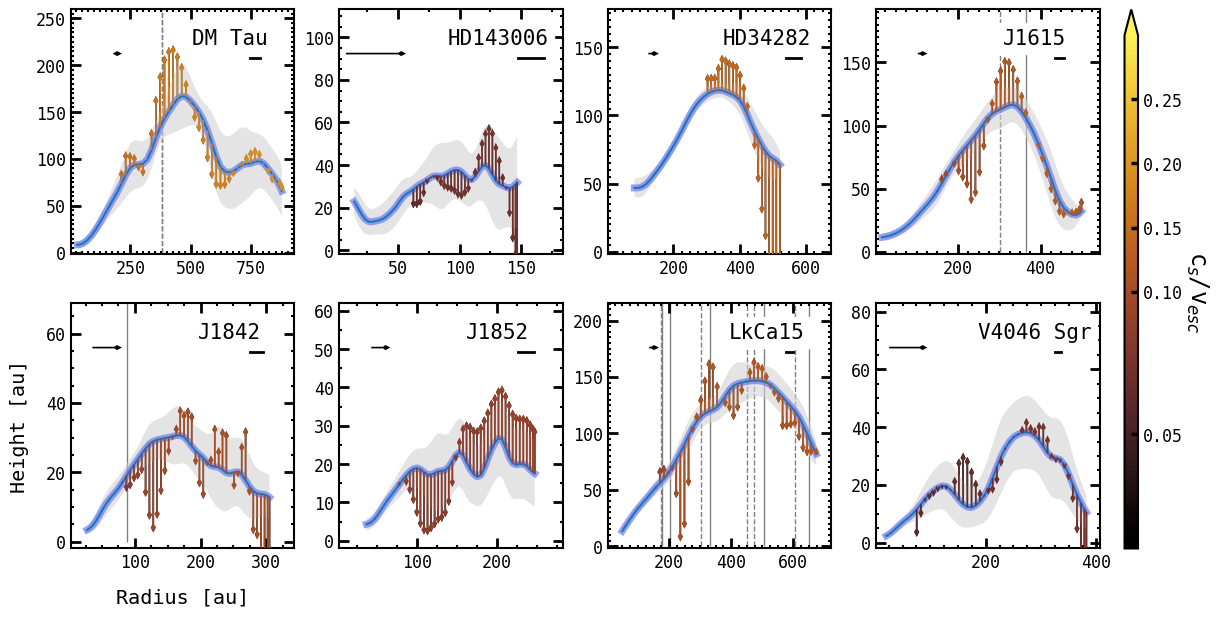}
    \caption{Vertical gas flows (arrows) along the \twCO{} non-parametric surfaces (blue). \vz{} Arrows are scaled to the local sound speed at each radius, where the reference arrow in the top left corner represents 0.1$c_s$. The arrows are colored by $c_s$/$v_{esc}$. Full (dashed) lines indicate \twCO{} and \thCO{} intensity peaks (troughs) obtained from  \citet{Galloway2025}. The beamsize is shown in the top right corners.}
    \label{fig:quiverplot}
\end{figure*}

We compare the perturbations in the azimuthally averaged velocity profiles, $v_z$ and $v_r$, to the molecular emission surface and temperature profile (traced through the peak intensity) structures identified in \cite{Galloway2025}. For this analysis, we focus on the disks for which we run an axisymmetric analysis (Figures \ref{fig:vz1D}, \ref{fig:vz1D_2} and \ref{fig:vz1Dno}), and do not include disks with large-scale spirals. HD135344B and J1604 being nearly face-on, no surface could be retrieved and we exclude these two disks from our analysis, even though they both show  vertical motions. We also exclude PDS66 that shows a flat surface. This leaves a sample of 9 disks. As mentioned previously, the radial locations of the peaks and valleys for the $v_z$ profiles are shown in Table \ref{tab:vz_structure}. We find that in over half of the sources (six out of nine analyzed), perturbations in \vz{} spatially correlate within a beam size with substructures in either the emission surfaces or peak intensity profiles in one of the two lines. We illustrate this in Figure \ref{fig:quiverplot} by showing the \twCO{} \vz{} as vertical arrows, overlaid on the non-parametric \twCO{} surfaces by \citet{Galloway2025}. Peak intensity substructures found in \twCO{} and \thCO{} in the same work are also indicated. The arrows are given in units of the local sound speed, computed at the $^{12}$CO emission layer by using the peak intensity values. 

We note that the presence of substructures in the surface layer does not significantly bias the extraction of vertical motions measured on smooth parametric surfaces. Such surface perturbations are mirror-symmetric and disappear when folding residuals maps and averaging, as in our methodology \citep[see ][their Figure B15; and exoALMA XX/ Izquierdo et al. subm.]{Izquierdo_ea_2023}. In addition, \cite{Galloway2023} found that there was no significant changes in the velocity residuals of AS209 when comparing a smooth surface to a structure surface.

The observed correlation is most pronounced in LkCa15, which exhibits multiple vertical velocity perturbations in $^{12}$CO. The upward motion at 330 au is co-incident with a  emission surface gap at 342 au and a peak intensity bump at 330 au. Upward motion at 472\,au is also co-located with peak intensity perturbations seen in both $^{12}$CO and $^{13}$CO. J1615 has prominent vertical velocity structure at 234 au (downward motion), co-located with a $^{13}$CO surface gap, and at 316 au (upward motion), co-located with a dip in the $^{12}$CO peak intensity. In V4046 Sgr, the only vertical velocity peak at 161 au is spatially co-incident with a prominent surface gap seen in both $^{12}$CO and $^{13}$CO.

Three disks have surface and/or intensity substructure correlated with radial velocity profile perturbations. In all three cases, the co-located structure is associated where the \vrad{} profile crosses zero. In LkCa15, dips in the peak intensity are at the same locations as where \vrad{} crosses zero ($\sim$ 270 au and $\sim$ 450 au). The same is seen in J1842 at $\sim$ 170 au, and in J1615 at $\sim$ 220 au.

Spatial correlations between velocity gradients and emission surface structures have been reported in previous studies, notably in HD163296, MWC480, and AS209 that we re-analyzed in this paper \citep{Teague_ea_2019b, Izquierdo_ea_2023, paneque+2023}. Variations in the emission surface likely reflect underlying changes in the local temperature structure and, potentially, surface density, as for example in the case of a gap. This is the case for AS209 for which \citet{Galloway2023} speculated that winds driven by ambipolar diffusion are favored in the depleted regions where the gas density is very low. Focusing on the intensity and surface gaps observed in \twCO{}, we find that DM Tau, J1615, and LkCa15 exhibit upward \vz{} associated with an intensity gap, similar to AS209, though at lower velocities and with less pronounced gaps. In contrast, J1615, J1842, and J1852 display downward \vz{} linked to a surface gap, while J1852, LkCa15, and V4046 Sgr show the opposite trend, that is, upward \vz{} associated with a surface gap.

\begin{table*}[t]
\centering
\caption{Max upward and downward \vz{} values in \twCO{} compared to the local sound speed $c_s$ at the corresponding locations ($R_\uparrow$ and $R_\downarrow$). For disks for which a surface could not be extracted in \citet{Galloway2025}, we do not derive any $v_z$/$c_s$. 
For the spirals, we do not report any error as we extract the maximum velocity along the spiral spines, while for the  azimuthal averaged \vz{}, the error on \vz{} represents the standard deviation along the azimuth, and is propagated here, along with the standard deviation of the brightness temperature from \citet{Galloway2025}. We also report the observed \vz{} pattern.
\label{tab:cs}}
\begin{tabular}{l|cccc|l}
\hline
Source & $\uparrow$ $v_z$/$c_s$ & $R_\uparrow$ [au] & $\downarrow$ $v_z$/$c_s$ & $R_\downarrow$ [au] &  Pattern\\
\hline
HD34282  & $0.05 \pm 0.06$ & 348 &  $0.25 \pm 0.10$ & 533 & single broad $\downarrow$ region\\
HD135344B & - & 132 & - & 66 & oscillatory \vz{} \\
LkCa15   & $0.08 \pm 0.03$& 330 & $0.14 \pm 0.08$ & 236 & oscillatory \vz{} \\
J1615    & $0.08 \pm 0.02$ & 316 & $0.10 \pm 0.04$ & 234 & oscillatory \vz{} \\
DM Tau   & $0.11 \pm 0.02$ &410 & $0.06 \pm 0.03$ & 583 & oscillatory \vz{} \\
J1604  & - & 201 & -  & 157 & single transition $\downarrow$ to $\uparrow$ \\
J1842    & $0.03 \pm 0.02$ & 266 & $0.05 \pm 0.02$ & 130 & single broad $\downarrow$ region \\
J1852    & $0.04 \pm 0.02$ &215  & $0.05 \pm 0.02$ & 110 & single transition $\downarrow$ to $\uparrow$\\
HD143006 & $0.01 \pm 0.03$ &125  & $0.02 \pm 0.03$ & 157 & weak/no \vz{} \\
PDS66   & - & 103 & - & 44 & weak/no \vz{} \\
V4046 Sgr   & $0.02 \pm 0.01$& 161 & $0.02 \pm 0.02$ & 370 &  weak/no \vz{} \\
MWC758  & $\sim 1.82$  & $>$220& - & & $\uparrow$ \vz{} [wind]  \\
\hline
\hline
CQ Tau & $\sim 0.70$ & &$\sim 0.63$  &&  Spirals $\uparrow$ and $\downarrow$  \\
MWC758 & $\sim 0.26$ & &  $\sim 0.73$  &&  Spirals $\uparrow$ and $\downarrow$\\
SY Cha & $\sim 0.70$ &&  - &&  Spirals $\uparrow$ and $\downarrow$\\
\hline
\end{tabular}
\end{table*}

\subsection{Origin of the vertical motions}
As illustrated in Sec.\,\ref{sec:benchmark}, planet-disk interactions and the three instabilities that we consider produce vertical motions of the order of 50-100 m/s, with oscillatory behavior across the disk radial extent and stronger amplitude for the planet case at the edge of the gaps \citep[e.g.,][]{Bi2021,Hu2025}. 
Winds, on the other hand, can produce high amplitude vertical motions \citep[e.g.,][]{Hu2025b}. The amplitude of the motions inferred from such models strongly depend on the assumed thermodynamics and vertical temperature structure \citep{Yun2025}, as well as on the disk geometry itself, preventing to infer a single, coherent, mechanism as the origin of the vertical motions for each disk. In addition, the processes at play are not mutually exclusive and multiple processes can occur at different locations of the disk.

For all targets, we compare the maximum vertical velocities to the local sound speed, c$_s$ and report the values in Table \ref{tab:cs}. Taking the absolute maximum values of \vz{}, we find that they range between a few \% to $\sim$25\% of the local sound speed (HD34282),  reaching $\sim$1.8 times the local sound speed for the outer wind in MWC758.

\paragraph{Fast vertical motions}  MWC 758, CQ Tau, and SY Cha show the strongest vertical motions, with non-Keplerian velocities appearing as spiral features and in the case of MWC 758, a fast wind in the outer disk. HD34282 and HD135344B exhibit the next highest vertical velocities, both mostly downward motion. Interestingly, all four of these disks are around Herbig Ae/F stars, suggesting a possible correlation between vertical gas motions and stellar mass. However, this trend is not consistent across the full exoALMA and MAPS sample: HD143006 (a G6 star) shows no significant vertical motion, while SY Cha or AS209, lower-mass K-type stars display some of the most prominent vertical velocities. 

It is also unclear whether the presence of strong vertical motions is directly related to the disk substructure, like cavities, gaps and rings, as the measured motions extend beyond any continuum structure. Nevertheless, the four targets with the strong vertical motions are transition disks with significant asymmetry in the continuum \citep{Curone2025}, spiral arms in the infrared scattered light and strong near-infrared excess \citep{PPVII_Benisty}, indicating very dynamically active disks. 

\paragraph{Spirals} While the current search for embedded companions within cavities has not yielded any candidate within the cavities of MWC758, CQ Tau, and SY Cha \cite[e.g.,][]{Ren2023}, the large spirals seen in velocity residuals might be tracing indirect signatures of these companions. Particularly relevant for these two disks, \citet{Ragusa2024} studied the non-Keplerian features arising in a circumbinary disk with eccentric motions. They consider a massive companion with a mass of 10\% of that of the central star and a semi-major axis of 15 au. Such a companion carves a 40\,au wide cavity with an eccentricity of 0.2, which in turn induces disk eccentricity as commonly seen in binary-disk interaction \citep[e.g.,][]{regaly2011,Miranda2017,Ragusa2020}. The change in the vertical component of gravity along the eccentric orbit produces vertical oscillations in the disk, inducing changes in the disk's local scale height. This oscillatory motion gives rise to strong vertical velocities, up to 400 m/s, and leads to two spiral arms, blue- and red-shifted, strikingly similar to what is observed in MWC 758 and CQ Tau. 
They also found that the amplitude of the vertical motions scales with height, which supports a stronger \vz{} for \twCO{} than \thCO{}. While the range of disk and companion parameters that would lead to such eccentric and vertical motions is too large to provide narrow constraints on the nature of the companions possibly embedded within the cavities of MWC 758 and CQ Tau, this work provides strong support for their presence. An alternative explanation for the observed spirals is that they are tracing the \textit{inner} spirals driven by an outer companion. This idea was originally proposed by \citet{Dong_2015} to reproduce the infrared scattered-light spirals in MWC 758 as a planetary-mass perturber in the cavity would induce \textit{outer} spirals that would be too tightly wound to match the pitch angle of IR spirals. However, the vertical motions in the inner spirals induced by massive planets do not show strong enough amplitudes to match our observations \citep{Rabago2021}, compared to the scenario of an eccentric stellar companion \citep{Ragusa2024}.  Further support  for an eccentric companion in the cavity comes from the measured eccentricity of the dust cavity in MWC 758 (e$\sim$0.1; \citealt{Dong2018}) and the tentative detection of spiral arms in the continuum of CQ Tau \citep{Zagaria2025}.
Interestingly, another object in our sample, HD135344B, shows similar disk properties but does not show such high velocity vertical motions. The differences might originate in the nature (planetary vs. stellar) and orbital parameters of the companion.

Large non-Keplerian velocity $m=2$ spirals can also be induced by temperature variations in the disks, induced by shadows. \citet{Zhang2024} and \citet{Ziampras2025} investigated the dynamical impact of shadows in transition disks that have misaligned inner disks and find that as the shadowed regions are cooler, asymmetric pressure gradients are generated, leading to spiral density waves detectable in their kinematical signatures with velocity about 100 m/s. However, MWC758 and CQ Tau residuals show $m=1$ spirals, hence this scenario does not appear to be likely. 

Large scale infall can also trigger spiral arms with large opening angles \citep{Hennebelle2017, Calcino2025} and result in disk eccentricity, naturally inducing vertical motions. While there is so far no indication of large scale streamer in the vicinity of CQ Tau and MWC758, we note that in MWC758, the outer disk seems to extend slightly further in a spiral-like shape in lower resolution \twCO{} cubes (see Figure 1 of \citealt{Galloway2025}). 

\paragraph{Radially varying \vz{}} For the targets analyzed with an axisymmetric method, as previously discussed, we see different behaviors. 
First, in both the exoALMA and MAPS targets (e.g., LkCa15, MWC480), we detect oscillatory up-and-down motions with amplitudes of a few tens of m/s. This behavior is reminiscent of VSI simulations \citep[see Figure \ref{fig:hydro};][]{Barraza2025}, although we cannot rule out that multiple mechanisms coherently contribute to the observed vertical motions. However, the relatively low vertical velocities, below 10\% of the sound speed at the emission heights traced by the \twCO{} and \thCO{} lines (roughly 2–3 scale heights), suggest that the disks are more dynamically quiet than expected from simulations. For example, large-scale upward flows associated with fully developed MRI can reach \vz{}~$> 0.1 c_s$ at these heights \citep{Flock2011}. It is however very likely that the limited angular resolution of our observations contributes to the low measured \vz{}, as small-scale turbulent eddies are washed out. A detailed comparison between the observed vertical motions and simulations is beyond the scope of this paper.
The second behavior, observed in e.g., J1604, J1852, and AS 209, is a transition from downward to upward vertical motions around 150\,au, which would be consistent with the launching of a disk wind in the outer regions. J1615, J1842, J1852 have downward motion associated with a surface gap, but J1604, HD34282 and J1615 show downward motion without a clear counterpart in surface nor intensity profiles. The origin of such downward motion over extended regions (100-200 au) is unclear.  

\paragraph{Winds} We consider whether the upward motion may be driven by a thermal pressure gradients. Fully ionised thermal winds would not be expected to have an appreciable molecular content \citep[e.g.][]{Hollenbach_ea_1993, Hollenback_ea_1994, Alexander_ea_2006}. However, this does not need  be the case for winds launched from the outer disk. In this case, winds may be driven by X-ray heating and advection can allow molecules to survive in the unbound gas \citep{Owen_ea_2010,Wang_Goodman_2017,Sellek_ea_2024}. Thermal winds are approximately launched when: 

\begin{equation}
    c_\mathrm{s} \gtrsim \sqrt{\frac{\gamma - 1}{4\gamma}
    } v_\mathrm{esc} \approx 0.3 v_\mathrm{esc},
\end{equation}where $c_\mathrm{s}$ is the
sound speed, $v_\mathrm{esc} = \sqrt{2G M_* /r}$ is the escape velocity and assuming $\gamma = 5/3$ \citep{Liffman_2003}. We can then consider how close to meeting this criterion the gas at the CO emission surface is for the disks in our sample. In Figure~\ref{fig:quiverplot}, we show that in several of the disks $c_\mathrm{s}/v_\mathrm{esc}$ is substantially lower than $0.3$. This would be in line with simulations that show that for Class II disks little CO is expected in the molecular component of the wind, unlike H$_2$ which is expected to be abundant \citep[][supported by evidence of H$_2$ winds in JWST data,  \citealt{Schwarz_ea_2025, Schwarz_ea_2025b}]{Sellek_ea_2024}. Indeed, any photoevaporative or MHD-driven wind might be active at higher altitudes above 4-5 scale heights \citep{Nakatani2018,Gressel2020} with much larger vertical velocities, however not contributing to the observed \twCO{} emission. Even in the outer regions of MWC758, where vertical velocities reach 1.8 times the sound speed, $c_\mathrm{s}/v_\mathrm{esc}$ is $\le$0.1, meaning that the material is likely located at the very base of a wind. However, in a couple of cases,  most evidently DM Tau and HD 34282, the highest point of the emission surface exhibits upward flow with $c_\mathrm{s}/v_\mathrm{esc}$ close to $0.3$. The coincidence of maximum emission height and strongest vertical velocity could plausibly exhibit CO at or close to the wind launching location. No work has yet explored the vertical velocity of molecular CO in a steady-state photoevaporative wind, and particularly whether $\sim 100$~m~s$^{-1}$ velocities may be reached under certain conditions. Although \citet{Sellek_ea_2024} suggested minimal CO advection into the wind, a systematic exploration of the conditions and degree to which CO can survive in the wind region is motivated by our findings.

\section{Summary and conclusions} \label{sec:concl}
In this paper, we present a comprehensive and uniform analysis of the kinematics of 14 exoALMA targets using their \twCO{} and \thCO{} emission lines, with the aim of characterizing their vertical motions, and include three MAPS targets for comparison. We find that the majority of the disks exhibit ubiquitous vertical motions, although their amplitudes and morphologies vary from one disk to another. We summarize our findings as follows:

\begin{itemize}
\item 
We derive vertical velocities from the line-of-sight velocity residuals, after subtracting a smooth Keplerian disk model using \discminer{}. We validate our \vz{} retrieval technique using hydrodynamical models of planet–disk interactions and three classes of disk instabilities (VSI, GI, and MRI) and find that, in those simulations, the complex kinematics does not lead to artificial vertical motions. We therefore demonstrate that in these simulations vertical velocities can be reliably retrieved down to amplitudes of about 10 m/s even if the disk dynamical state is high.

\item 
Depending on the level of asymmetry observed in the line-of-sight residual maps of our disks, we adopt two different approaches. For targets displaying clear spiral residual patterns (MWC758, CQ Tau, and SY Cha), we extract velocities in the 2D residual maps directly along the spiral spines. In MWC758 and CQ Tau, we detect strong upward and downward motions, reaching up to approximately 300 m/s, which appear as red- and blue-shifted spiral structures. We interpret these features as signatures of disk eccentricity induced by massive companions. 

\item 
For the remaining targets, we follow the methodology employed in previous studies and compute azimuthally averaged radial profiles. In those disks, two general behaviors emerge: (1) a radially oscillatory pattern of upward and downward motions, which likely trace meridional flows induced by disk instabilities, and (2) a smooth radial transition from downward to upward motion, which we interpret as evidence of disk winds in the outer regions. In general, these disks exhibit maximum vertical velocity amplitudes ranging from 30 to 100 m/s, corresponding to approximately 1 to 25\% of the local sound speed, except for MWC758 that has a high-velocity disk wind in the outer disk, with velocities up to 500 m/s, that is $\sim$1.8 times the local sound speed. The \thCO{} observations are noisier and cover a smaller radial extent than the \twCO{} line data, but in several targets they display similar amplitudes and radial trends to those derived from the \twCO{} line. 
\end{itemize}

In conclusion, based on the analysis of 14 exoALMA targets and 3 MAPS disks, we find that vertical motions are ubiquitous, as revealed by the \twCO{} and \thCO{} emission lines, indicating  meridional circulations in disks, although our angular resolution might not be high enough to catch small eddies. In objects that show strong dynamical sculpting, such as transition disks like MWC758 and CQ Tau, these vertical motions may provide an additional constraint for identifying the origin of their cavities. Strong molecular winds appear to be relatively rare in \twCO{} and \thCO{}, and the measured vertical motions might be probing the base of such winds.  Overall, the velocity structure traced in our discs is complex and highly diverse, suggesting that it is unlikely that a single physical mechanism is responsible for the vertical motions observed across all systems. This diversity motivates the need for further theoretical and numerical studies capable of explaining the full range of observed kinematics.

\section*{Acknowledgments}
We thank Zhaohuan Zhu and Ilaria Pascucci for insightful discussions, as well as the anonymous referee for constructive feedback. MB thanks Paulina Palma Bifani for her support. 
This paper makes use of the following ALMA data: ADS/JAO.ALMA\#2021.1.01123.L. ALMA is a partnership of ESO (representing its member states), NSF (USA) and NINS (Japan), together with NRC (Canada), MOST and ASIAA (Taiwan), and KASI (Republic of Korea), in cooperation with the Republic of Chile. The Joint ALMA Observatory is operated by ESO, AUI/NRAO and NAOJ. The National Radio Astronomy Observatory is a facility of the National Science Foundation operated under cooperative agreement by Associated Universities, Inc. We thank the North American ALMA Science Center (NAASC) for their generous support including providing computing facilities and financial support for student attendance at workshops and publications. JB acknowledges support from NASA XRP grant No. 80NSSC23K1312. MB, DF, JS have received funding from the European Research Council (ERC) under the European Union’s Horizon 2020 research and innovation programme (PROTOPLANETS, grant agreement No. 101002188). Computations by JS have been performed on the `Mesocentre SIGAMM' machine, hosted by Observatoire de la Cote d’Azur. PC acknowledges support by the Italian Ministero dell'Istruzione, Universit\`a e Ricerca through the grant Progetti Premiali 2012 – iALMA (CUP C52I13000140001) and by the ANID BASAL project FB210003. SF is funded by the European Union (ERC, UNVEIL, 101076613), and acknowledges financial contribution from PRIN-MUR 2022YP5ACE. MF is supported by a Grant-in-Aid from the Japan Society for the Promotion of Science (KAKENHI: No. JP22H01274). JDI acknowledges support from an STFC Ernest Rutherford Fellowship (ST/W004119/1) and a University Academic Fellowship from the University of Leeds. Support for AFI was provided by NASA through the NASA Hubble Fellowship grant No. HST-HF2-51532.001-A awarded by the Space Telescope Science Institute, which is operated by the Association of Universities for Research in Astronomy, Inc., for NASA, under contract NAS5-26555. CL has received funding from the European Union's Horizon 2020 research and innovation program under the Marie Sklodowska-Curie grant agreement No. 823823 (DUSTBUSTERS) and by the UK Science and Technology research Council (STFC) via the consolidated grant ST/W000997/1. GL has received funding from the European Union's Horizon 2020 research and innovation program under the Marie Sklodowska-Curie grant agreement No. 823823 (DUSTBUSTERS) and from PRIN-MUR 20228JPA3A. CP acknowledges Australian Research Council funding via FT170100040, DP18010423, DP220103767, and DP240103290. DP acknowledges Australian Research Council funding via DP18010423, DP220103767, and DP240103290. GR acknowledges funding from the Fondazione Cariplo, grant no. 2022-1217, and the European Research Council (ERC) under the European Union’s Horizon Europe Research \& Innovation Programme under grant agreement no. 101039651 (DiscEvol). FMe received funding from the European Research Council (ERC) under the European Union’s Horizon Europe research and innovation program (grant agreement No. 101053020, project Dust2Planets). H-WY acknowledges support from National Science and Technology Council (NSTC) in Taiwan through grant NSTC 113-2112-M-001-035- and from the Academia Sinica Career Development Award (AS-CDA-111-M03). GWF acknowledges support from the European Research Council (ERC) under the European Union Horizon 2020 research and innovation program (Grant agreement no. 815559 (MHDiscs)). GWF was granted access to the HPC resources of IDRIS under the allocation A0120402231 made by GENCI. Support for BZ was provided by The Brinson Foundation. Views and opinions expressed by ERC-funded scientists are however those of the author(s) only and do not necessarily reflect those of the European Union or the European Research Council. Neither the European Union nor the granting authority can be held responsible for them.

\bibliography{mybib}{}

\begin{thebibliography}{}
\expandafter\ifx\csname natexlab\endcsname\relax\def\natexlab#1{#1}\fi
\providecommand{\url}[1]{\href{#1}{#1}}
\providecommand{\dodoi}[1]{doi:~\href{http://doi.org/#1}{\nolinkurl{#1}}}
\providecommand{\doeprint}[1]{\href{http://ascl.net/#1}{\nolinkurl{http://ascl.net/#1}}}
\providecommand{\doarXiv}[1]{\href{https://arxiv.org/abs/#1}{\nolinkurl{https://arxiv.org/abs/#1}}}

\bibitem[{{Alexander} {et~al.}(2014){Alexander}, {Pascucci}, {Andrews},
  {Armitage}, \& {Cieza}}]{Alexander2014}
{Alexander}, R., {Pascucci}, I., {Andrews}, S., {Armitage}, P., \& {Cieza}, L.
  2014, in Protostars and Planets VI, ed. H.~{Beuther}, R.~S. {Klessen}, C.~P.
  {Dullemond}, \& T.~{Henning}, 475--496,
  \dodoi{10.2458/azu_uapress_9780816531240-ch021}

\bibitem[{{Alexander} {et~al.}(2006){Alexander}, {Clarke}, \&
  {Pringle}}]{Alexander_ea_2006}
{Alexander}, R.~D., {Clarke}, C.~J., \& {Pringle}, J.~E. 2006, \mnras, 369,
  216, \dodoi{10.1111/j.1365-2966.2006.10293.x}

\bibitem[{{Andrews} {et~al.}(2024){Andrews}, {Teague}, {Wirth}, {Huang}, \&
  {Zhu}}]{Andrews2024}
{Andrews}, S.~M., {Teague}, R., {Wirth}, C.~P., {Huang}, J., \& {Zhu}, Z. 2024,
  \apj, 970, 153, \dodoi{10.3847/1538-4357/ad5285}

\bibitem[{{Bacciotti} {et~al.}(2025){Bacciotti}, {Nony}, {Podio}, {Dougados},
  {Garufi}, {Cabrit}, {Codella}, {Zimniak}, \& {Ferreira}}]{Bacciotti2025}
{Bacciotti}, F., {Nony}, T., {Podio}, L., {et~al.} 2025, arXiv e-prints,
  arXiv:2501.03920, \dodoi{10.48550/arXiv.2501.03920}

\bibitem[{{Bae} {et~al.}(2025){Bae}, {Flock}, {Izquierdo}, {Kanagawa}, {Ono},
  {Pinte}, {Price}, {Rosotti}, {Wafflard-Fernandez}, {Lesur}, {Masset},
  {Andrews}, {Barraza-Alfaro}, {Benisty}, {Cataldi}, {Cuello}, {Curone},
  {Czekala}, {Facchini}, {Fasano}, {Galloway-Sprietsma}, {Hall}, {Hammond},
  {Huang}, {Lodato}, {Longarini}, {Stadler}, {Teague}, {Wilner}, {Winter},
  {W{\"o}lfer}, \& {Yoshida}}]{Bae2025}
{Bae}, J., {Flock}, M., {Izquierdo}, A., {et~al.} 2025, \apjl, 984, L12,
  \dodoi{10.3847/2041-8213/adc436}

\bibitem[{{Banzatti} {et~al.}(2022){Banzatti}, {Abernathy}, {Brittain},
  {Bosman}, {Pontoppidan}, {Boogert}, {Jensen}, {Carr}, {Najita}, {Grant},
  {Sigler}, {Sanchez}, {Kern}, \& {Rayner}}]{Banzatti2022}
{Banzatti}, A., {Abernathy}, K.~M., {Brittain}, S., {et~al.} 2022, \aj, 163,
  174, \dodoi{10.3847/1538-3881/ac52f0}

\bibitem[{{Barraza-Alfaro} {et~al.}(2024){Barraza-Alfaro}, {Flock}, \&
  {Henning}}]{Barraza2024}
{Barraza-Alfaro}, M., {Flock}, M., \& {Henning}, T. 2024, \aap, 683, A16,
  \dodoi{10.1051/0004-6361/202347726}

\bibitem[{{Barraza-Alfaro} {et~al.}(2025){Barraza-Alfaro}, {Flock},
  {B{\'e}thune}, {Teague}, {Bae}, {Benisty}, {Cataldi}, {Curone}, {Czekala},
  {Facchini}, {Fasano}, {Fukagawa}, {Galloway-Sprietsma}, {Garg}, {Hall},
  {Huang}, {Ilee}, {Izquierdo}, {Kanagawa}, {Koch}, {Lesur}, {Longarini},
  {Loomis}, {Orihara}, {Pinte}, {Price}, {Rosotti}, {Stadler},
  {Wafflard-Fernandez}, {Winter}, {W{\"o}lfer}, {Yen}, {Yoshida}, \&
  {Zawadzki}}]{Barraza2025}
{Barraza-Alfaro}, M., {Flock}, M., {B{\'e}thune}, W., {et~al.} 2025, \apjl,
  984, L21, \dodoi{10.3847/2041-8213/adc42d}

\bibitem[{{Bast} {et~al.}(2011){Bast}, {Brown}, {Herczeg}, {van Dishoeck}, \&
  {Pontoppidan}}]{Bast2011}
{Bast}, J.~E., {Brown}, J.~M., {Herczeg}, G.~J., {van Dishoeck}, E.~F., \&
  {Pontoppidan}, K.~M. 2011, \aap, 527, A119,
  \dodoi{10.1051/0004-6361/201015225}

\bibitem[{{Benisty} {et~al.}(2023){Benisty}, {Dominik}, {Follette}, {Garufi},
  {Ginski}, {Hashimoto}, {Keppler}, {Kley}, \& {Monnier}}]{PPVII_Benisty}
{Benisty}, M., {Dominik}, C., {Follette}, K., {et~al.} 2023, in Astronomical
  Society of the Pacific Conference Series, Vol. 534, Protostars and Planets
  VII, ed. S.~{Inutsuka}, Y.~{Aikawa}, T.~{Muto}, K.~{Tomida}, \& M.~{Tamura},
  605, \dodoi{10.48550/arXiv.2203.09991}

\bibitem[{{Bi} {et~al.}(2021){Bi}, {Lin}, \& {Dong}}]{Bi2021}
{Bi}, J., {Lin}, M.-K., \& {Dong}, R. 2021, \apj, 912, 107,
  \dodoi{10.3847/1538-4357/abef6b}

\bibitem[{{Calcino} {et~al.}(2025){Calcino}, {Price}, {Hilder}, {Christiaens},
  {Speedie}, \& {Ormel}}]{Calcino2025}
{Calcino}, J., {Price}, D.~J., {Hilder}, T., {et~al.} 2025, \mnras, 537, 2695,
  \dodoi{10.1093/mnras/staf135}

\bibitem[{{Casassus} {et~al.}(2022){Casassus}, {C{\'a}rcamo}, {Hales}, {Weber},
  \& {Dent}}]{Casassus_ea_2022}
{Casassus}, S., {C{\'a}rcamo}, M., {Hales}, A., {Weber}, P., \& {Dent}, B.
  2022, \apjl, 933, L4, \dodoi{10.3847/2041-8213/ac75e8}

\bibitem[{{Cridland} {et~al.}(2025){Cridland}, {Lega}, \&
  {Benisty}}]{Cridland2025}
{Cridland}, A.~J., {Lega}, E., \& {Benisty}, M. 2025, \aap, 693, A86,
  \dodoi{10.1051/0004-6361/202451140}

\bibitem[{{Curone} {et~al.}(2025){Curone}, {Facchini}, {Andrews}, {Testi},
  {Benisty}, {Czekala}, {Huang}, {Ilee}, {Isella}, {Lodato}, {Loomis},
  {Stadler}, {Winter}, {Bae}, {Barraza-Alfaro}, {Cataldi}, {Cuello}, {Fasano},
  {Flock}, {Fukagawa}, {Galloway-Sprietsma}, {Garg}, {Hall}, {Izquierdo},
  {Kanagawa}, {Lesur}, {Longarini}, {Menard}, {Orihara}, {Pinte}, {Price},
  {Rosotti}, {Teague}, {Wafflard-Fernandez}, {Wilner}, {W{\"o}lfer}, {Yen},
  {Yoshida}, \& {Zawadzki}}]{Curone2025}
{Curone}, P., {Facchini}, S., {Andrews}, S.~M., {et~al.} 2025, \apjl, 984, L9,
  \dodoi{10.3847/2041-8213/adc438}

\bibitem[{{de Valon} {et~al.}(2020){de Valon}, {Dougados}, {Cabrit}, {Louvet},
  {Zapata}, \& {Mardones}}]{devalon2020}
{de Valon}, A., {Dougados}, C., {Cabrit}, S., {et~al.} 2020, \aap, 634, L12,
  \dodoi{10.1051/0004-6361/201936950}

\bibitem[{{Dong} {et~al.}(2015){Dong}, {Zhu}, {Rafikov}, \&
  {Stone}}]{Dong_2015}
{Dong}, R., {Zhu}, Z., {Rafikov}, R.~R., \& {Stone}, J.~M. 2015, \apjl, 809,
  L5, \dodoi{10.1088/2041-8205/809/1/L5}

\bibitem[{{Dong} {et~al.}(2018){Dong}, {Liu}, {Eisner}, {Andrews}, {Fung},
  {Zhu}, {Chiang}, {Hashimoto}, {Liu}, {Casassus}, {Esposito}, {Hasegawa},
  {Muto}, {Pavlyuchenkov}, {Wilner}, {Akiyama}, {Tamura}, \&
  {Wisniewski}}]{Dong2018}
{Dong}, R., {Liu}, S.-y., {Eisner}, J., {et~al.} 2018, \apj, 860, 124,
  \dodoi{10.3847/1538-4357/aac6cb}

\bibitem[{{Dullemond} {et~al.}(2012){Dullemond}, {Juhasz}, {Pohl}, {Sereshti},
  {Shetty}, {Peters}, {Commercon}, \& {Flock}}]{Dullemond_2012}
{Dullemond}, C.~P., {Juhasz}, A., {Pohl}, A., {et~al.} 2012, {RADMC-3D: A
  multi-purpose radiative transfer tool}, Astrophysics Source Code Library,
  record ascl:1202.015.
\newblock \doeprint{1202.015}

\bibitem[{{Dutrey} {et~al.}(1998){Dutrey}, {Guilloteau}, {Prato}, {Simon},
  {Duvert}, {Schuster}, \& {Menard}}]{Dutrey1998}
{Dutrey}, A., {Guilloteau}, S., {Prato}, L., {et~al.} 1998, \aap, 338, L63

\bibitem[{{Fern{\'a}ndez-L{\'o}pez} {et~al.}(2020){Fern{\'a}ndez-L{\'o}pez},
  {Zapata}, {Rodr{\'\i}guez}, {Vazzano}, {Guzm{\'a}n}, \&
  {L{\'o}pez}}]{Fernandez2020}
{Fern{\'a}ndez-L{\'o}pez}, M., {Zapata}, L.~A., {Rodr{\'\i}guez}, L.~F.,
  {et~al.} 2020, \aj, 159, 171, \dodoi{10.3847/1538-3881/ab7a10}

\bibitem[{{Flock} {et~al.}(2011){Flock}, {Dzyurkevich}, {Klahr}, {Turner}, \&
  {Henning}}]{Flock2011}
{Flock}, M., {Dzyurkevich}, N., {Klahr}, H., {Turner}, N.~J., \& {Henning}, T.
  2011, \apj, 735, 122, \dodoi{10.1088/0004-637X/735/2/122}

\bibitem[{{Flock} {et~al.}(2015){Flock}, {Ruge}, {Dzyurkevich}, {Henning},
  {Klahr}, \& {Wolf}}]{Flock2015}
{Flock}, M., {Ruge}, J.~P., {Dzyurkevich}, N., {et~al.} 2015, \aap, 574, A68,
  \dodoi{10.1051/0004-6361/201424693}

\bibitem[{{Galloway-Sprietsma} {et~al.}(2023){Galloway-Sprietsma}, {Bae},
  {Teague}, {Benisty}, {Facchini}, {Aikawa}, {Alarc{\'o}n}, {Andrews},
  {Bergin}, {Cataldi}, {Cleeves}, {Czekala}, {Guzm{\'a}n}, {Huang}, {Law}, {Le
  Gal}, {Liu}, {Long}, {M{\'e}nard}, {{\"O}berg}, {Walsh}, \&
  {Wilner}}]{Galloway2023}
{Galloway-Sprietsma}, M., {Bae}, J., {Teague}, R., {et~al.} 2023, \apj, 950,
  147, \dodoi{10.3847/1538-4357/accae4}

\bibitem[{{Galloway-Sprietsma} {et~al.}(2025){Galloway-Sprietsma}, {Bae},
  {Izquierdo}, {Stadler}, {Longarini}, {Teague}, {Andrews}, {Winter},
  {Benisty}, {Facchini}, {Rosotti}, {Zawadzki}, {Pinte}, {Fasano},
  {Barraza-Alfaro}, {Cataldi}, {Cuello}, {Curone}, {Czekala}, {Flock},
  {Fukagawa}, {Gardner}, {Garg}, {Hall}, {Huang}, {Ilee}, {Kanagawa}, {Lesur},
  {Lodato}, {Loomis}, {Menard}, {Orihara}, {Price}, {Wafflard-Fernandez},
  {Wilner}, {W{\"o}lfer}, {Yen}, \& {Yoshida}}]{Galloway2025}
{Galloway-Sprietsma}, M., {Bae}, J., {Izquierdo}, A.~F., {et~al.} 2025, \apjl,
  984, L10, \dodoi{10.3847/2041-8213/adc437}

\bibitem[{{Gressel} {et~al.}(2020){Gressel}, {Ramsey}, {Brinch}, {Nelson},
  {Turner}, \& {Bruderer}}]{Gressel2020}
{Gressel}, O., {Ramsey}, J.~P., {Brinch}, C., {et~al.} 2020, \apj, 896, 126,
  \dodoi{10.3847/1538-4357/ab91b7}

\bibitem[{{Guilloteau} \& {Dutrey}(1994)}]{Guilloteau1994}
{Guilloteau}, S., \& {Dutrey}, A. 1994, \aap, 291, L23

\bibitem[{{Hennebelle} {et~al.}(2017){Hennebelle}, {Lesur}, \&
  {Fromang}}]{Hennebelle2017}
{Hennebelle}, P., {Lesur}, G., \& {Fromang}, S. 2017, \aap, 599, A86,
  \dodoi{10.1051/0004-6361/201629779}

\bibitem[{{Hilder} {et~al.}(2025){Hilder}, {Casey}, {Price}, {Pinte},
  {Izquierdo}, {Hardiman}, {Bae}, {Barraza-Alfaro}, {Benisty}, {Cataldi},
  {Curone}, {Czekala}, {Facchini}, {Fasano}, {Flock}, {Fukagawa},
  {Galloway-Sprietsma}, {Garg}, {Hall}, {Hammond}, {Huang}, {Ilee}, {Kanagawa},
  {Lesur}, {Longarini}, {Loomis}, {Orihara}, {Rosotti}, {Stadler}, {Teague},
  {Yen}, {Wafflard}, {Winter}, {W{\"o}lfer}, {Yoshida}, \&
  {Zawadzki}}]{Hilder2025}
{Hilder}, T., {Casey}, A.~R., {Price}, D.~J., {et~al.} 2025, \apjl, 984, L13,
  \dodoi{10.3847/2041-8213/adc435}

\bibitem[{{Hollenbach} {et~al.}(1994){Hollenbach}, {Johnstone}, {Lizano}, \&
  {Shu}}]{Hollenback_ea_1994}
{Hollenbach}, D., {Johnstone}, D., {Lizano}, S., \& {Shu}, F. 1994, \apj, 428,
  654, \dodoi{10.1086/174276}

\bibitem[{{Hollenbach} {et~al.}(1993){Hollenbach}, {Johnstone}, \&
  {Shu}}]{Hollenbach_ea_1993}
{Hollenbach}, D., {Johnstone}, D., \& {Shu}, F. 1993, in Astronomical Society
  of the Pacific Conference Series, Vol.~35, Massive Stars: Their Lives in the
  Interstellar Medium, ed. J.~P. {Cassinelli} \& E.~B. {Churchwell}, 26

\bibitem[{{Hu} {et~al.}(2024){Hu}, {Bae}, {Zhu}, \& {Wang}}]{Hu2024}
{Hu}, X., {Bae}, J., {Zhu}, Z., \& {Wang}, L. 2024, arXiv e-prints,
  arXiv:2412.15371, \dodoi{10.48550/arXiv.2412.15371}

\bibitem[{{Hu} {et~al.}(2025{\natexlab{a}}){Hu}, {Bae}, {Zhu}, \&
  {Wang}}]{Hu2025}
---. 2025{\natexlab{a}}, \apj, 986, 161, \dodoi{10.3847/1538-4357/add300}

\bibitem[{{Hu} {et~al.}(2025{\natexlab{b}}){Hu}, {Li}, {Bae}, \&
  {Zhu}}]{Hu2025b}
{Hu}, X., {Li}, Z.-Y., {Bae}, J., \& {Zhu}, Z. 2025{\natexlab{b}}, \mnras, 536,
  1374, \dodoi{10.1093/mnras/stae2681}

\bibitem[{{Isella} {et~al.}(2007){Isella}, {Testi}, {Natta}, {Neri}, {Wilner},
  \& {Qi}}]{Isella2007}
{Isella}, A., {Testi}, L., {Natta}, A., {et~al.} 2007, \aap, 469, 213,
  \dodoi{10.1051/0004-6361:20077385}

\bibitem[{{Izquierdo} {et~al.}(2022){Izquierdo}, {Facchini}, {Rosotti}, {van
  Dishoeck}, \& {Testi}}]{Izquierdo_ea_2022}
{Izquierdo}, A.~F., {Facchini}, S., {Rosotti}, G.~P., {van Dishoeck}, E.~F., \&
  {Testi}, L. 2022, \apj, 928, 2, \dodoi{10.3847/1538-4357/ac474d}

\bibitem[{{Izquierdo} {et~al.}(2021){Izquierdo}, {Testi}, {Facchini},
  {Rosotti}, \& {van Dishoeck}}]{Izquierdo_ea_2021}
{Izquierdo}, A.~F., {Testi}, L., {Facchini}, S., {Rosotti}, G.~P., \& {van
  Dishoeck}, E.~F. 2021, \aap, 650, A179, \dodoi{10.1051/0004-6361/202140779}

\bibitem[{{Izquierdo} {et~al.}(2023){Izquierdo}, {Testi}, {Facchini},
  {Rosotti}, {van Dishoeck}, {W{\"o}lfer}, \&
  {Paneque-Carre{\~n}o}}]{Izquierdo_ea_2023}
{Izquierdo}, A.~F., {Testi}, L., {Facchini}, S., {et~al.} 2023, \aap, 674,
  A113, \dodoi{10.1051/0004-6361/202245425}

\bibitem[{{Izquierdo} {et~al.}(2025){Izquierdo}, {Stadler},
  {Galloway-Sprietsma}, {Benisty}, {Pinte}, {Bae}, {Teague}, {Facchini},
  {W{\"o}lfer}, {Longarini}, {Curone}, {Andrews}, {Barraza-Alfaro}, {Cataldi},
  {Cuello}, {Czekala}, {Fasano}, {Flock}, {Fukagawa}, {Garg}, {Hall},
  {Hammond}, {Hilder}, {Huang}, {Ilee}, {Isella}, {Kanagawa}, {Lesur},
  {Lodato}, {Loomis}, {Orihara}, {Price}, {Rosotti}, {Testi}, {Yen},
  {Wafflard-Fernandez}, {Wilner}, {Winter}, {Yoshida}, \&
  {Zawadzki}}]{Izquierdo2025}
{Izquierdo}, A.~F., {Stadler}, J., {Galloway-Sprietsma}, M., {et~al.} 2025,
  \apjl, 984, L8, \dodoi{10.3847/2041-8213/adc439}

\bibitem[{{Koch} \& {Rosolowsky}(2015)}]{Koch2015}
{Koch}, E.~W., \& {Rosolowsky}, E.~W. 2015, \mnras, 452, 3435,
  \dodoi{10.1093/mnras/stv1521}

\bibitem[{{Koerner} {et~al.}(1993){Koerner}, {Sargent}, \&
  {Beckwith}}]{Koerner1993}
{Koerner}, D.~W., {Sargent}, A.~I., \& {Beckwith}, S.~V.~W. 1993, \icarus, 106,
  2, \dodoi{10.1006/icar.1993.1154}

\bibitem[{{Lega} {et~al.}(2024){Lega}, {Benisty}, {Cridland}, {Morbidelli},
  {Schulik}, \& {Lambrechts}}]{Lega2024}
{Lega}, E., {Benisty}, M., {Cridland}, A., {et~al.} 2024, \aap, 690, A183,
  \dodoi{10.1051/0004-6361/202450899}

\bibitem[{{Lesur} {et~al.}(2023){Lesur}, {Flock}, {Ercolano}, {Lin}, {Yang},
  {Barranco}, {Benitez-Llambay}, {Goodman}, {Johansen}, {Klahr}, {Laibe},
  {Lyra}, {Marcus}, {Nelson}, {Squire}, {Simon}, {Turner}, {Umurhan}, \&
  {Youdin}}]{Lesur2023}
{Lesur}, G., {Flock}, M., {Ercolano}, B., {et~al.} 2023, in Astronomical
  Society of the Pacific Conference Series, Vol. 534, Protostars and Planets
  VII, ed. S.~{Inutsuka}, Y.~{Aikawa}, T.~{Muto}, K.~{Tomida}, \& M.~{Tamura},
  465, \dodoi{10.48550/arXiv.2203.09821}

\bibitem[{{Liffman}(2003)}]{Liffman_2003}
{Liffman}, K. 2003, \pasa, 20, 337, \dodoi{10.1071/AS03019}

\bibitem[{{Longarini} {et~al.}(2025){Longarini}, {Lodato}, {Rosotti},
  {Andrews}, {Winter}, {Stadler}, {Izquierdo}, {Galloway-Sprietsma},
  {Facchini}, {Curone}, {Benisty}, {Teague}, {Bae}, {Barraza-Alfaro},
  {Cataldi}, {Czekala}, {Cuello}, {Fasano}, {Flock}, {Fukagawa}, {Garg},
  {Hall}, {Hammond}, {Hardiman}, {Hilder}, {Huang}, {Ilee}, {Isella},
  {Kanagawa}, {Lesur}, {Loomis}, {M{\'e}nard}, {Orihara}, {Pinte}, {Price},
  {Testi}, {Fernandez}, {W{\"o}lfer}, {Yen}, {Yoshida}, \&
  {Zawadzki}}]{Longarini2025}
{Longarini}, C., {Lodato}, G., {Rosotti}, G., {et~al.} 2025, \apjl, 984, L17,
  \dodoi{10.3847/2041-8213/adc431}

\bibitem[{{Loomis} {et~al.}(2025){Loomis}, {Facchini}, {Benisty}, {Curone},
  {Ilee}, {Cataldi}, {Yen}, {Teague}, {Pinte}, {Huang}, {Garg}, {Orihara},
  {Czekala}, {Zawadzki}, {Andrews}, {Wilner}, {Bae}, {Barraza-Alfaro},
  {Fasano}, {Flock}, {Fukagawa}, {Galloway-Sprietsma}, {Izquierdo}, {Kanagawa},
  {Lesur}, {Longarini}, {Menard}, {Price}, {Rosotti}, {Stadler},
  {Wafflard-Fernandez}, {W{\"o}lfer}, \& {Yoshida}}]{Loomis2025}
{Loomis}, R.~A., {Facchini}, S., {Benisty}, M., {et~al.} 2025, \apjl, 984, L7,
  \dodoi{10.3847/2041-8213/adc43a}

\bibitem[{{Louvet} {et~al.}(2018){Louvet}, {Dougados}, {Cabrit}, {Mardones},
  {M{\'e}nard}, {Tabone}, {Pinte}, \& {Dent}}]{Louvet2018}
{Louvet}, F., {Dougados}, C., {Cabrit}, S., {et~al.} 2018, \aap, 618, A120,
  \dodoi{10.1051/0004-6361/201731733}

\bibitem[{{Mignone} {et~al.}(2012){Mignone}, {Zanni}, {Tzeferacos}, {van
  Straalen}, {Colella}, \& {Bodo}}]{Mignone2012}
{Mignone}, A., {Zanni}, C., {Tzeferacos}, P., {et~al.} 2012, \apjs, 198, 7,
  \dodoi{10.1088/0067-0049/198/1/7}

\bibitem[{{Miranda} {et~al.}(2017){Miranda}, {Mu{\~n}oz}, \&
  {Lai}}]{Miranda2017}
{Miranda}, R., {Mu{\~n}oz}, D.~J., \& {Lai}, D. 2017, \mnras, 466, 1170,
  \dodoi{10.1093/mnras/stw3189}

\bibitem[{{Morbidelli} {et~al.}(2014){Morbidelli}, {Szul{\'a}gyi}, {Crida},
  {Lega}, {Bitsch}, {Tanigawa}, \& {Kanagawa}}]{Morbidelli2014}
{Morbidelli}, A., {Szul{\'a}gyi}, J., {Crida}, A., {et~al.} 2014, \icarus, 232,
  266, \dodoi{10.1016/j.icarus.2014.01.010}

\bibitem[{{Nakatani} {et~al.}(2018){Nakatani}, {Hosokawa}, {Yoshida}, {Nomura},
  \& {Kuiper}}]{Nakatani2018}
{Nakatani}, R., {Hosokawa}, T., {Yoshida}, N., {Nomura}, H., \& {Kuiper}, R.
  2018, \apj, 865, 75, \dodoi{10.3847/1538-4357/aad9fd}

\bibitem[{{Norfolk} {et~al.}(2022){Norfolk}, {Pinte}, {Calcino}, {Hammond},
  {van der Marel}, {Price}, {Maddison}, {Christiaens}, {Gonzalez}, {Blakely},
  {Rosotti}, \& {Ginski}}]{Norfolk2022}
{Norfolk}, B.~J., {Pinte}, C., {Calcino}, J., {et~al.} 2022, \apjl, 936, L4,
  \dodoi{10.3847/2041-8213/ac85ed}

\bibitem[{{{\"O}berg} {et~al.}(2021){{\"O}berg}, {Guzm{\'a}n}, {Walsh},
  {Aikawa}, {Bergin}, {Law}, {Loomis}, {Alarc{\'o}n}, {Andrews}, {Bae},
  {Bergner}, {Boehler}, {Booth}, {Bosman}, {Calahan}, {Cataldi}, {Cleeves},
  {Czekala}, {Furuya}, {Huang}, {Ilee}, {Kurtovic}, {Le Gal}, {Liu}, {Long},
  {M{\'e}nard}, {Nomura}, {P{\'e}rez}, {Qi}, {Schwarz}, {Sierra}, {Teague},
  {Tsukagoshi}, {Yamato}, {van't Hoff}, {Waggoner}, {Wilner}, \&
  {Zhang}}]{oberg2021}
{{\"O}berg}, K.~I., {Guzm{\'a}n}, V.~V., {Walsh}, C., {et~al.} 2021, \apjs,
  257, 1, \dodoi{10.3847/1538-4365/ac1432}

\bibitem[{{Owen} {et~al.}(2010){Owen}, {Ercolano}, {Clarke}, \&
  {Alexander}}]{Owen_ea_2010}
{Owen}, J.~E., {Ercolano}, B., {Clarke}, C.~J., \& {Alexander}, R.~D. 2010,
  \mnras, 401, 1415, \dodoi{10.1111/j.1365-2966.2009.15771.x}

\bibitem[{{Paneque-Carre{\~n}o} {et~al.}(2023){Paneque-Carre{\~n}o},
  {Miotello}, {van Dishoeck}, {Tabone}, {Izquierdo}, \&
  {Facchini}}]{paneque+2023}
{Paneque-Carre{\~n}o}, T., {Miotello}, A., {van Dishoeck}, E.~F., {et~al.}
  2023, \aap, 669, A126, \dodoi{10.1051/0004-6361/202244428}

\bibitem[{{Pascucci} {et~al.}(2023){Pascucci}, {Cabrit}, {Edwards}, {Gorti},
  {Gressel}, \& {Suzuki}}]{Pascucci2023}
{Pascucci}, I., {Cabrit}, S., {Edwards}, S., {et~al.} 2023, in Astronomical
  Society of the Pacific Conference Series, Vol. 534, Protostars and Planets
  VII, ed. S.~{Inutsuka}, Y.~{Aikawa}, T.~{Muto}, K.~{Tomida}, \& M.~{Tamura},
  567, \dodoi{10.48550/arXiv.2203.10068}

\bibitem[{{Pascucci} {et~al.}(2020){Pascucci}, {Banzatti}, {Gorti}, {Fang},
  {Pontoppidan}, {Alexander}, {Ballabio}, {Edwards}, {Salyk}, {Sacco},
  {Flaccomio}, {Blake}, {Carmona}, {Hall}, {Kamp}, {K{\"a}ufl}, {Meeus},
  {Meyer}, {Pauly}, {Steendam}, \& {Sterzik}}]{Pascucci2020}
{Pascucci}, I., {Banzatti}, A., {Gorti}, U., {et~al.} 2020, \apj, 903, 78,
  \dodoi{10.3847/1538-4357/abba3c}

\bibitem[{{Pascucci} {et~al.}(2025){Pascucci}, {Beck}, {Cabrit}, {Bajaj},
  {Edwards}, {Louvet}, {Najita}, {Skinner}, {Gorti}, {Salyk}, {Brittain},
  {Krijt}, {Muzerolle Page}, {Ruaud}, {Schwarz}, {Semenov}, {Duch{\^e}ne}, \&
  {Villenave}}]{Pascucci2025}
{Pascucci}, I., {Beck}, T.~L., {Cabrit}, S., {et~al.} 2025, Nature Astronomy,
  9, 81, \dodoi{10.1038/s41550-024-02385-7}

\bibitem[{{Pinte} {et~al.}(2018){Pinte}, {Price}, {M{\'e}nard}, {Duch{\^e}ne},
  {Dent}, {Hill}, {de Gregorio-Monsalvo}, {Hales}, \&
  {Mentiplay}}]{Pinte_ea_2018b}
{Pinte}, C., {Price}, D.~J., {M{\'e}nard}, F., {et~al.} 2018, \apjl, 860, L13,
  \dodoi{10.3847/2041-8213/aac6dc}

\bibitem[{{Rabago} \& {Zhu}(2021)}]{Rabago2021}
{Rabago}, I., \& {Zhu}, Z. 2021, \mnras, 502, 5325,
  \dodoi{10.1093/mnras/stab447}

\bibitem[{{Ragusa} {et~al.}(2020){Ragusa}, {Alexander}, {Calcino}, {Hirsh}, \&
  {Price}}]{Ragusa2020}
{Ragusa}, E., {Alexander}, R., {Calcino}, J., {Hirsh}, K., \& {Price}, D.~J.
  2020, \mnras, 499, 3362, \dodoi{10.1093/mnras/staa2954}

\bibitem[{{Ragusa} {et~al.}(2024){Ragusa}, {Lynch}, {Laibe}, {Longarini}, \&
  {Ceppi}}]{Ragusa2024}
{Ragusa}, E., {Lynch}, E., {Laibe}, G., {Longarini}, C., \& {Ceppi}, S. 2024,
  \aap, 686, A264, \dodoi{10.1051/0004-6361/202449583}

\bibitem[{{Reg{\'a}ly} {et~al.}(2011){Reg{\'a}ly}, {S{\'a}ndor}, {Dullemond},
  \& {Kiss}}]{regaly2011}
{Reg{\'a}ly}, Z., {S{\'a}ndor}, Z., {Dullemond}, C.~P., \& {Kiss}, L.~L. 2011,
  \aap, 528, A93, \dodoi{10.1051/0004-6361/201016152}

\bibitem[{{Ren} {et~al.}(2023){Ren}, {Benisty}, {Ginski}, {Tazaki}, {Wallack},
  {Milli}, {Garufi}, {Bae}, {Facchini}, {M{\'e}nard}, {Pinilla}, {Swastik},
  {Teague}, \& {Wahhaj}}]{Ren2023}
{Ren}, B.~B., {Benisty}, M., {Ginski}, C., {et~al.} 2023, \aap, 680, A114,
  \dodoi{10.1051/0004-6361/202347353}

\bibitem[{{Riols} {et~al.}(2020){Riols}, {Lesur}, \& {Menard}}]{Riols2020}
{Riols}, A., {Lesur}, G., \& {Menard}, F. 2020, \aap, 639, A95,
  \dodoi{10.1051/0004-6361/201937418}

\bibitem[{{Schwarz} {et~al.}(2025{\natexlab{a}}){Schwarz}, {Samland},
  {Olofsson}, {Henning}, {Sellek}, {G{\"u}del}, {Tabone}, {Kamp}, {Lagage},
  {van Dishoeck}, {Caratti o Garatti}, {Glauser}, {Ray}, {Arabhavi},
  {Christiaens}, {Franceschi}, {Gasman}, {Grant}, {Kanwar}, {Kaeufer},
  {Kurtovic}, {Perotti}, {Temmink}, \& {Vlasblom}}]{Schwarz_ea_2025}
{Schwarz}, K.~R., {Samland}, M., {Olofsson}, G., {et~al.} 2025{\natexlab{a}},
  \apj, 980, 148, \dodoi{10.3847/1538-4357/adaa79}

\bibitem[{{Schwarz} {et~al.}(2025{\natexlab{b}}){Schwarz}, {Samland},
  {Olofsson}, {Henning}, {Sellek}, {G{\"u}del}, {Tabone}, {Kamp}, {Lagage},
  {van Dishoeck}, {Caratti o Garatti}, {Glauser}, {Ray}, {Arabhavi},
  {Christiaens}, {Franceschi}, {Gasman}, {Grant}, {Kanwar}, {Kaeufer},
  {Kurtovic}, {Perotti}, {Temmink}, \& {Vlasblom}}]{Schwarz_ea_2025b}
---. 2025{\natexlab{b}}, \apj, 991, 232, \dodoi{10.3847/1538-4357/ae0603}

\bibitem[{{Sellek} {et~al.}(2024){Sellek}, {Grassi}, {Picogna}, {Rab},
  {Clarke}, \& {Ercolano}}]{Sellek_ea_2024}
{Sellek}, A.~D., {Grassi}, T., {Picogna}, G., {et~al.} 2024, \aap, 690, A296,
  \dodoi{10.1051/0004-6361/202450171}

\bibitem[{{Stadler} {et~al.}(2025){Stadler}, {Benisty}, {Winter}, {Izquierdo},
  {Longarini}, {Galloway-Sprietsma}, {Curone}, {Andrews}, {Bae}, {Facchini},
  {Rosotti}, {Teague}, {Barraza-Alfaro}, {Cataldi}, {Cuello}, {Czekala},
  {Fasano}, {Flock}, {Fukagawa}, {Garg}, {Hall}, {Hammond}, {Hilder}, {Huang},
  {Ilee}, {Kanagawa}, {Lesur}, {Lodato}, {Loomis}, {Menard}, {Orihara},
  {Pinte}, {Price}, {Yen}, {Wafflard-Fernandez}, {Wilner}, {W{\"o}lfer},
  {Yoshida}, \& {Zawadzki}}]{Stadler2025}
{Stadler}, J., {Benisty}, M., {Winter}, A.~J., {et~al.} 2025, \apjl, 984, L11,
  \dodoi{10.3847/2041-8213/adb152}

\bibitem[{{Tabone} {et~al.}(2017){Tabone}, {Cabrit}, {Bianchi}, {Ferreira},
  {Pineau des For{\^e}ts}, {Codella}, {Gusdorf}, {Gueth}, {Podio}, \&
  {Chapillon}}]{tabone2017}
{Tabone}, B., {Cabrit}, S., {Bianchi}, E., {et~al.} 2017, \aap, 607, L6,
  \dodoi{10.1051/0004-6361/201731691}

\bibitem[{{Tabone} {et~al.}(2020){Tabone}, {Cabrit}, {Pineau des For{\^e}ts},
  {Ferreira}, {Gusdorf}, {Podio}, {Bianchi}, {Chapillon}, {Codella}, \&
  {Gueth}}]{tabone2020}
{Tabone}, B., {Cabrit}, S., {Pineau des For{\^e}ts}, G., {et~al.} 2020, \aap,
  640, A82, \dodoi{10.1051/0004-6361/201834377}

\bibitem[{{Tabone} {et~al.}(2025){Tabone}, {Rosotti}, {Trapman}, {Pinilla},
  {Pascucci}, {Somigliana}, {Alexander}, {Vioque}, {Anania}, {Kuznetsova},
  {Zhang}, {P{\'e}rez}, {Cieza}, {Carpenter}, {Deng}, {Agurto-Gangas},
  {Ruiz-Rodriguez}, {Sierra}, {Kurtovic}, {Miley}, {Gonz{\'a}lez-Ruilova},
  {TorresVillanueva}, {Hogerheijde}, {Schwarz}, {Toci}, {Testi}, \&
  {Lodato}}]{Tabone2025}
{Tabone}, B., {Rosotti}, G.~P., {Trapman}, L., {et~al.} 2025, \apj, 989, 7,
  \dodoi{10.3847/1538-4357/adc7b1}

\bibitem[{{Teague} {et~al.}(2019){Teague}, {Bae}, \&
  {Bergin}}]{Teague_ea_2019b}
{Teague}, R., {Bae}, J., \& {Bergin}, E.~A. 2019, \nat, 574, 378,
  \dodoi{10.1038/s41586-019-1642-0}

\bibitem[{{Teague} {et~al.}(2018){Teague}, {Bae}, {Bergin}, {Birnstiel}, \&
  {Foreman-Mackey}}]{Teague_ea_2018b}
{Teague}, R., {Bae}, J., {Bergin}, E.~A., {Birnstiel}, T., \& {Foreman-Mackey},
  D. 2018, \apjl, 860, L12, \dodoi{10.3847/2041-8213/aac6d7}

\bibitem[{{Teague} {et~al.}(2021){Teague}, {Bae}, {Aikawa}, {Andrews},
  {Bergin}, {Bergner}, {Boehler}, {Booth}, {Bosman}, {Cataldi}, {Czekala},
  {Guzm{\'a}n}, {Huang}, {Ilee}, {Law}, {Le Gal}, {Long}, {Loomis},
  {M{\'e}nard}, {{\"O}berg}, {P{\'e}rez}, {Schwarz}, {Sierra}, {Walsh},
  {Wilner}, {Yamato}, \& {Zhang}}]{Teague2021}
{Teague}, R., {Bae}, J., {Aikawa}, Y., {et~al.} 2021, \apjs, 257, 18,
  \dodoi{10.3847/1538-4365/ac1438}

\bibitem[{{Teague} {et~al.}(2022){Teague}, {Bae}, {Andrews}, {Benisty},
  {Bergin}, {Facchini}, {Huang}, {Longarini}, \& {Wilner}}]{Teague_ea_2022b}
{Teague}, R., {Bae}, J., {Andrews}, S.~M., {et~al.} 2022, \apj, 936, 163,
  \dodoi{10.3847/1538-4357/ac88ca}

\bibitem[{{Teague} {et~al.}(2025){Teague}, {Benisty}, {Facchini}, {Fukagawa},
  {Pinte}, {Andrews}, {Bae}, {Barraza-Alfaro}, {Cataldi}, {Cuello}, {Curone},
  {Czekala}, {Fasano}, {Flock}, {Galloway-Sprietsma}, {Garg}, {Hall},
  {Hammond}, {Hilder}, {Huang}, {Ilee}, {Izquierdo}, {Kanagawa}, {Lesur},
  {Lodato}, {Longarini}, {Loomis}, {Masset}, {Menard}, {Orihara}, {Price},
  {Rosotti}, {Stadler}, {Testi}, {Yen}, {Wafflard-Fernandez}, {Wilner},
  {Winter}, {W{\"o}lfer}, {Yoshida}, \& {Zawadzki}}]{Teague2025}
{Teague}, R., {Benisty}, M., {Facchini}, S., {et~al.} 2025, \apjl, 984, L6,
  \dodoi{10.3847/2041-8213/adc43b}

\bibitem[{{Wafflard-Fernandez} \& {Lesur}(2025)}]{Wafflard2025}
{Wafflard-Fernandez}, G., \& {Lesur}, G. 2025, \aap, 696, A8,
  \dodoi{10.1051/0004-6361/202453541}

\bibitem[{{Wang} \& {Goodman}(2017)}]{Wang_Goodman_2017}
{Wang}, L., \& {Goodman}, J. 2017, \apj, 847, 11,
  \dodoi{10.3847/1538-4357/aa8726}

\bibitem[{{W{\"o}lfer} {et~al.}(2025){W{\"o}lfer}, {Barraza-Alfaro}, {Teague},
  {Curone}, {Benisty}, {Fukagawa}, {Bae}, {Cataldi}, {Czekala}, {Facchini},
  {Fasano}, {Flock}, {Galloway-Sprietsma}, {Garg}, {Hall}, {Huang}, {Ilee},
  {Izquierdo}, {Kanagawa}, {Lesur}, {Longarini}, {Loomis}, {Menard}, {Nath},
  {Orihara}, {Pinte}, {Price}, {Rosotti}, {Stadler}, {Wafflard-Fernandez},
  {Winter}, {Yen}, {Yoshida}, \& {Zawadzki}}]{Wolfer2025}
{W{\"o}lfer}, L., {Barraza-Alfaro}, M., {Teague}, R., {et~al.} 2025, \apjl,
  984, L22, \dodoi{10.3847/2041-8213/adc42c}

\bibitem[{{Yu} {et~al.}(2021){Yu}, {Teague}, {Bae}, \& {{\"O}berg}}]{Yu2021}
{Yu}, H., {Teague}, R., {Bae}, J., \& {{\"O}berg}, K. 2021, \apjl, 920, L33,
  \dodoi{10.3847/2041-8213/ac283e}

\bibitem[{{Yun} {et~al.}(2025){Yun}, {Kim}, {Bae}, \& {Han}}]{Yun2025}
{Yun}, H.-G., {Kim}, W.-T., {Bae}, J., \& {Han}, C. 2025, \apj, 980, 15,
  \dodoi{10.3847/1538-4357/ad9f42}

\bibitem[{{Zagaria} {et~al.}(2025){Zagaria}, {Jiang}, {Cataldi}, {Facchini},
  {Benisty}, {Aikawa}, {Andrews}, {Bae}, {Barraza-Alfaro}, {Curone}, {Czekala},
  {Fasano}, {Hall}, {Hammond}, {Huang}, {Ilee}, {Izquierdo}, {Lawrence},
  {Lodato}, {M{\'e}nard}, {Pinte}, {Rosotti}, {Stadler}, {Teague}, {Testi},
  {Wilner}, {Winter}, \& {Yoshida}}]{Zagaria2025}
{Zagaria}, F., {Jiang}, H., {Cataldi}, G., {et~al.} 2025, arXiv e-prints,
  arXiv:2506.16481, \dodoi{10.48550/arXiv.2506.16481}

\bibitem[{{Zhang} \& {Zhu}(2024)}]{Zhang2024}
{Zhang}, S., \& {Zhu}, Z. 2024, \apjl, 974, L38,
  \dodoi{10.3847/2041-8213/ad815f}

\bibitem[{{Ziampras} {et~al.}(2025){Ziampras}, {Dullemond}, {Birnstiel},
  {Benisty}, \& {Nelson}}]{Ziampras2025}
{Ziampras}, A., {Dullemond}, C.~P., {Birnstiel}, T., {Benisty}, M., \&
  {Nelson}, R.~P. 2025, \mnras, 540, 1185, \dodoi{10.1093/mnras/staf785}

\end{thebibliography}
\bibliographystyle{aasjournal}

\appendix
\restartappendixnumbering

\section{Data table}
Table\,\ref{tab:asymmfactor} gives the information on the \twCO{}, \thCO{} cubes used in this paper, the method to compute the centroid map, the outer disk radius considered in the analysis, as well as the mean and standard deviation of the corresponding folded maps. Table \ref{tab:vz_structure} gives the location of \vz{} peaks and dips for \twCO{} and \thCO{}. 

\begin{table*}[!t]
\begin{centering}
    \begin{tabular}{c||ccc|cc||ccc|cc||c}
    Target  & \twCO{}  & Centroid &  Rout cut  & Mean  & Std & \thCO{}  & Centroid & Rout cut & Mean  & Std & Comment\\
     & res. &   &[\% Rout] & [km/s] &  [km/s]  & res. &  &[\% Rout]& [km/s]  & [km/s]& \\
    \hline
    MWC758   & 0.15" & G & 0.80 & 0.25 & 0.18 & 0.30" & G & 0.65& 0.08 & 0.03 & spirals\\
    SY\,Cha	  &  0.30" & DB & 1.0 & 0.22 & 0.18& 0.30" & DB & 1.0  &  0.32& 0.13   & spirals\\
    R$<$360\,au&        &   & 0.6 &  0.20 & 0.14 &     &  & 0.80 &  0.315& 0.13 & \\
    CQ\,Tau	  & 0.15" & G & 1.0 & 0.22 & 0.12& 0.15" & G & 0.85 & 0.18 & 0.10 & spirals\\
    AA\,Tau     & 0.30" & DB & 0.82 & 0.21 & 0.11& 0.30" & DB & 0.30 & 0.29 & 0.16 & excluded\\
    HD135344B  & 0.15" & G & 1.0 & 0.14 & 0.05 & 0.15" & G & 0.95 & 0.12 & 0.03 & 1D\\
    LkCa15	   & 0.30" & DB & 1.0 &  0.14 & 0.09& 0.30" & DB & 0.8 &  0.24 & 0.15 & 1D\\
    HD34282     & 0.30" &DB & 0.9 &  0.14 & 0.01& 0.30" & DB & 1.0 & 0.33 & 0.24 & 1D\\
    J1842  & 0.15" & DB &  0.92 & 0.12 & 0.07& 0.15" & DB & 0.50 & 0.11 & 0.05 & 1D\\
    DM Tau      & 0.30" & DB & 1.0 & 0.11 & 0.07& 0.30" & DB & 0.68 & 0.035 & 0.02 & 1D\\
    J1615 & 0.30" & DB & 0.95 & 0.10 & 0.07 & 0.30" & DB & 0.85 & 0.14 & 0.10 & 1D\\
    HD143006   & 0.15" & G & 1.0 & 0.06 & 0.03& 0.15" & G & 0.92 & 0.07 & 0.02&1D\\
    J1852  & 0.15" & G & 0.9 & 0.05 & 0.02& 0.15" & G & 0.6 & 0.06 & 0.03 & 1D\\
    J1604  & 0.15" & B & 1.0 &  0.05 & 0.02& 0.15" & B & 1.0 & 0.05 & 0.02 & 1D\\
    PDS66   & 0.15" & G & 0.92 & 0.04 & 0.01& 0.15" & G & 0.55 & 0.13& 0.06  & 1D\\
    V4046\,Sgr  & 0.30" & G & 1.0 & 0.02 & 0.01& 0.30" & G & 0.95 & 0.03 & 0.01 & 1D\\
    \hline
    HD163296  & 0.15" & DB & 0.95 & 0.21 &  0.13 & - & - & -  &  & & 1D \\
    AS209  & 0.15" & G & 0.82 &0.25 &  0.10& - & - & -  & -  & & 1D\\
    MWC480  & 0.15" & G  & 1.0 &0.09 & 0.04 & - & - & -  & -  & & 1D\\
    \end{tabular}
    \caption{\label{tab:asymmfactor} Data cubes used in this paper. For each line, we provide the angular resolution and the method used to retrieve the centroid velocity \citep[][G: Gaussian; B: Bell; DB: double Bell.]{Izquierdo2025}. We also report the outer radii considered in our analysis in fraction of the cube outer radius and the mean and standard deviations computed in the corresponding folded residual maps. The last three rows correspond to the MAPS cubes. For SY~Cha we report two lines, the second one corresponding to the folded map after we cut off the diffuse emission in the outer disk.}
    \end{centering}
\end{table*}

\begin{table}[t]
\centering
\caption{Radial locations (in au) of \vz{} peaks and dips for \twCO{} and \thCO{}. Bold numbers represent peaks and valleys associated with a surface or intensity substructure, within a beamsize. We exclude HD135344B, J1604 and PDS66 (see text). We refer to Table 4 of \cite{Galloway2025} for a complete list of surface and intensity structures.  
\label{tab:vz_structure}}
\begin{tabular}{c || c | c || c |c}
\hline
 & \multicolumn{2}{c||}{\twCO{}} & \multicolumn{2}{c}{\thCO{}} \\
Source &  Peaks & Dips & Peaks & Dips \\ 
\hline
DM\,Tau & 410 & 583  & -- & -- \\
HD135344B & 132 & \textbf{66} & 122 & 71 \\
HD143006 & -- & -- & -- & -- \\
HD34282 & -- & 533 & -- & 185, 347 \\
J1604 & 201 & 157 & -- & 81\\
J1615 & 316 & \textbf{234} & \textbf{250} & \textbf{94}, 386 \\
J1842 & 266 & \textbf{130}, 204  & -- & \textbf{79, 142}\\
J1852 & -- & \textbf{110} & -- & \textbf{121} \\
LkCa15 & \textbf{330, 472} & \textbf{236, 401, 566} & -- & 248\\
V4046 Sgr & \textbf{161} & - & - & - \\
\hline
\end{tabular}
\vspace{0.3cm}
\end{table}

\section{Beam convolution effect}
\label{sec:beamconvolution}
\begin{figure}[!h]
    \centering
    \includegraphics[width=0.8\linewidth]{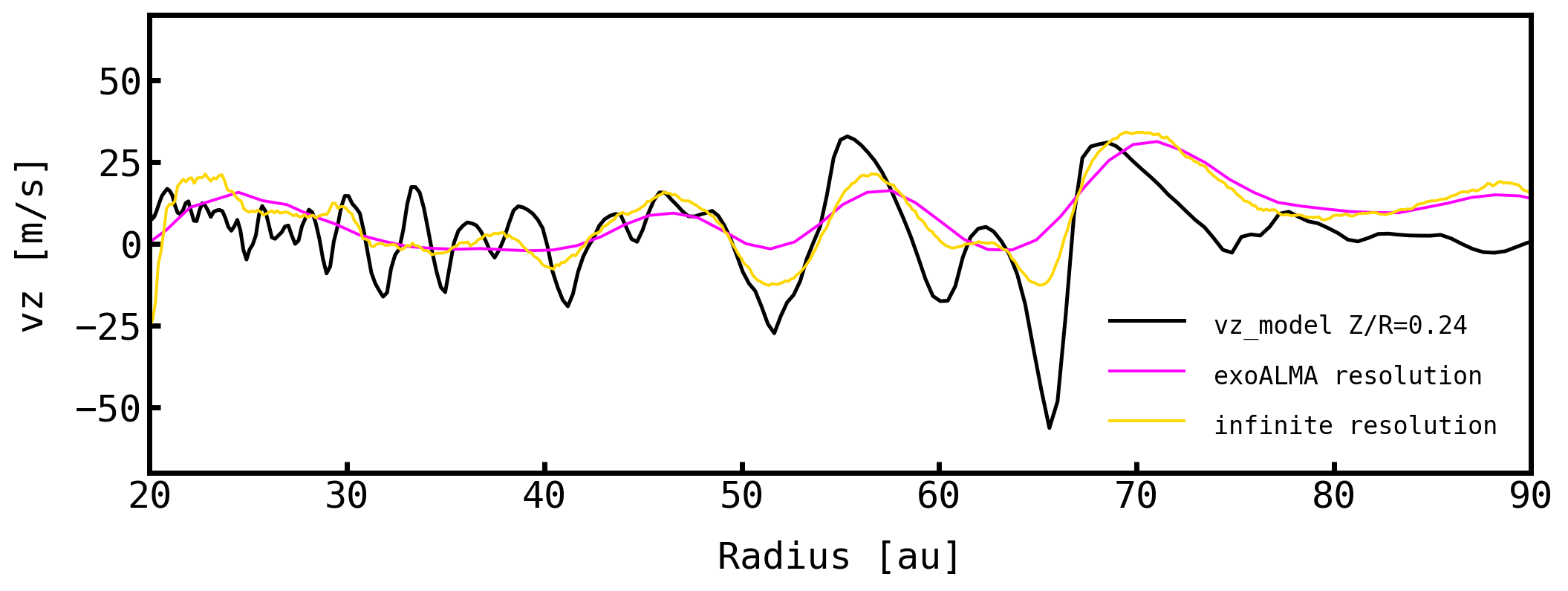}
    \caption{Comparison between the vertical velocities extracted from the VSI simulation at a given height with those extracted after post-processing and beam convolution.}
    \label{fig:beamconvolution}
\end{figure}

In Fig.\,\ref{fig:beamconvolution}, we illustrate the effects of radiative transfer and beam convolution on the simulations of the vertical shear instability presented in Sec.\,\ref{sec:benchmark}. We show, on the one hand, the vertical velocities extracted directly from the model at a given height, $Z/R = 0.24$ (black line), and, on the other hand, the velocities extracted from the post-processed simulated image after radiative transfer, both at infinite angular resolution (yellow line) and at the exoALMA angular resolution (pink line).

This comparison highlights two key aspects. First, the model velocities at $Z/R = 0.24$ have higher amplitudes than those obtained after radiative transfer, even at infinite angular resolution. This indicates that, in the radiative-transfer post-processed image, multiple contributions from velocities at different vertical layers around the $\tau = 1$ surface can smear out the observed signal. 

Second, by comparing the infinite angular resolution case with the exoALMA resolution, we find that beam convolution at the current exoALMA resolution can significantly reduce the amplitudes of the intrinsic velocities. This effect would be enhanced for processes in which multiple velocity components coexist at the same location, but minimized when the flow has a single coherent direction, as in the case of wind motions. Overall, this suggests that the vertical motions measured in our data are likely lower limits on the true velocity amplitudes.

\section{Additional figures}

\begin{figure*}[t]
    \centering
    \begin{tabular}{cccc}
        \includegraphics[width=0.25\textwidth]{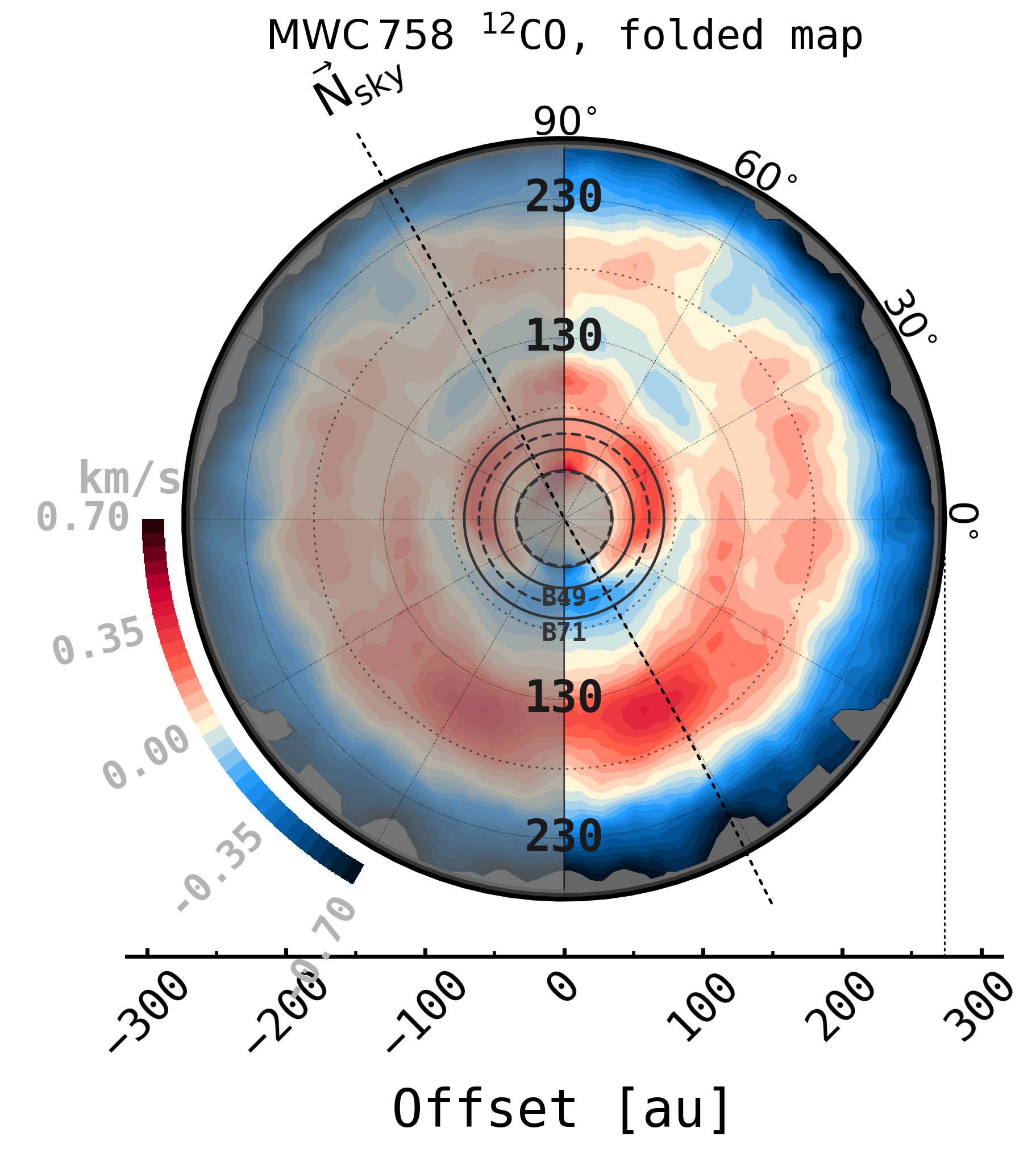} & 
        \includegraphics[width=0.25\textwidth]{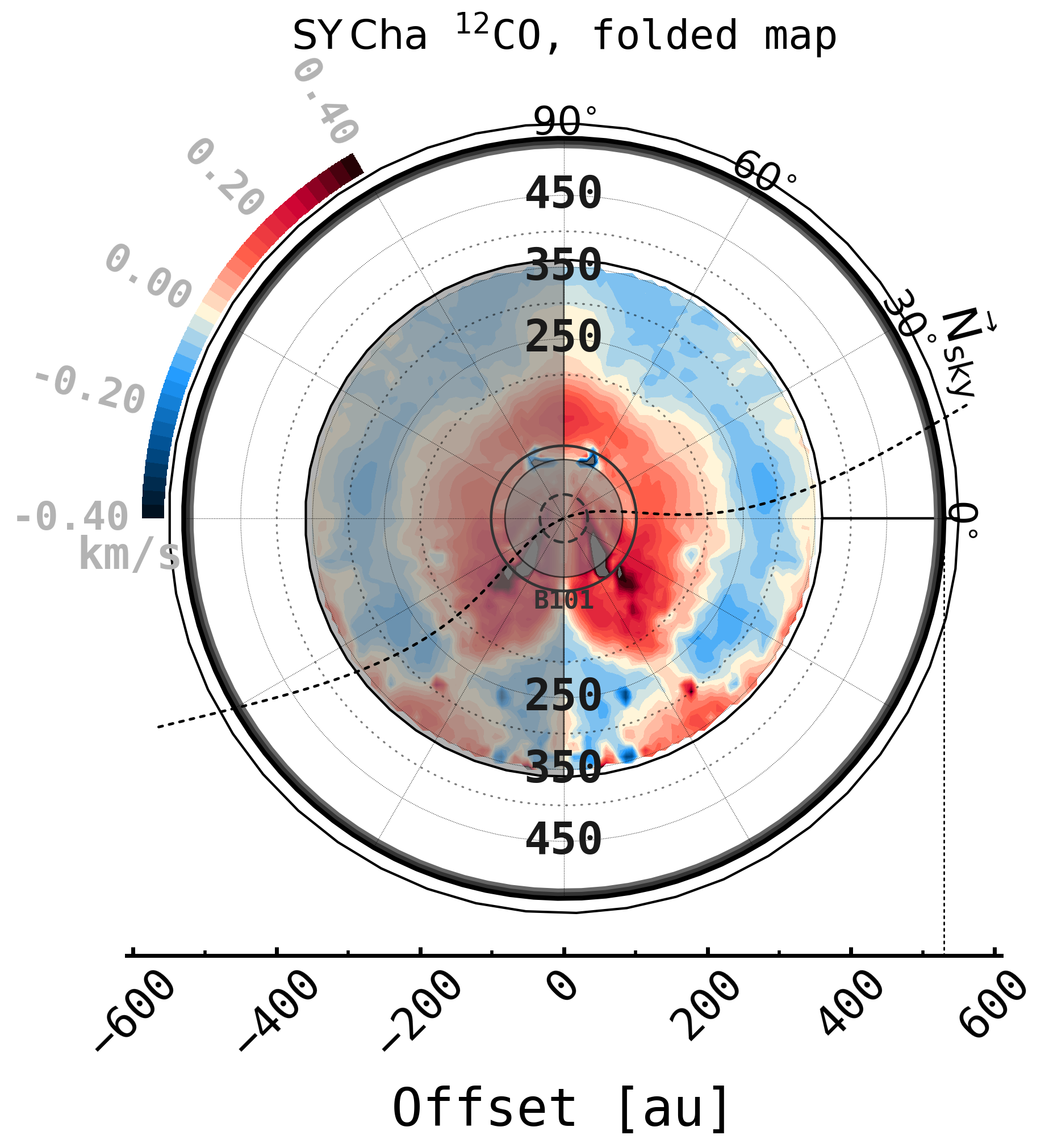} & 
        \includegraphics[width=0.25\textwidth]{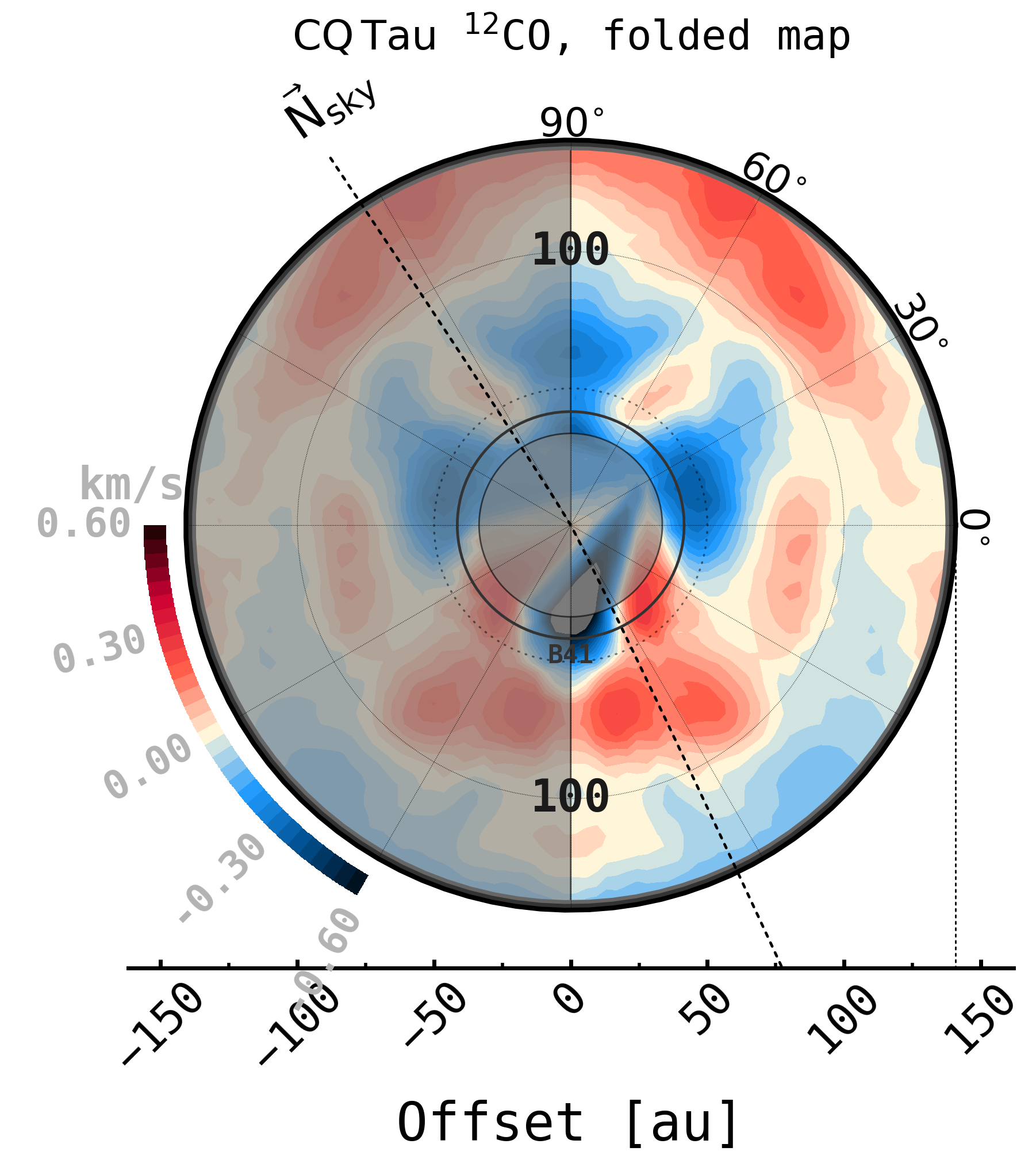} &
        \includegraphics[width=0.25\textwidth]{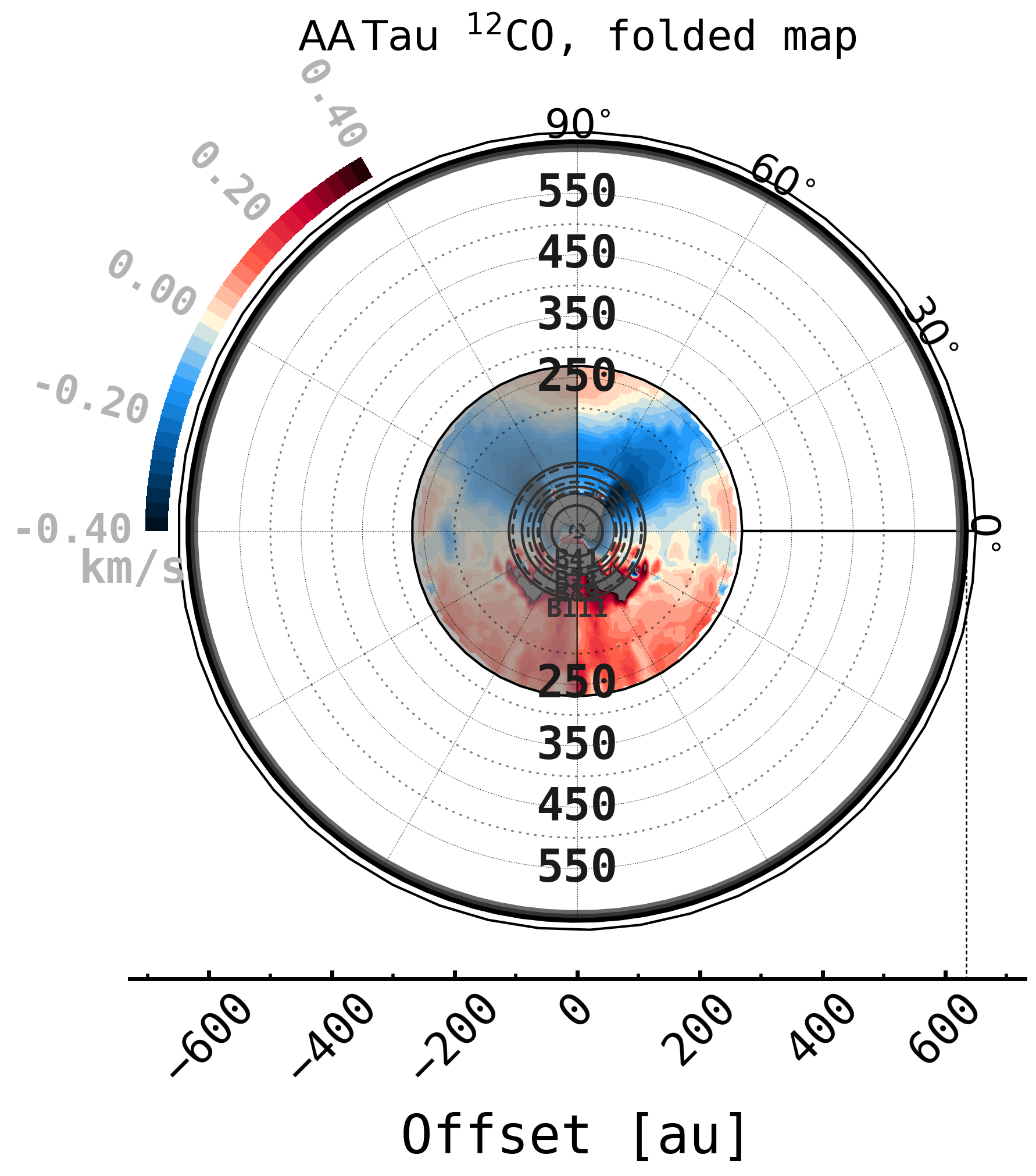} \\
        \includegraphics[width=0.25\textwidth]{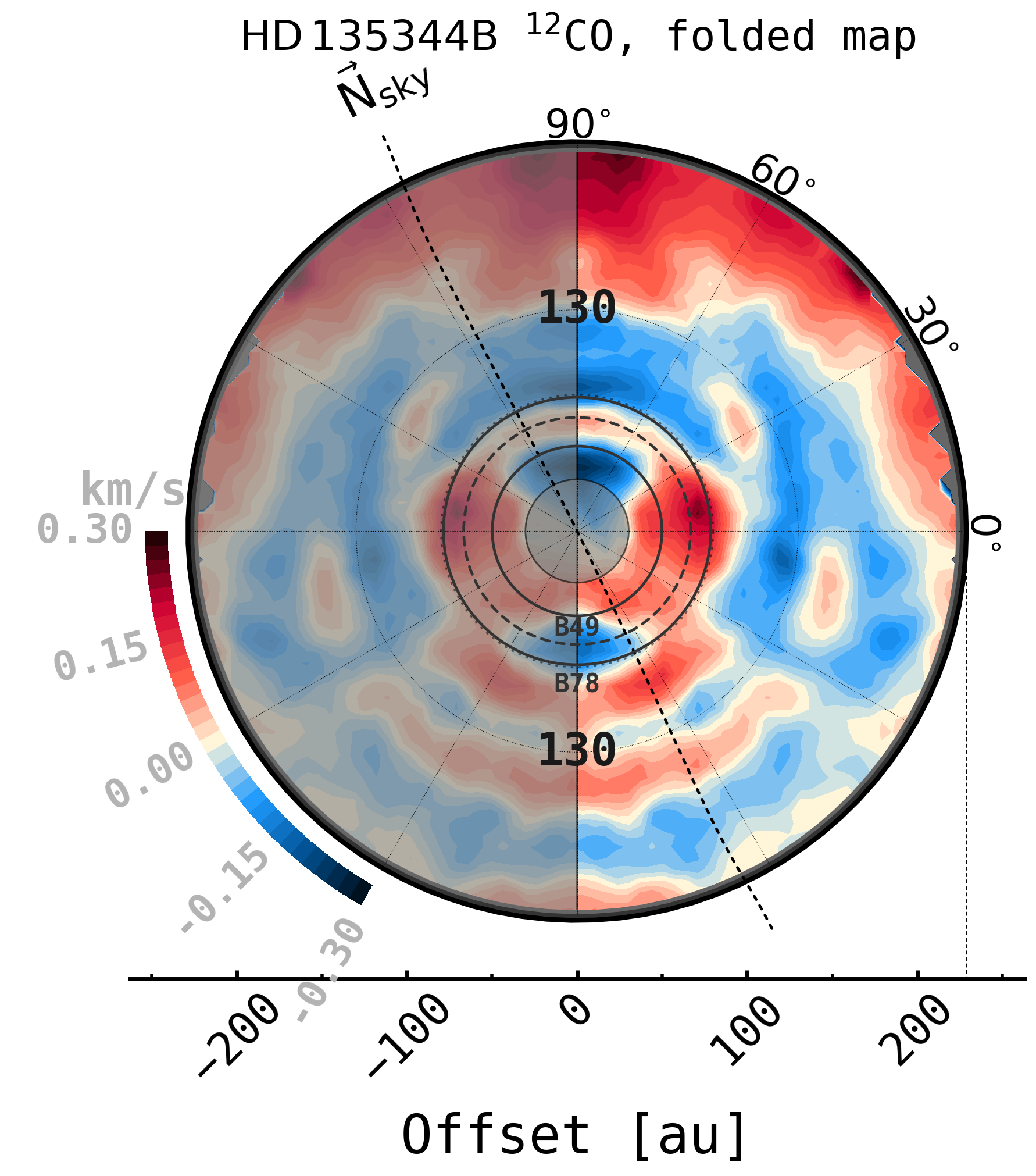} & 
        \includegraphics[width=0.25\textwidth]{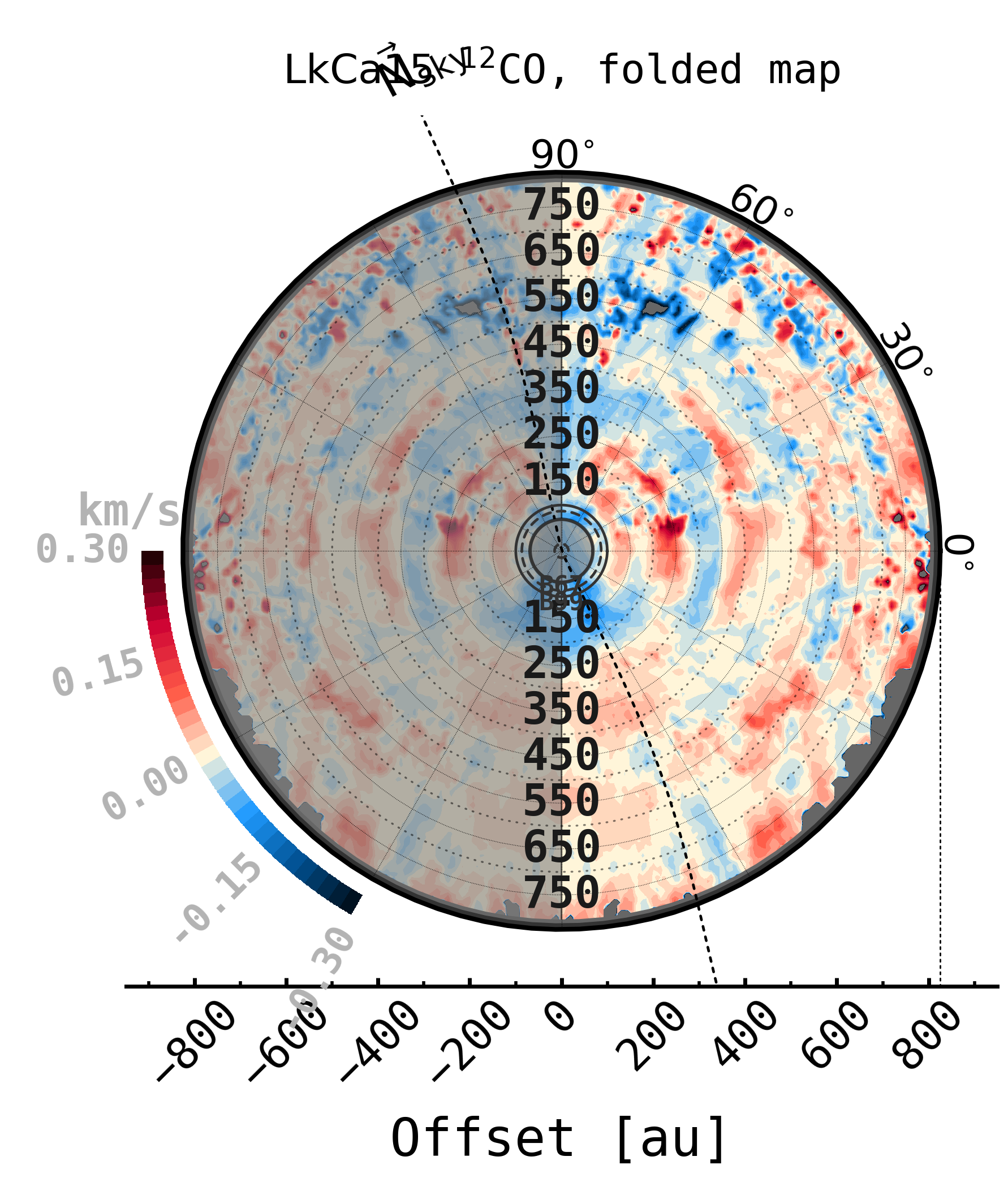} &        
        \includegraphics[width=0.25\textwidth]{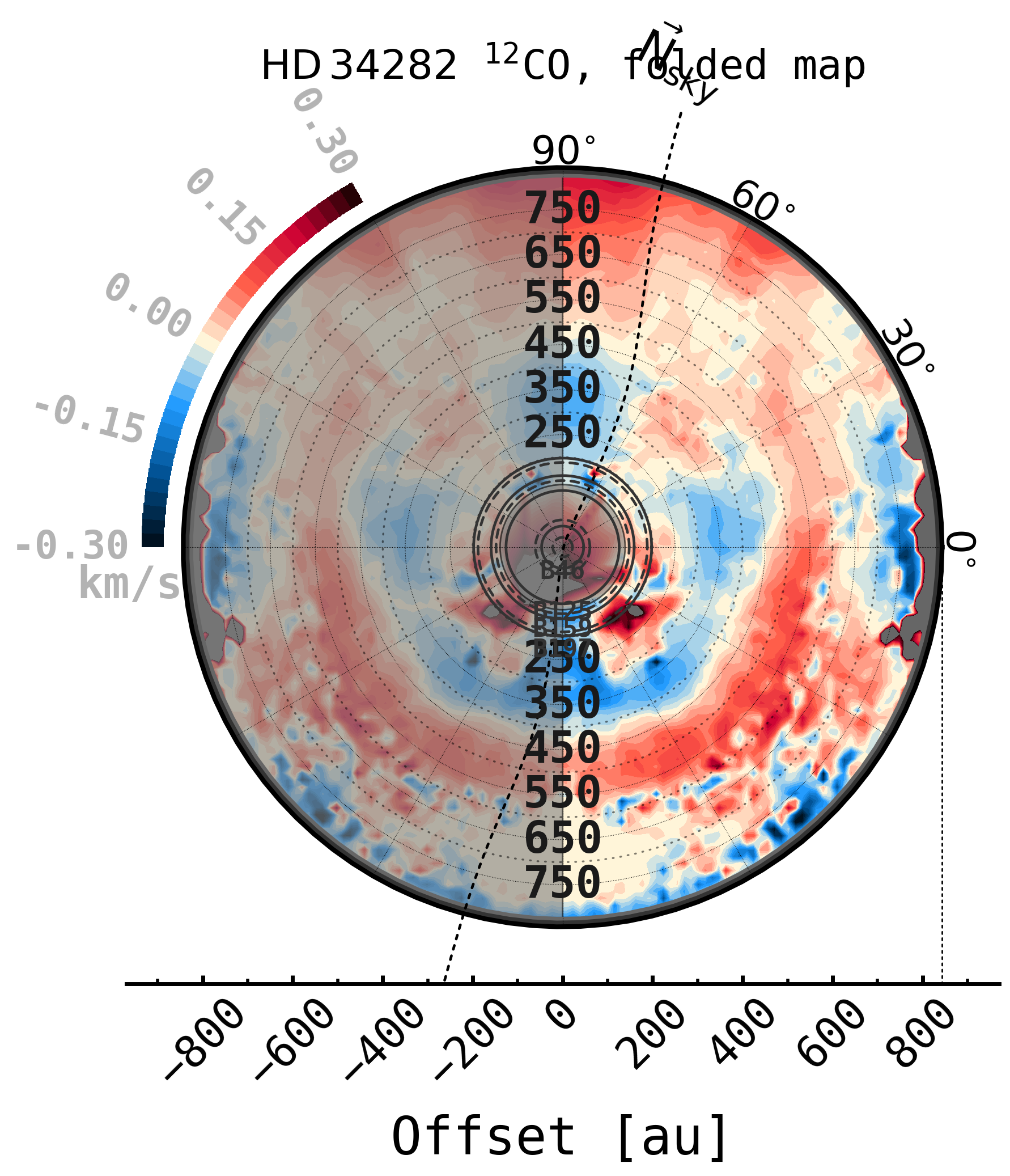} & 
        \includegraphics[width=0.25\textwidth]{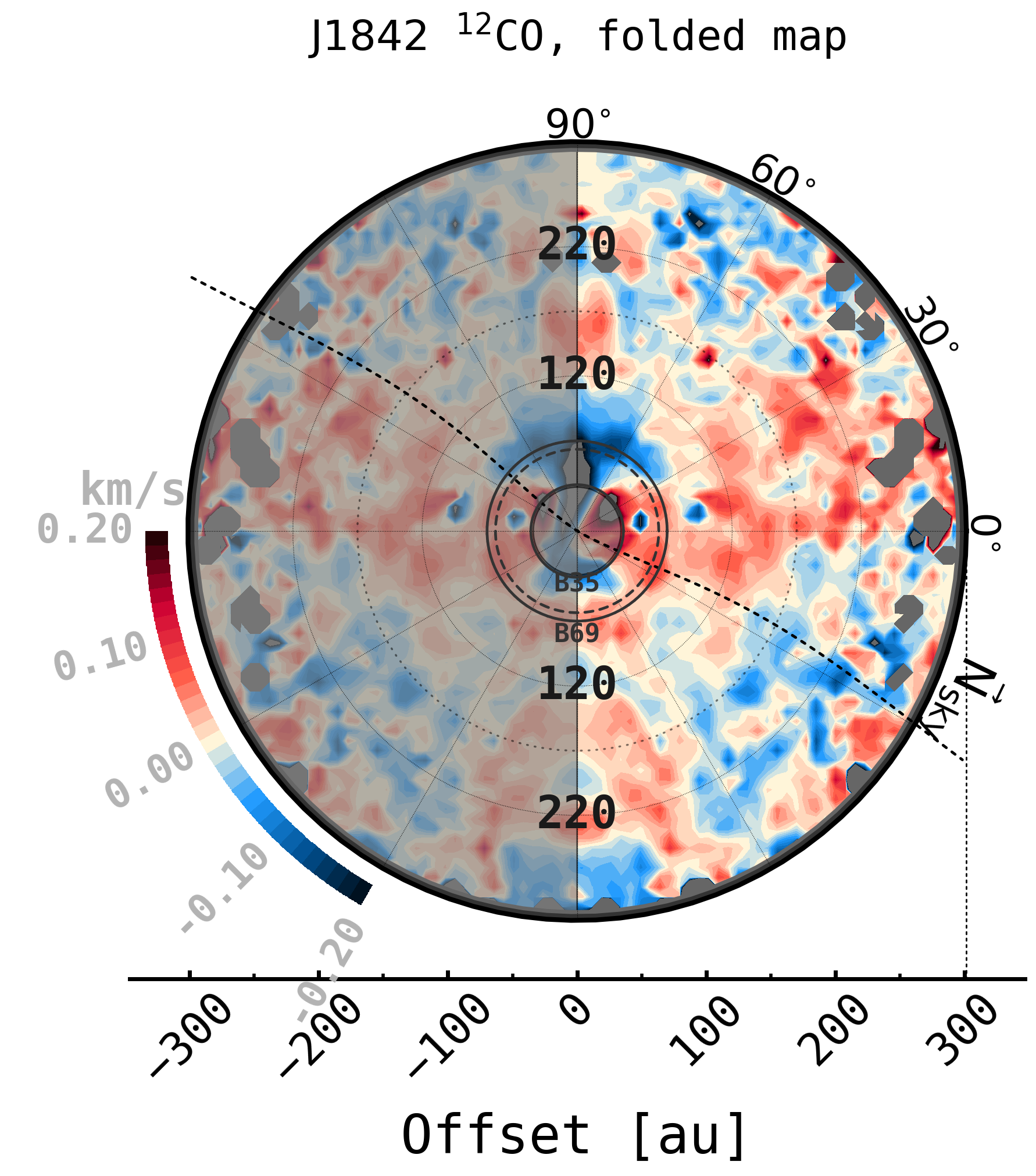} \\
        \includegraphics[width=0.25\textwidth]{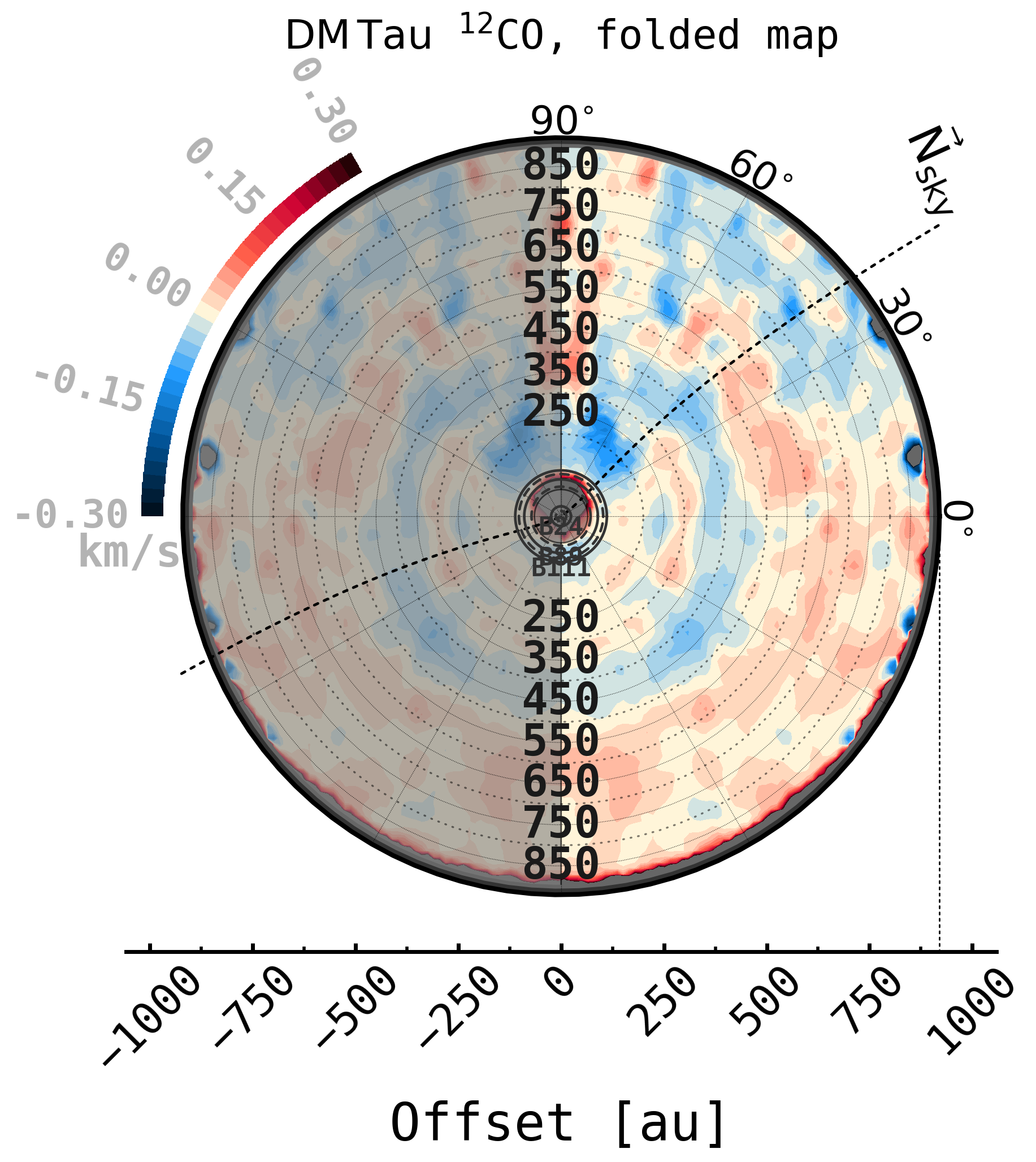} &
        \includegraphics[width=0.25\textwidth]{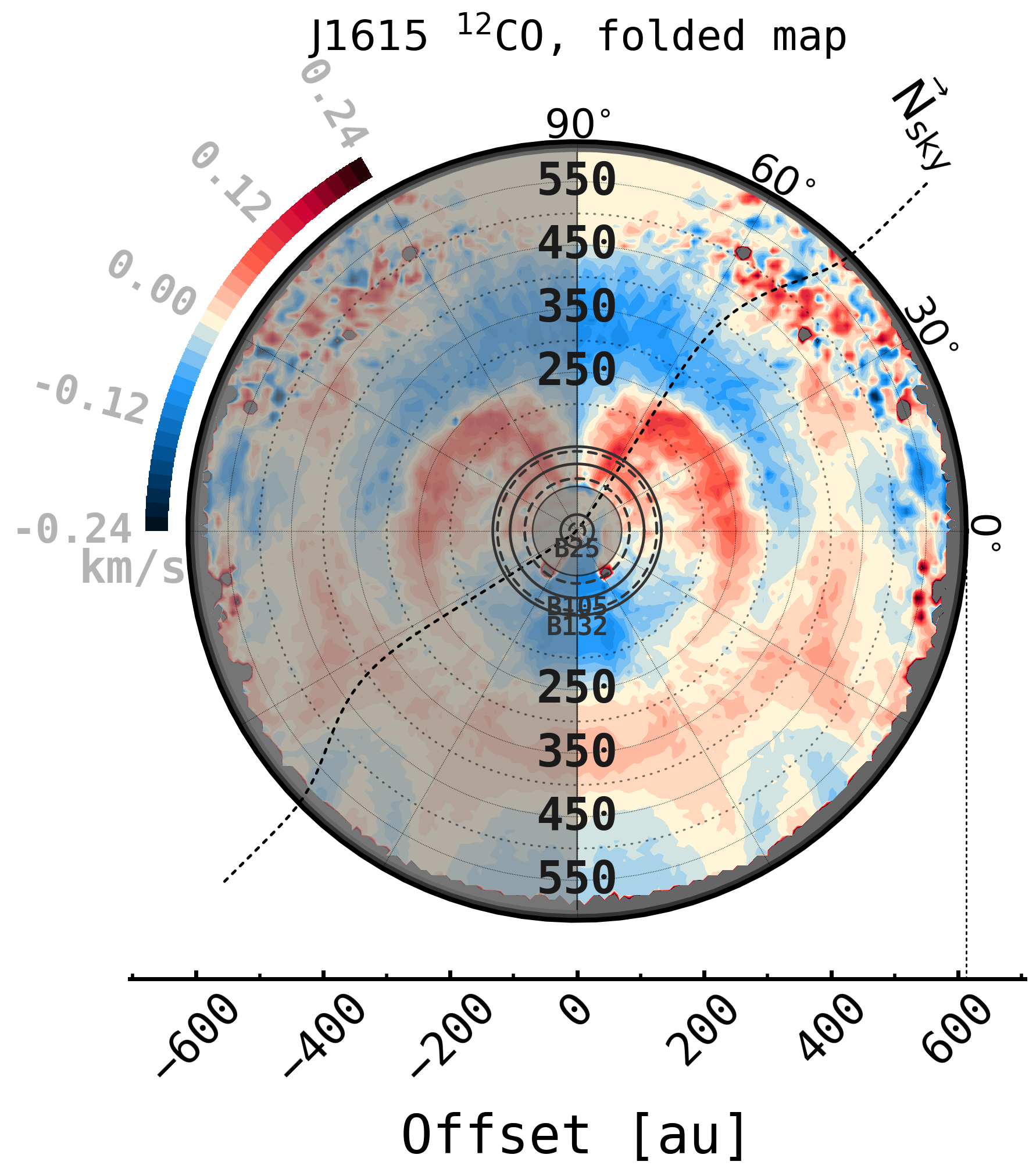} & 
        \includegraphics[width=0.25\textwidth]{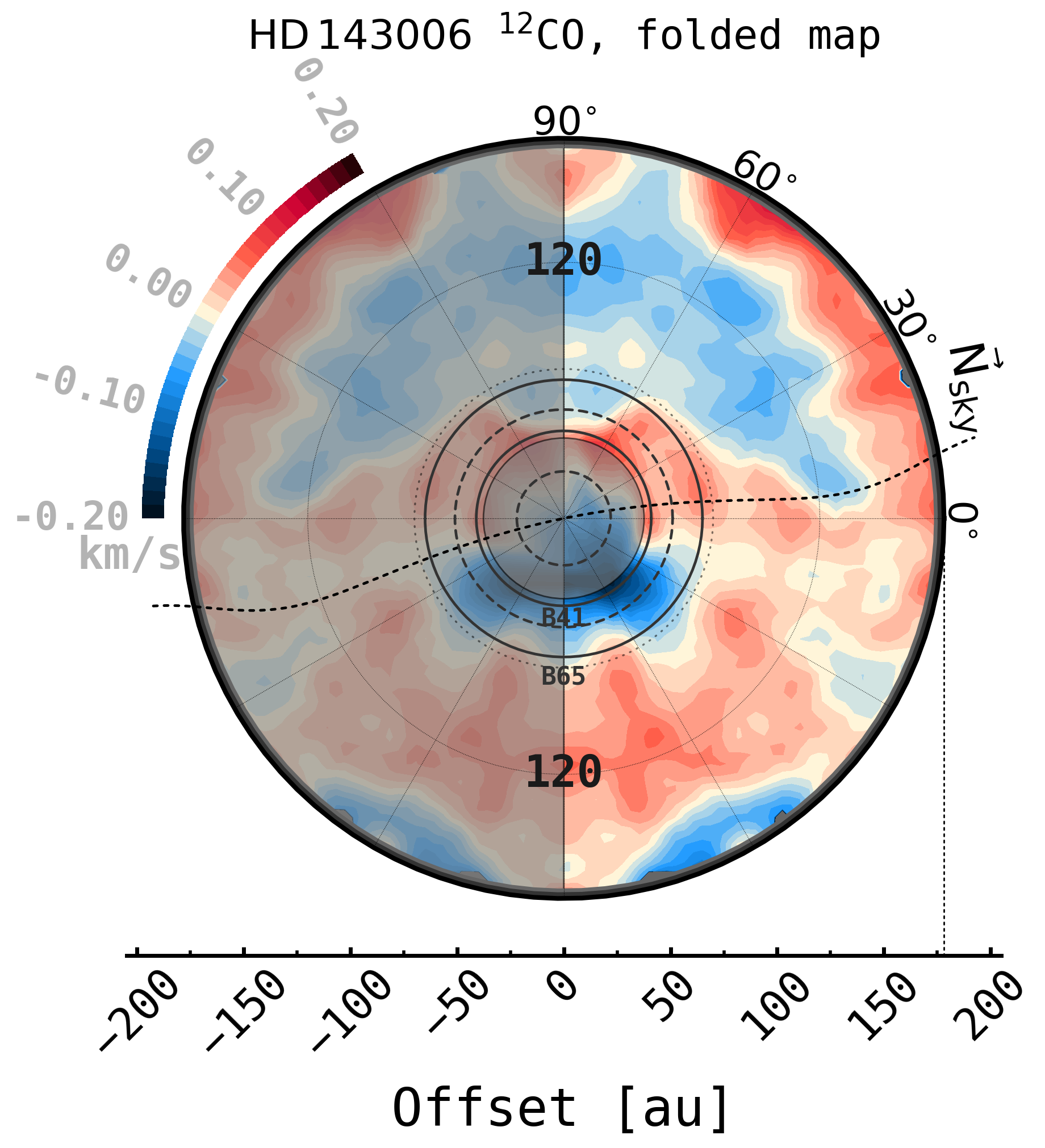} & 
        \includegraphics[width=0.25\textwidth]{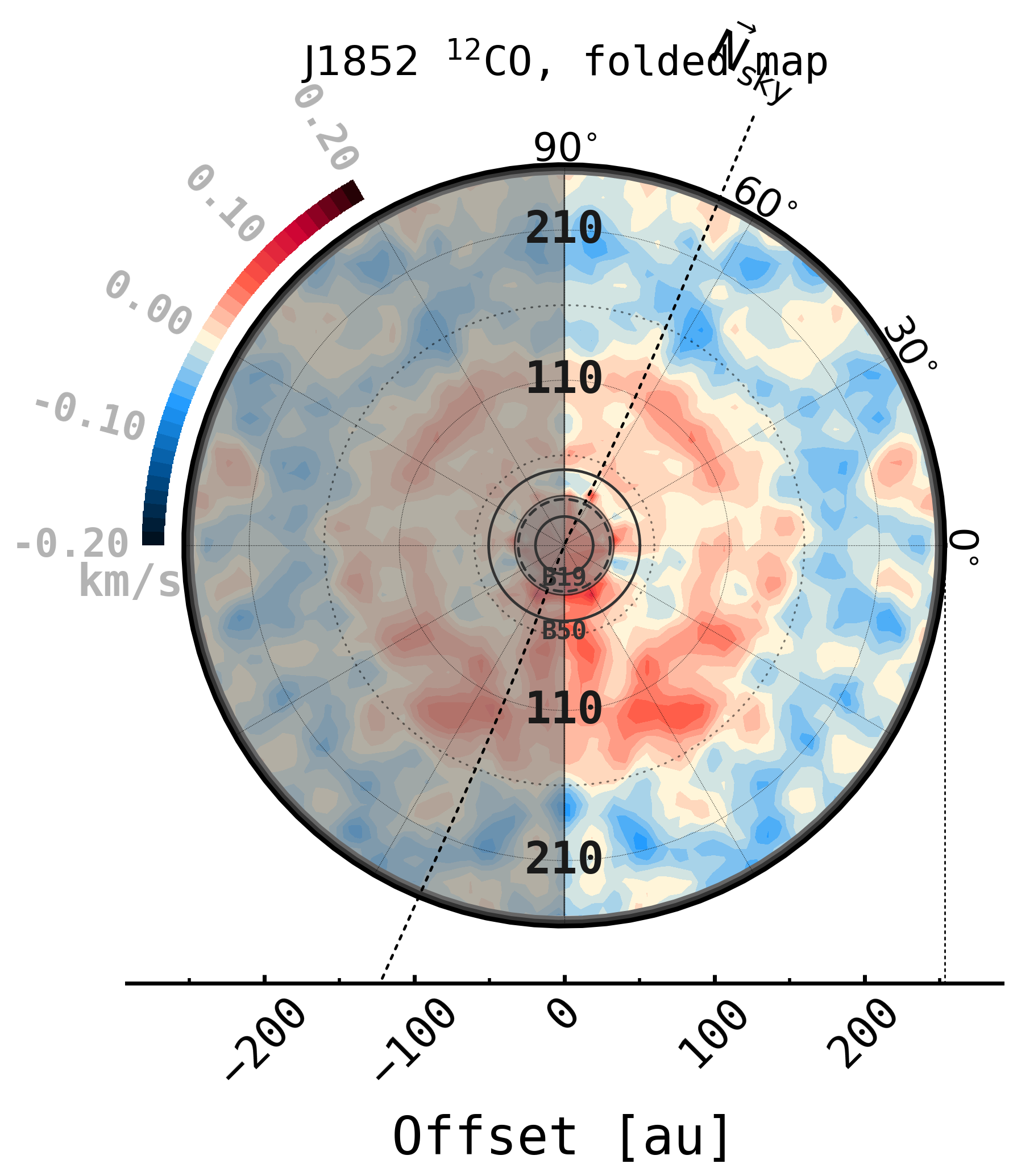} \\
        \includegraphics[width=0.25\textwidth]{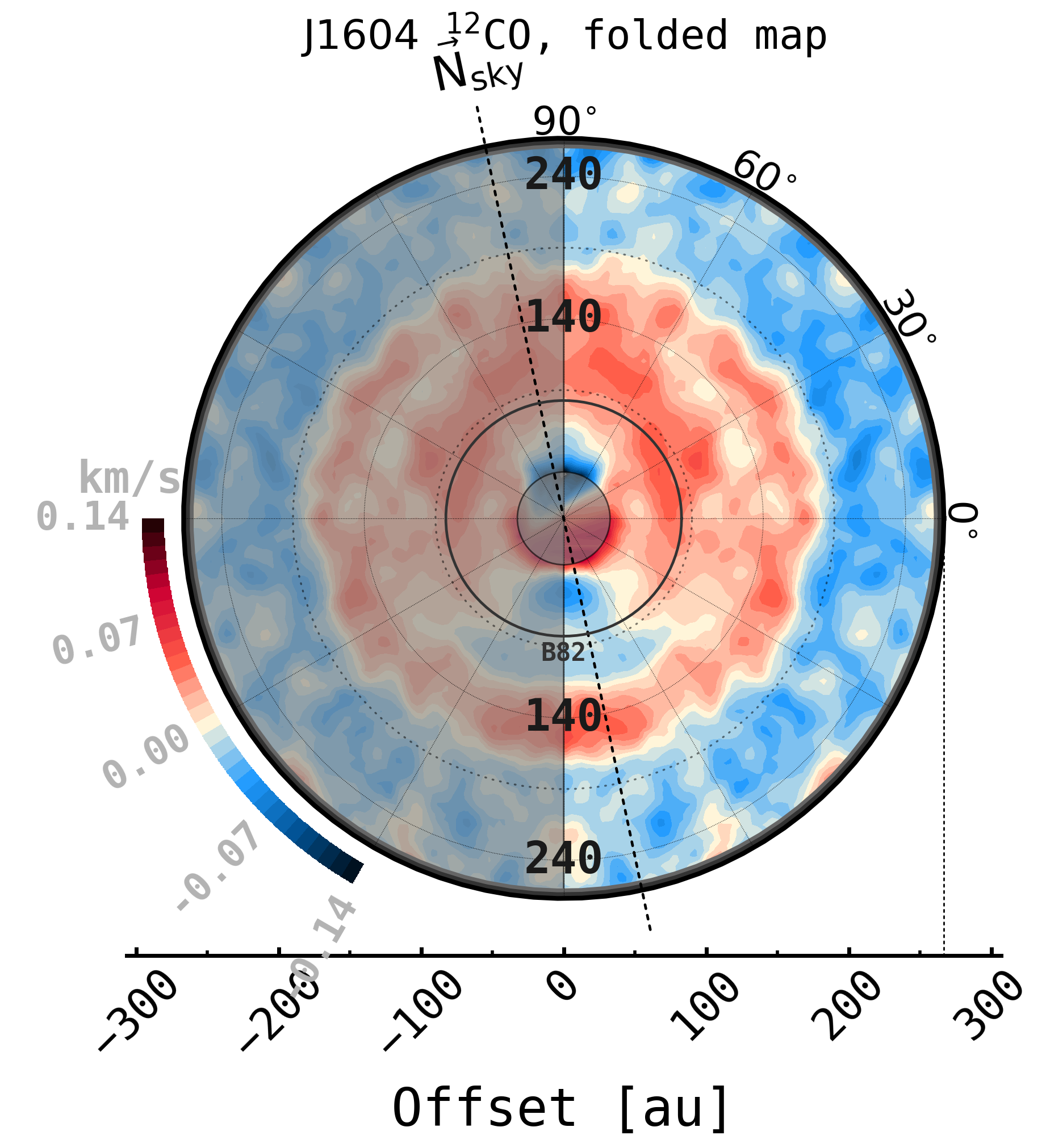} & 
        \includegraphics[width=0.25\textwidth]{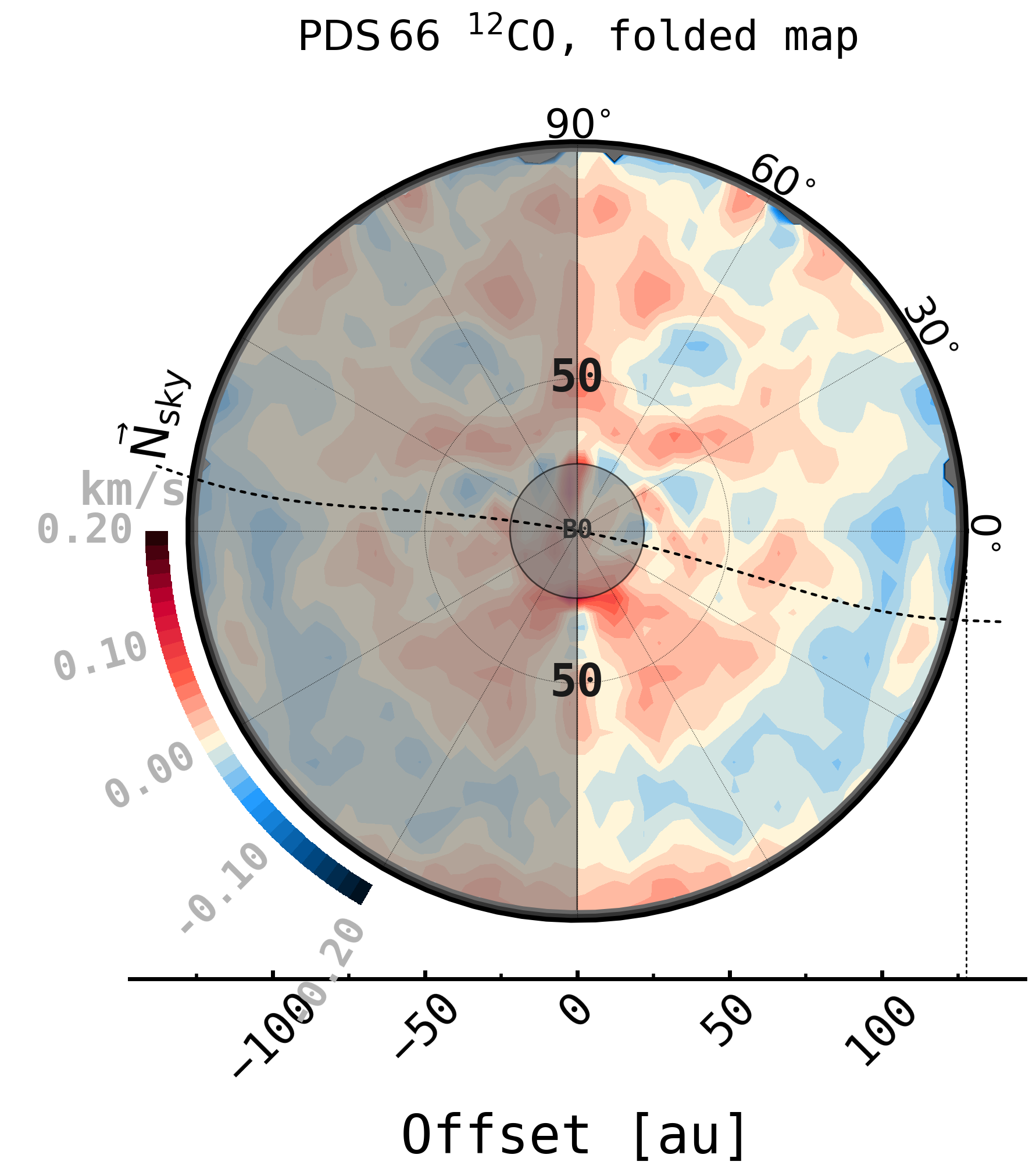} & 
        \includegraphics[width=0.25\textwidth]{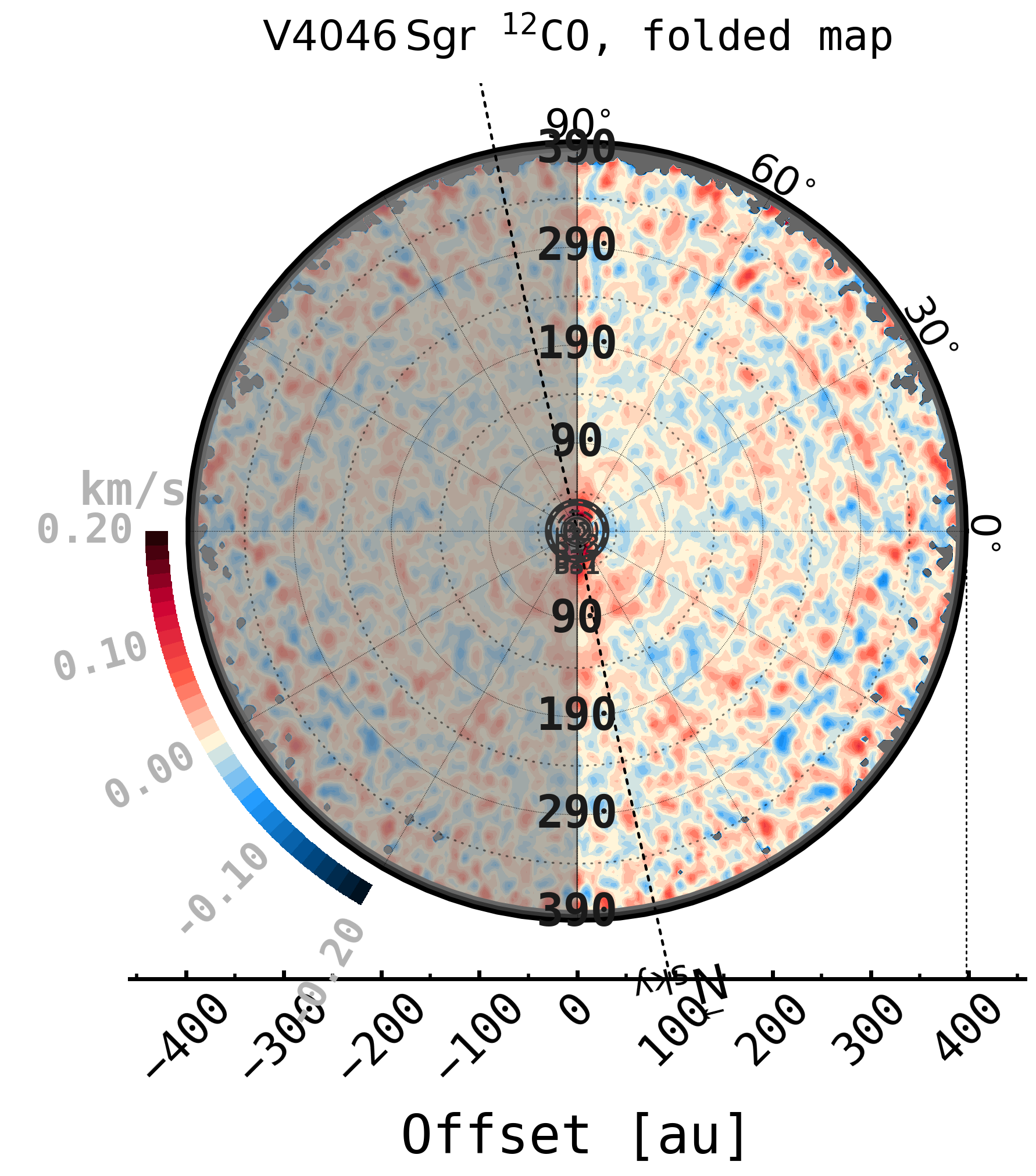} & \\
    \end{tabular}
\caption{Gallery of folded centroid residuals maps of the \twCO{} emission.} 
    \label{fig:folded_gallery}
\end{figure*}

\begin{figure*}[t]
    \centering
    \begin{tabular}{c}
        \includegraphics[width=0.7\linewidth]{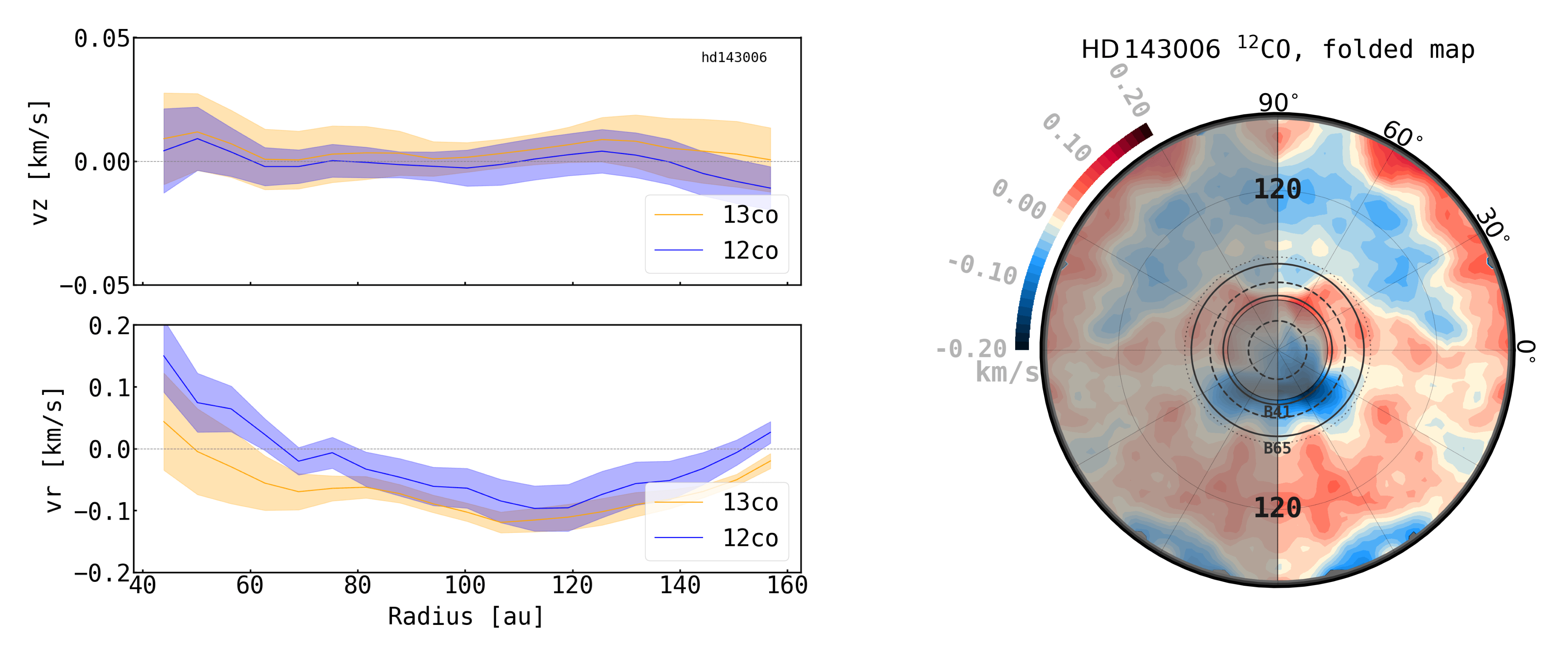} \\
        \includegraphics[width=0.7\linewidth]{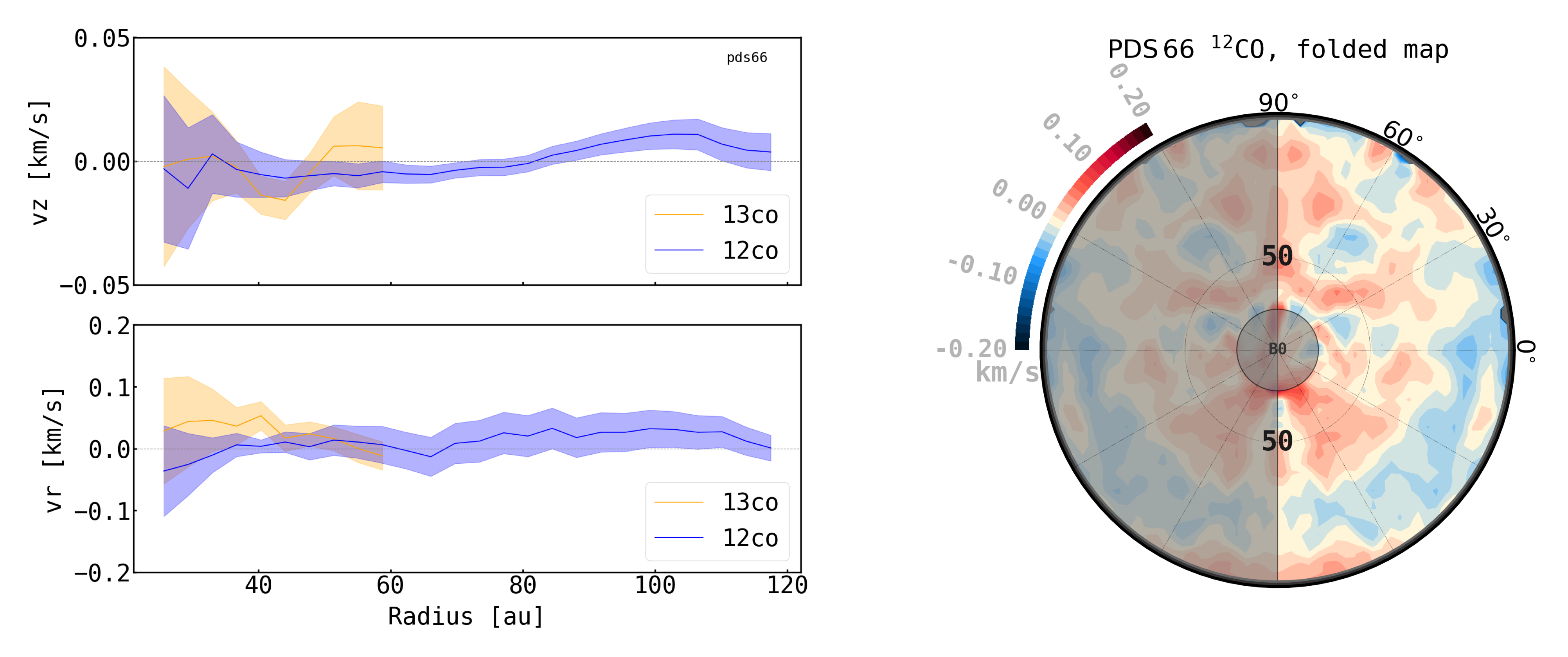} \\
        \includegraphics[width=0.7\linewidth]{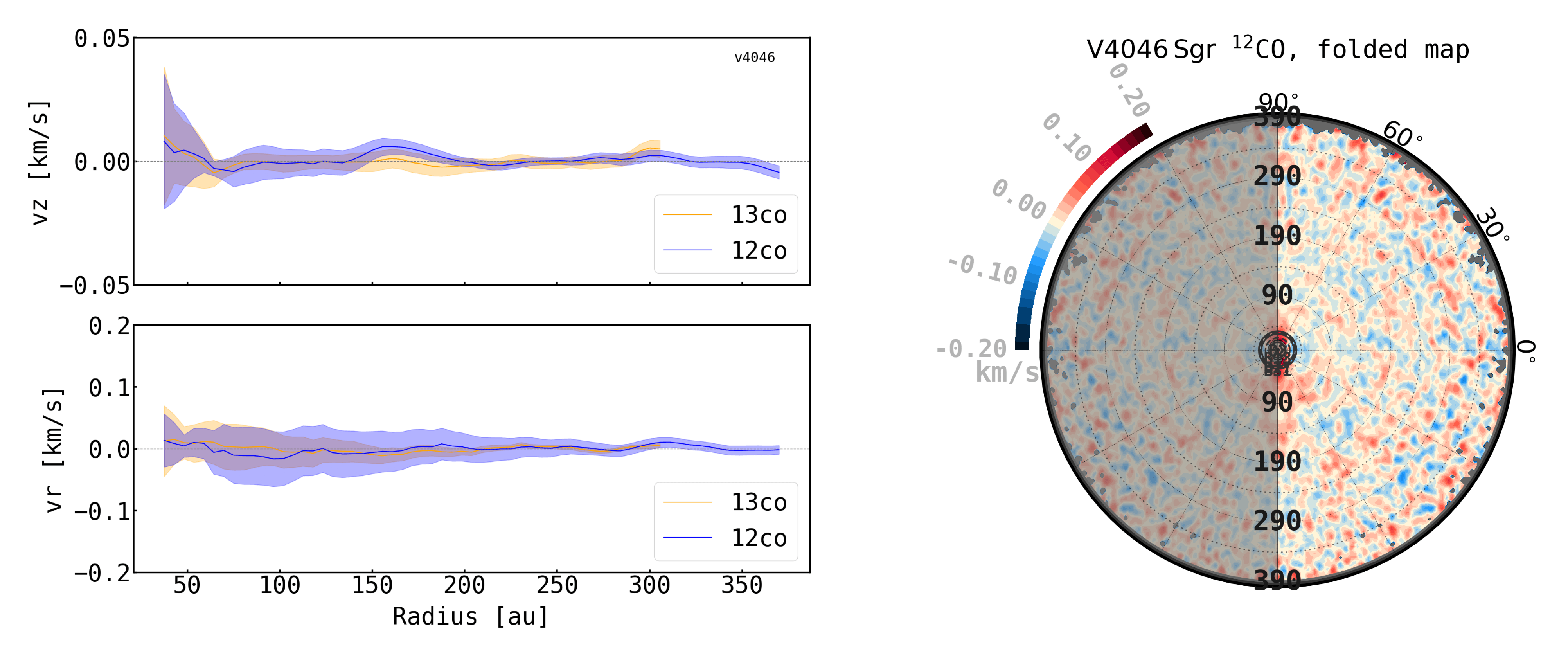} 
        \\
    \end{tabular}
    \caption{Same as Figure\,\ref{fig:vz1D}, for three targets that don't show strong vertical motion.}
    \label{fig:vz1Dno}
\end{figure*}

\begin{figure*}[t]
    \centering
    \begin{tabular}{c}  
         \includegraphics[width=0.7\linewidth]{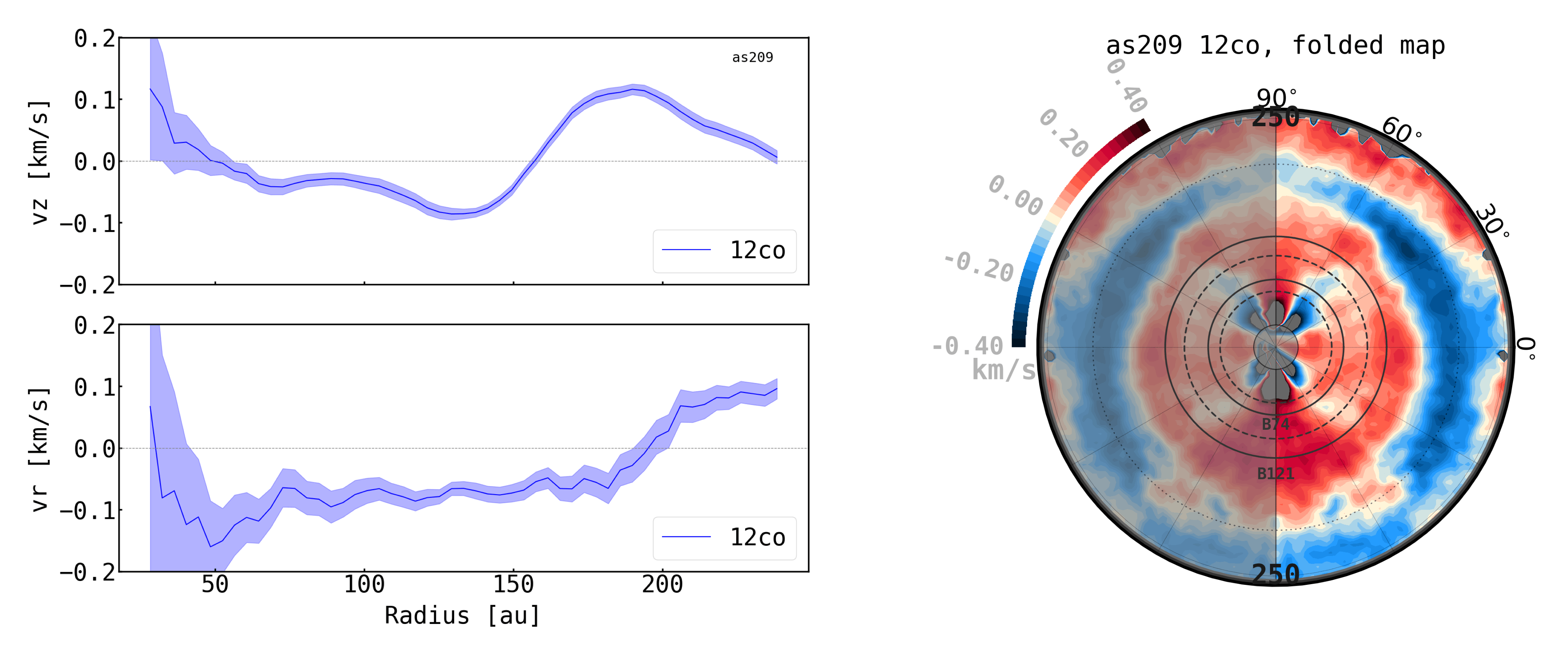} \\
        \includegraphics[width=0.7\linewidth]{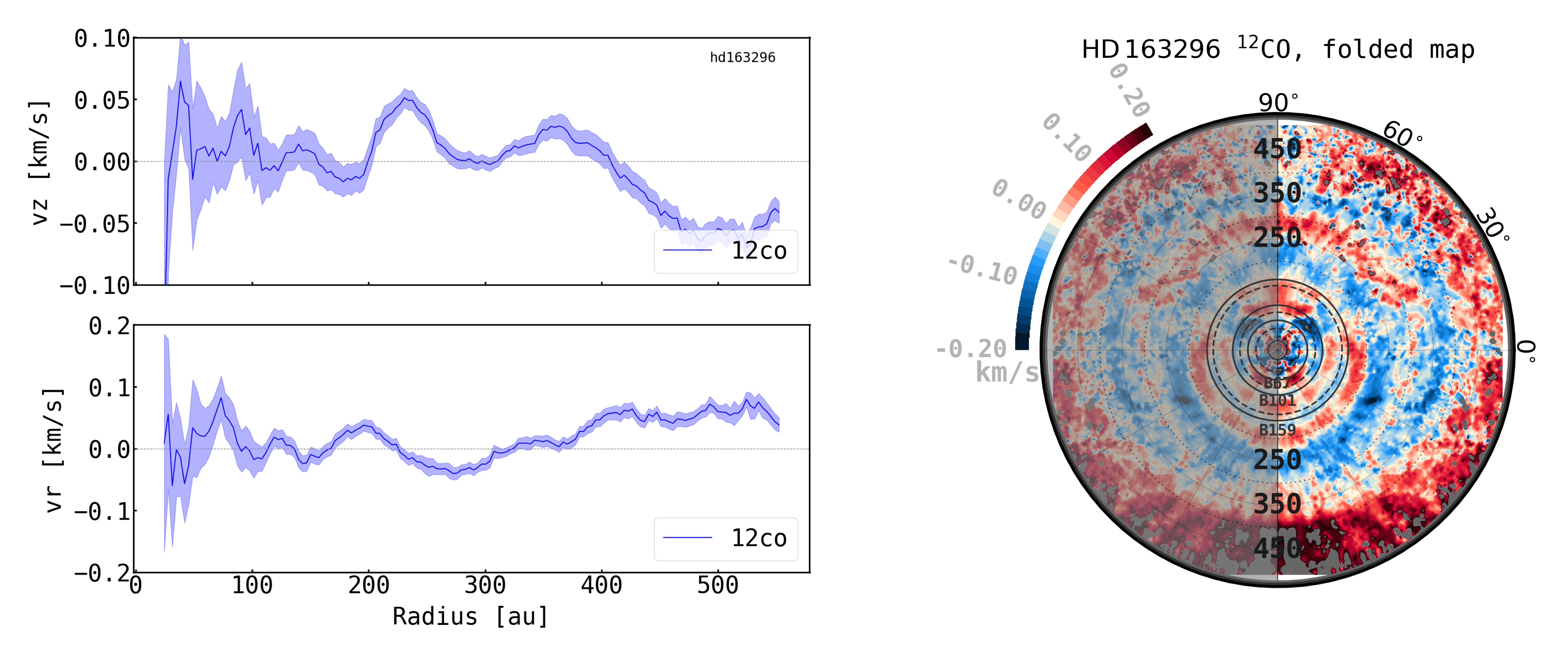} \\
        \includegraphics[width=0.7\linewidth]{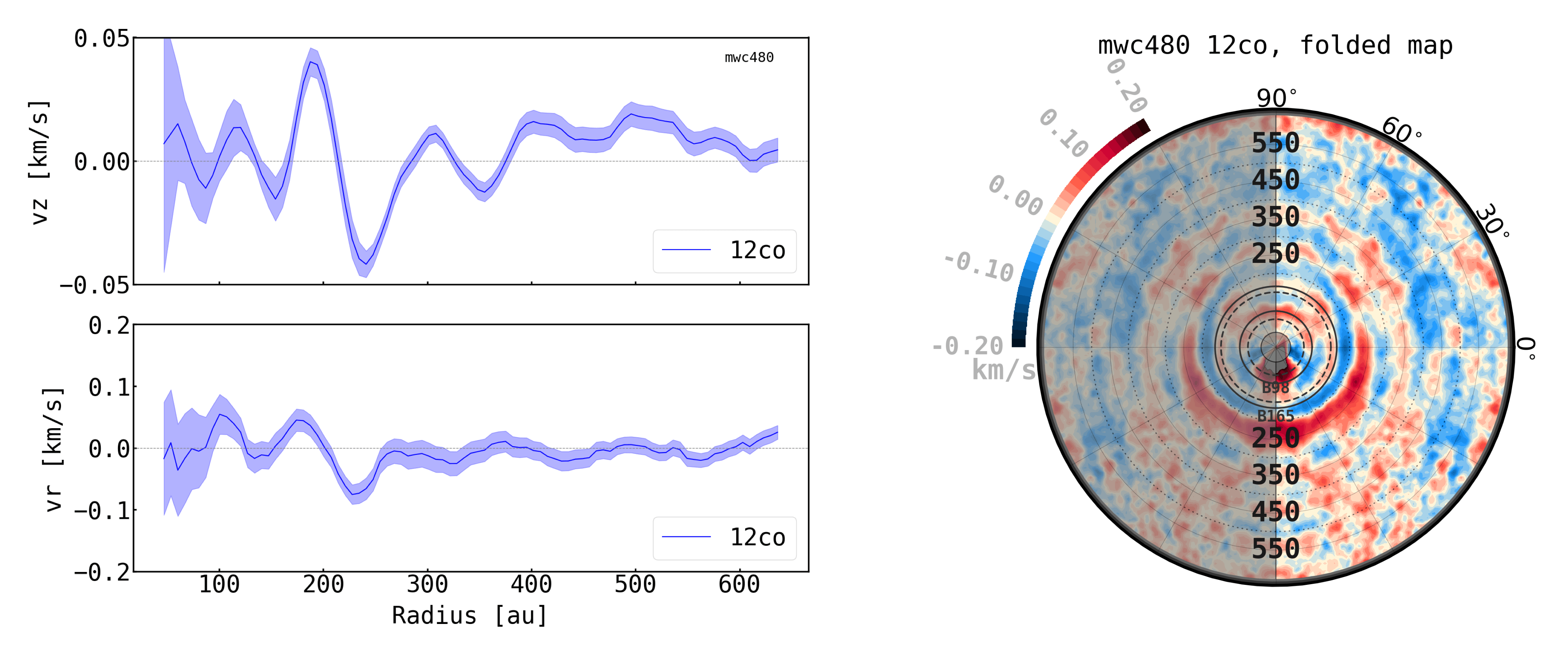} 
        \\
    \end{tabular}
    \caption{Same as Figure\,\ref{fig:vz1D} but for MAPS targets.}
    \label{fig:vz1DMAPS}
\end{figure*}

\begin{figure*}[t]
    \centering
    \begin{tabular}{cccc}
        \includegraphics[width=0.25\textwidth]{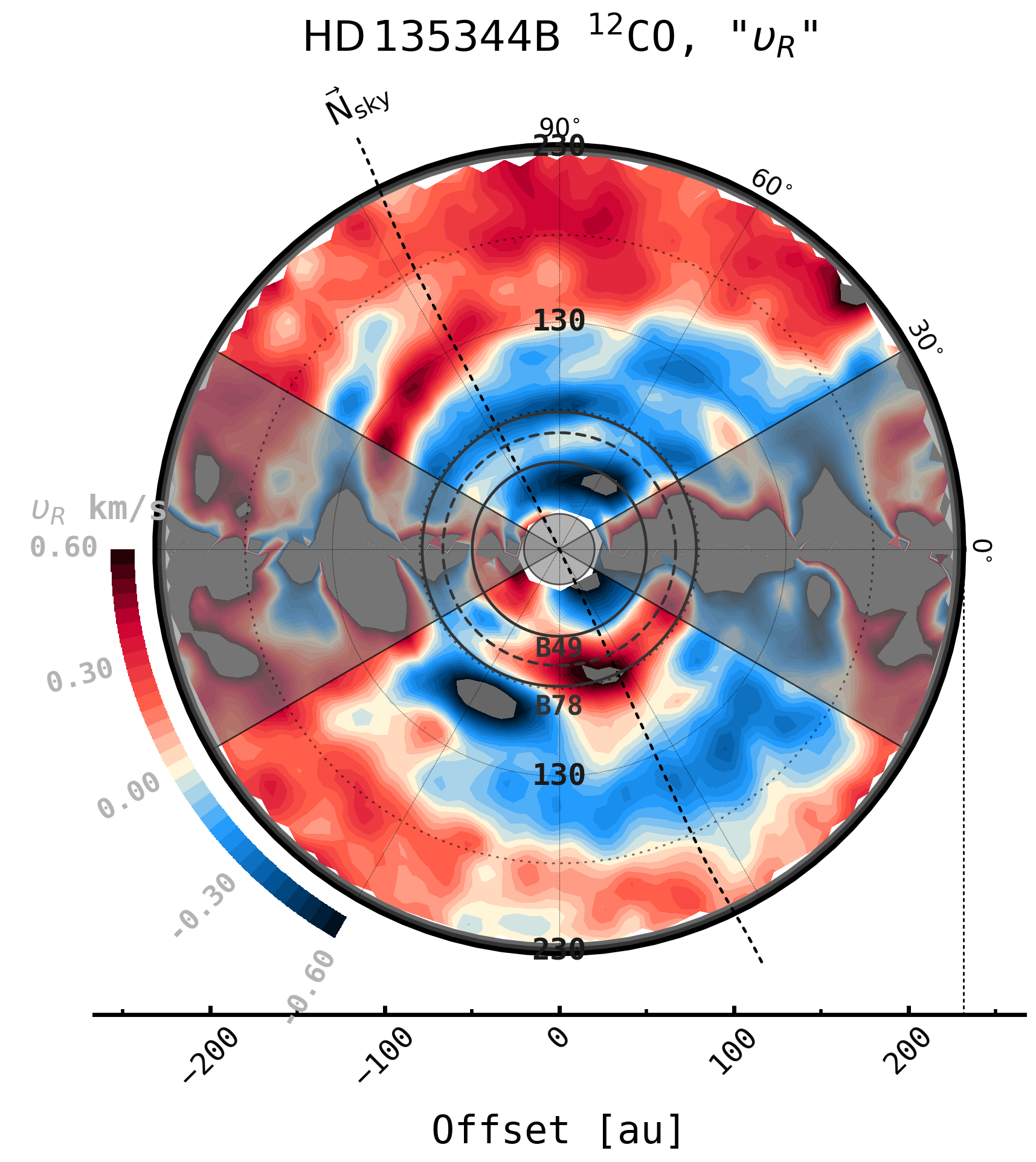} & 
        \includegraphics[width=0.25\textwidth]{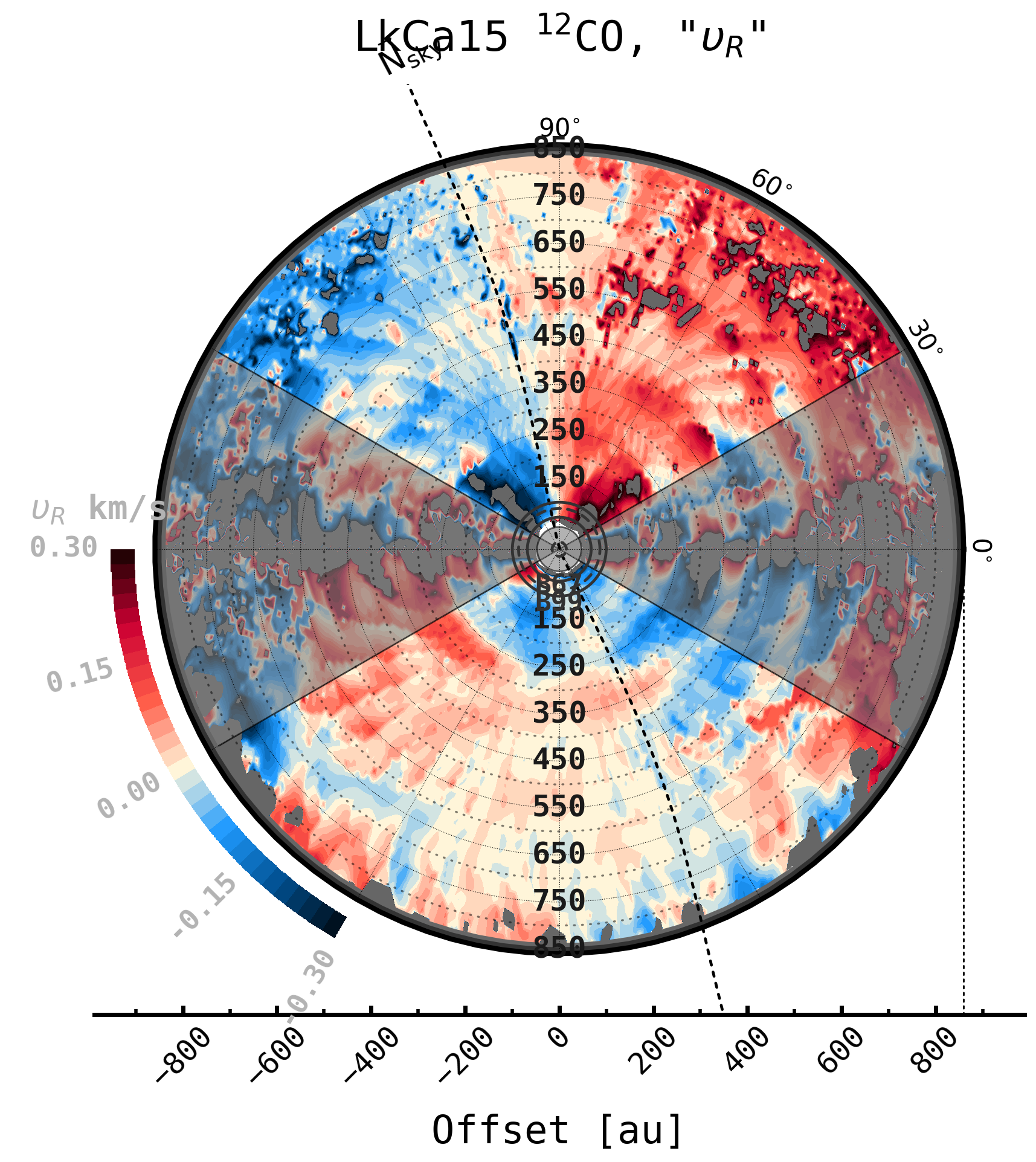} & 
        \includegraphics[width=0.25\textwidth]{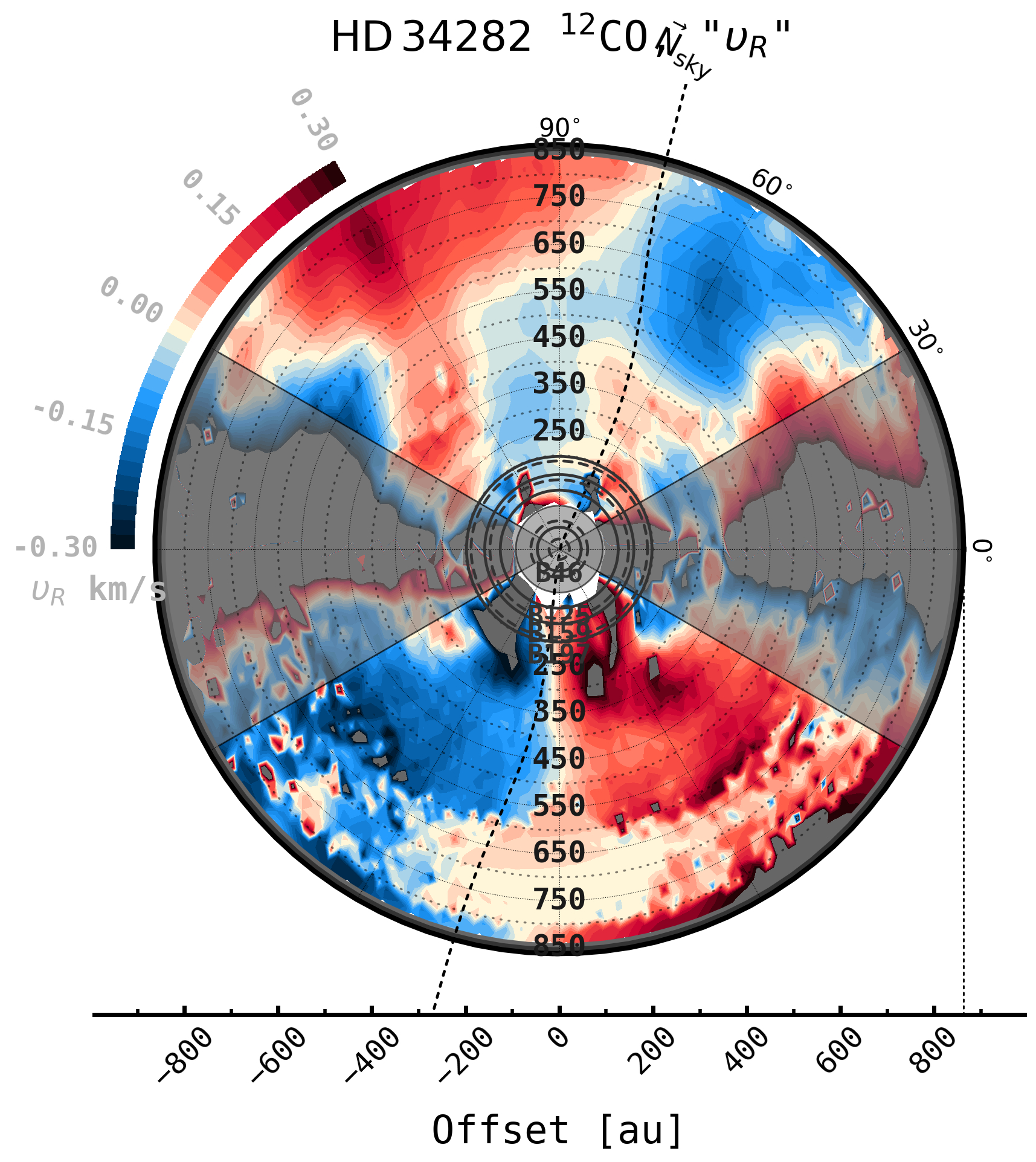} &
        \includegraphics[width=0.25\textwidth]{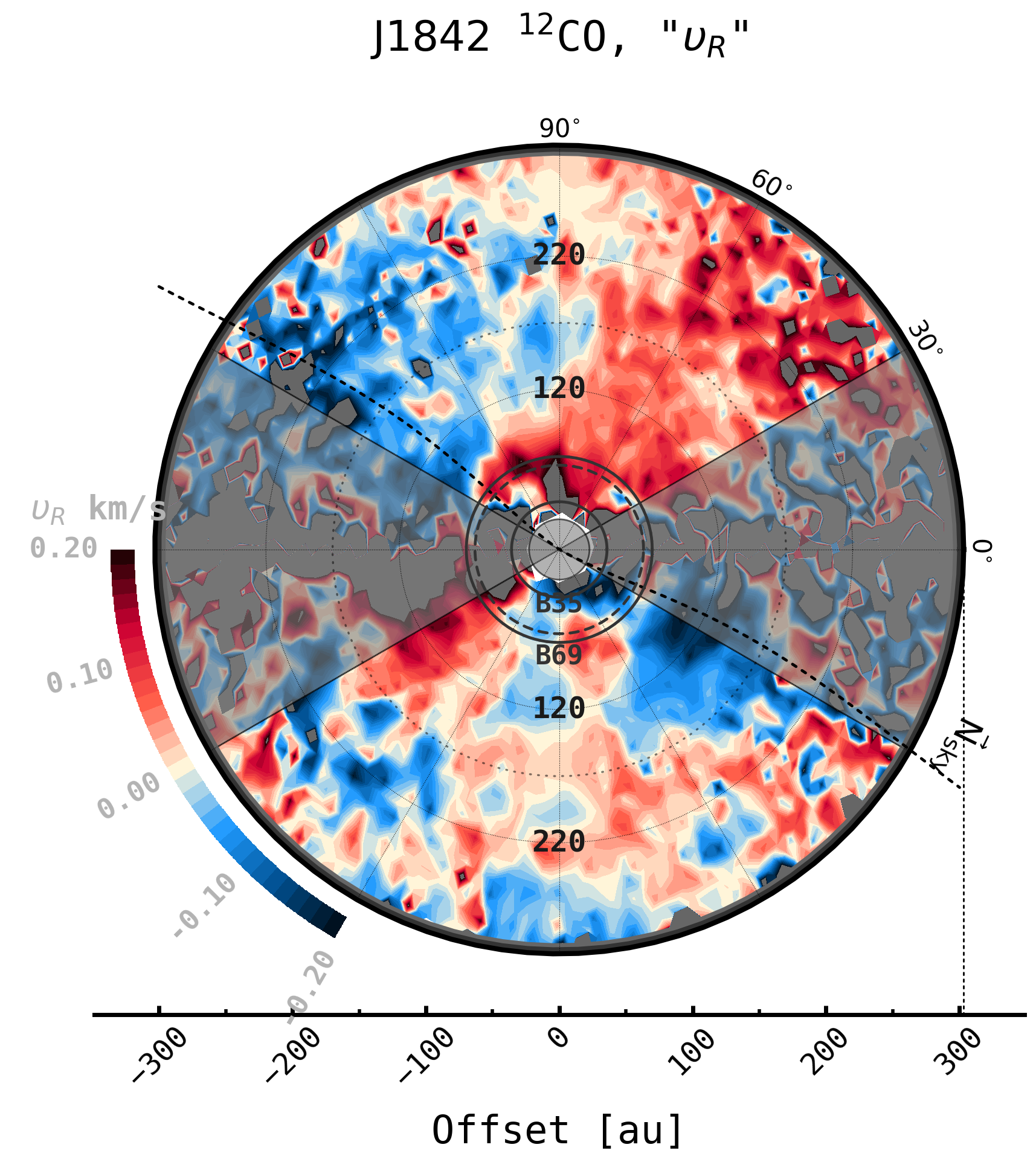} \\
        \includegraphics[width=0.25\textwidth]{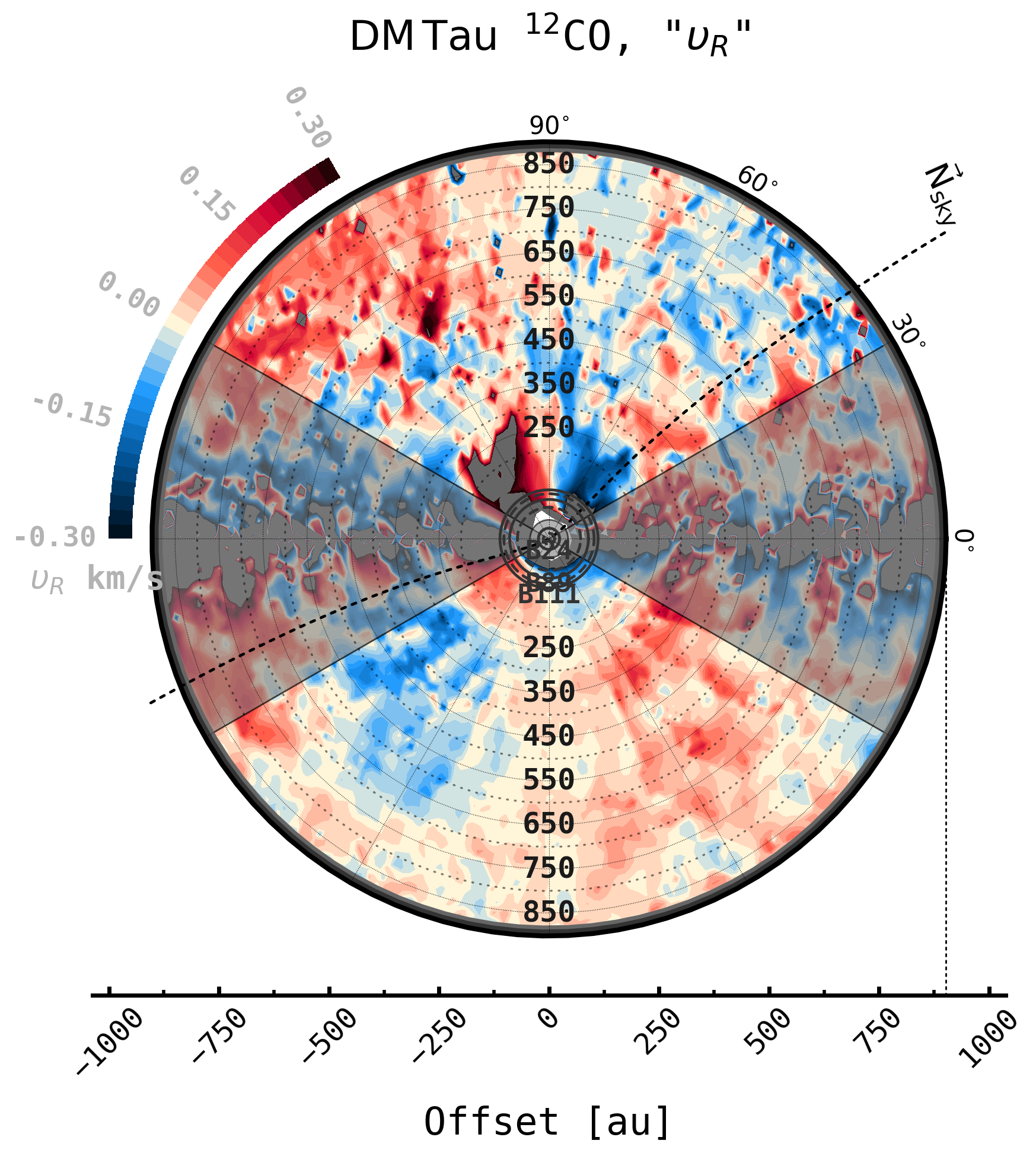} & 
        \includegraphics[width=0.25\textwidth]{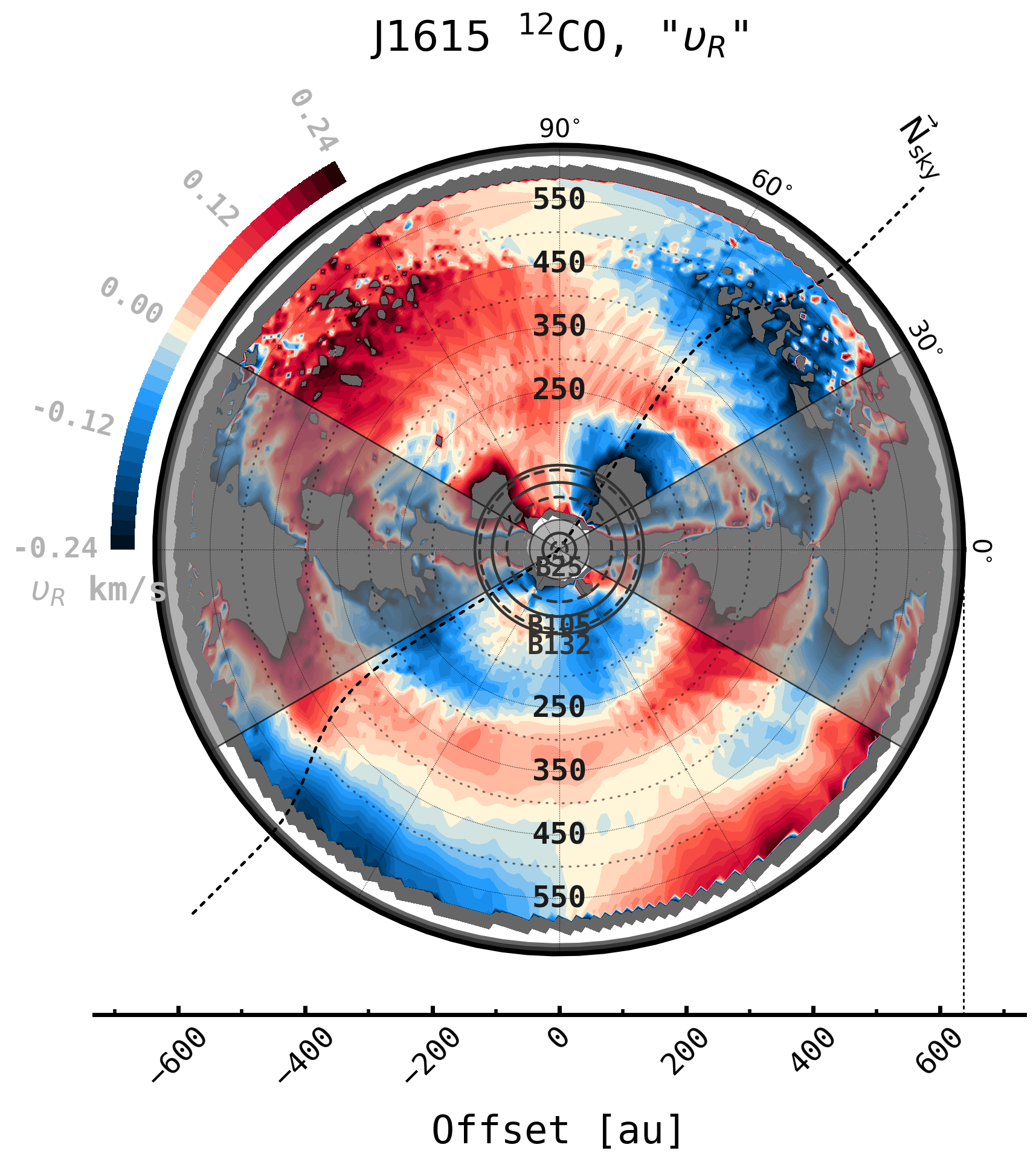} &        
        \includegraphics[width=0.25\textwidth]{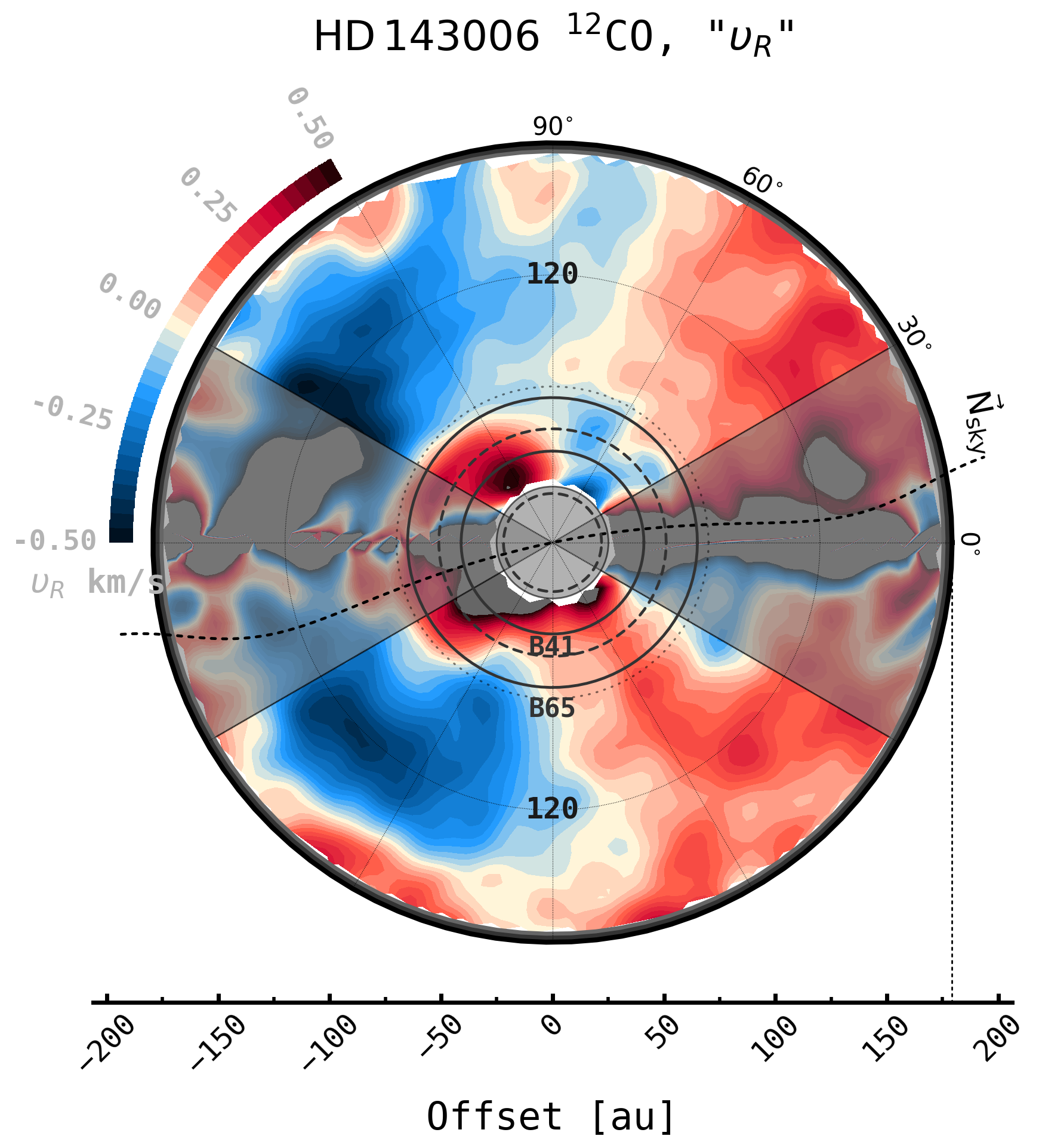} & 
        \includegraphics[width=0.25\textwidth]{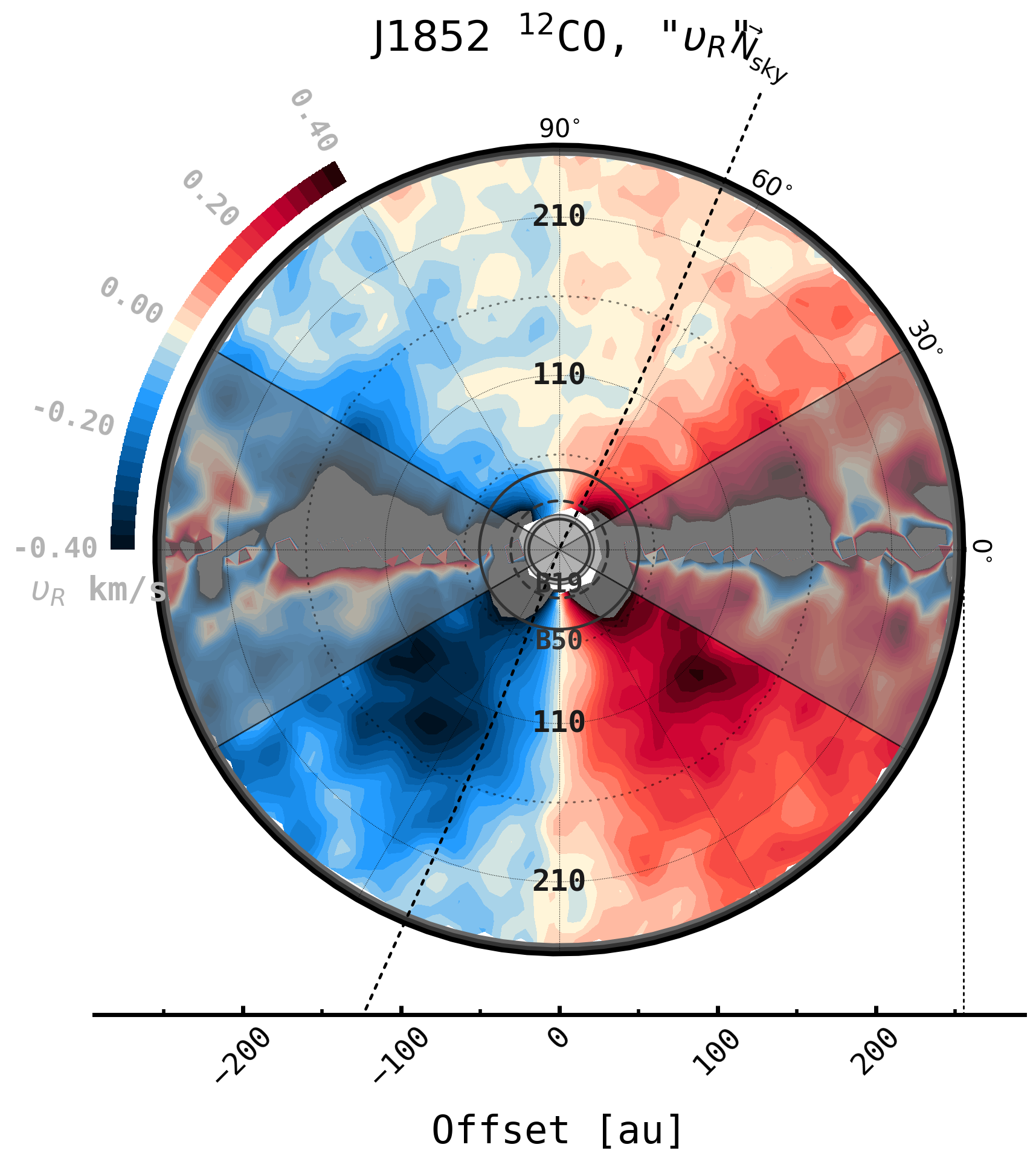} \\
        \includegraphics[width=0.25\textwidth]{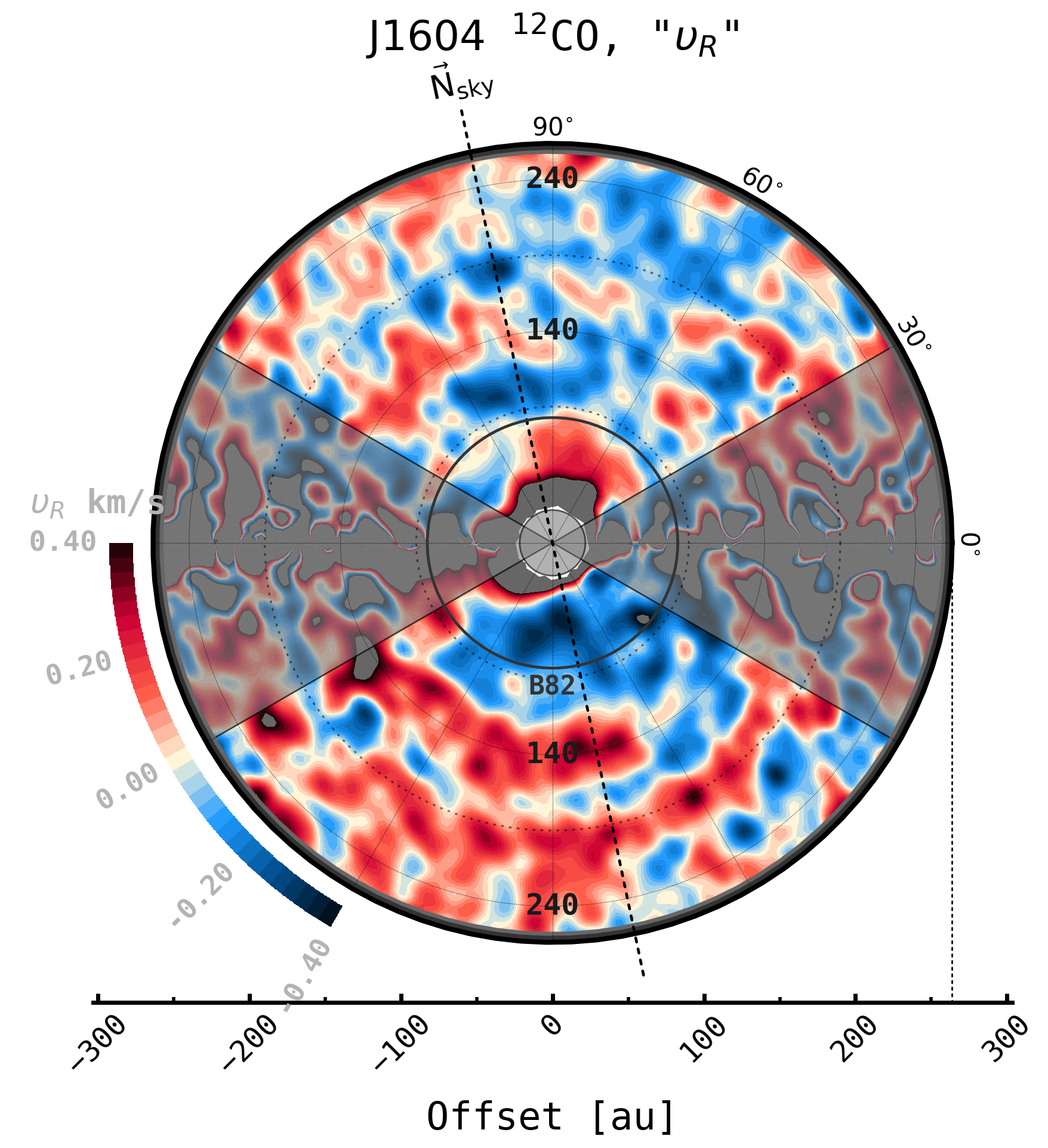} &
        \includegraphics[width=0.25\textwidth]{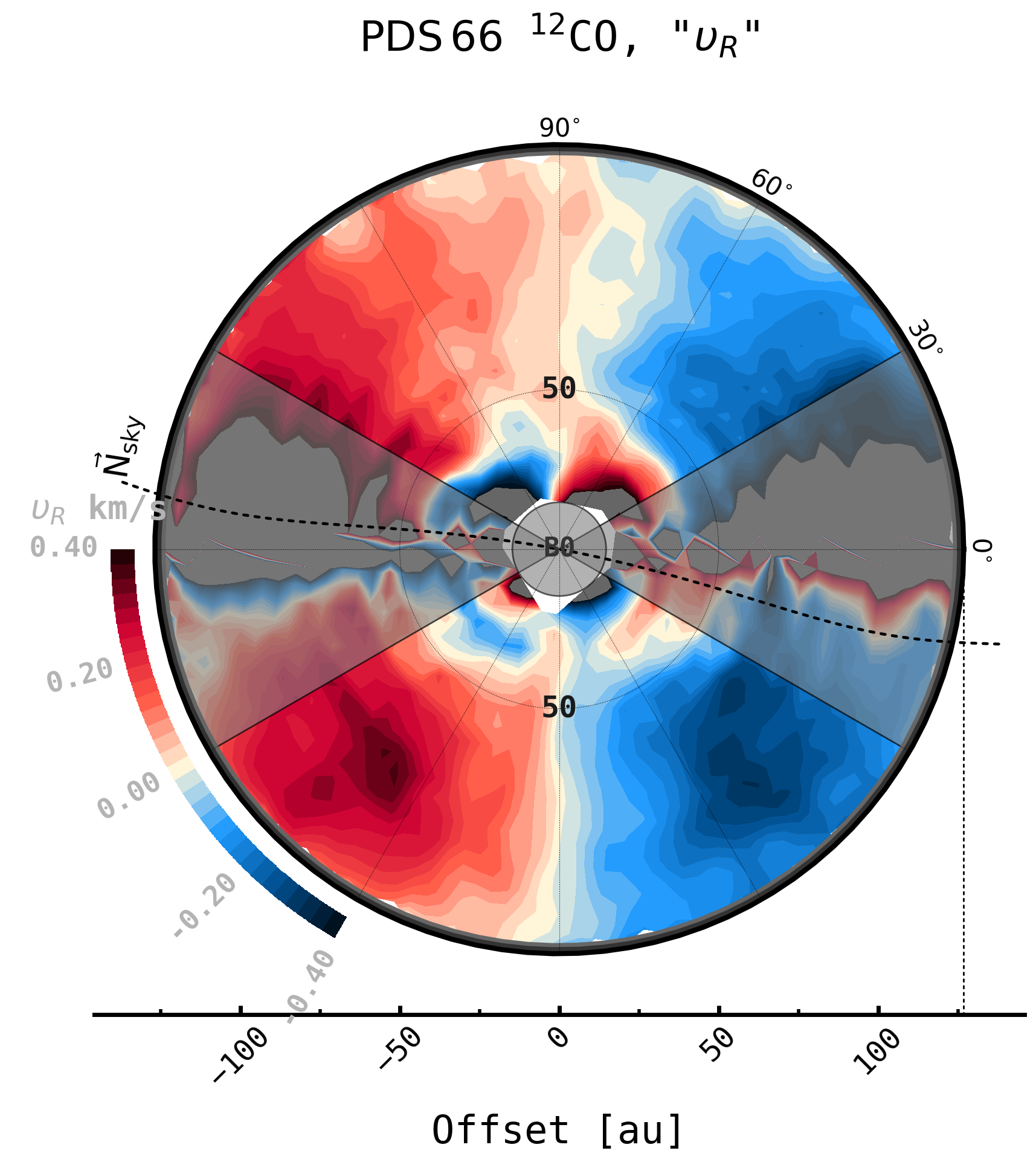} & 
        \includegraphics[width=0.25\textwidth]{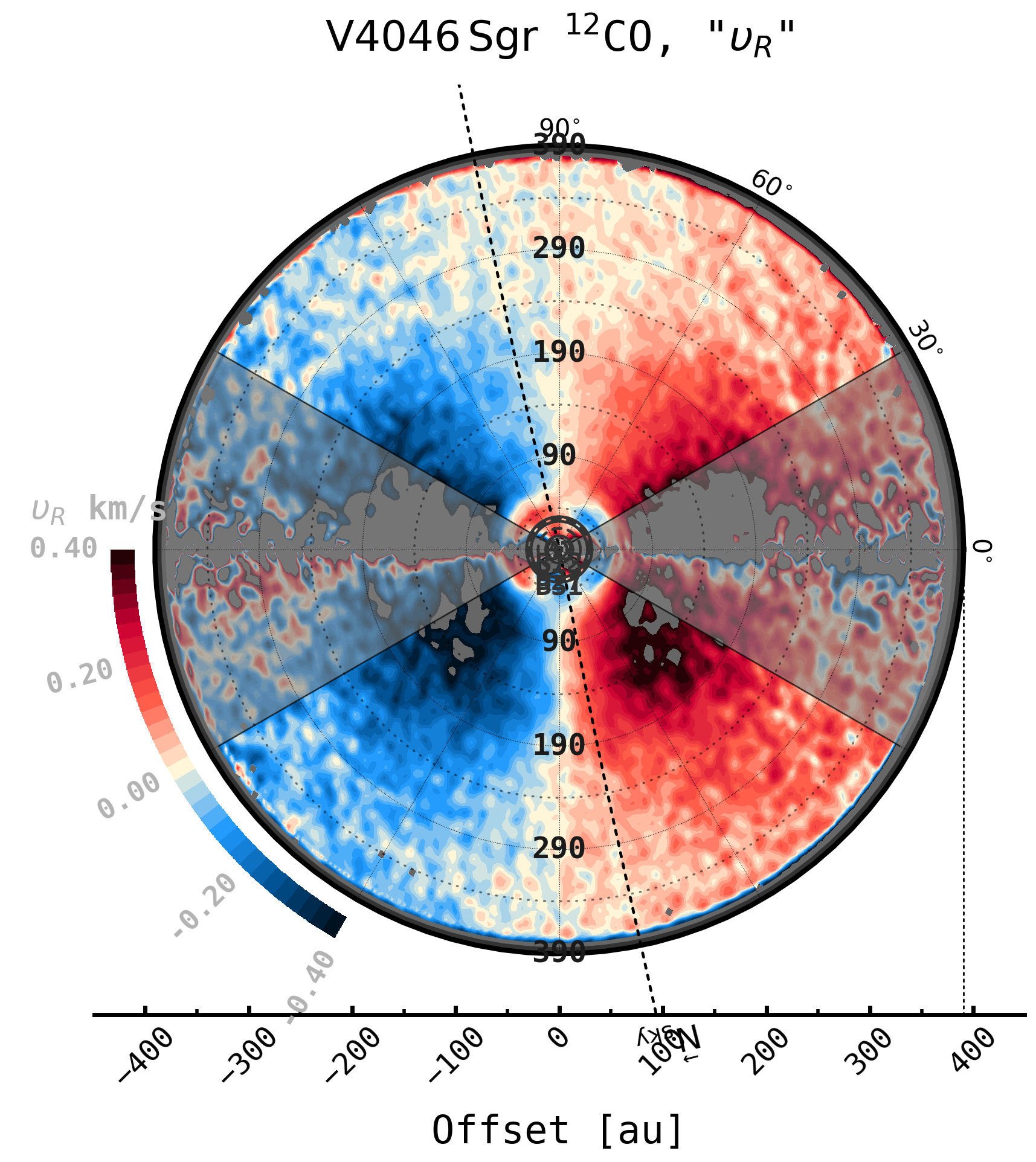} & 
         \\
    \end{tabular}
    \caption{Gallery of 2D radial velocity residual maps for all targets for which we applied the 1D analysis obtained following \citet{Izquierdo_ea_2023}. The grey area represent the major axis where radial velocities diverge. \label{fig:vr_leftover_gallery}}
\end{figure*}

\begin{figure*}[t]
    \centering
    \includegraphics[width=1\linewidth]{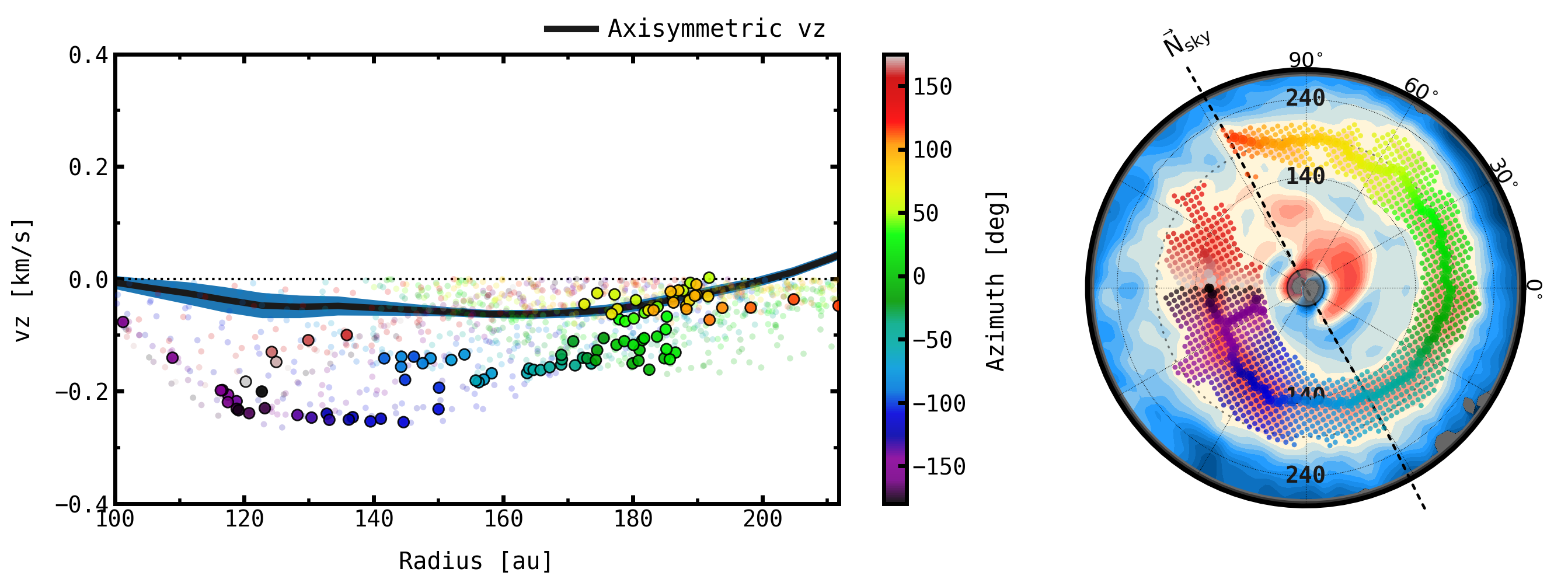}
    \caption{An axisymmetric analysis of the velocity in MWC758 (black line) would underestimate the vertical motions at the radial locations traced by the spiral spine.}
    \label{fig:comp1d}
\end{figure*}

Figure \ref{fig:folded_gallery} shows a gallery of folded residual maps. Figures \ref{fig:vz1Dno} and \ref{fig:vz1DMAPS}  provide the radial profiles of vertical and radial velocities, for the exoALMA targets for which we do not detect clear vertical motions, and for the MAPS targets.  Figure \ref{fig:vr_leftover_gallery} shows a gallery of residuals in radial velocity component, from which an azimuthal profile is extracted.  Figure \ref{fig:comp1d} shows how a \vz{} extraction following an azimuthally average would underestimate the velocities in the spirals in MWC758. 

\end{document}